\documentclass[longauth]{aa} 
\usepackage{graphicx}
\usepackage[varg]{txfonts}
\usepackage{color}
\usepackage{ulem}
\usepackage{natbib}
\bibpunct{(}{)}{;}{a}{}{,} 

\def\mpcoh{\,h^{-1}{\rm Mpc}}
\def\kms{\,{\rm km\, s}^{-1}}

\newcommand{\bS}{\mathsf{S} } 
 
\newcommand{\bs}{{\bf s} } 
 
\newcommand{\bSL}{\mathsf{S}_L }

\begin{document}
  \title{The VIMOS Public Extragalactic Redshift
Survey (VIPERS)\thanks{Based on observations collected at the European Southern
Observatory, Cerro Paranal, Chile, using the Very Large Telescope
under programs 182.A-0886 and partly 070.A-9007.  Also based on
observations obtained with MegaPrime/MegaCam, a joint project of CFHT
and CEA/DAPNIA, at the Canada-France-Hawaii Telescope (CFHT), which is
operated by the National Research Council (NRC) of Canada, the
Institut National des Sciences de l'Univers, of the Centre
National de la Recherche Scientifique (CNRS) of France, and the
University of Hawaii. This work is based in part on data products
produced at TERAPIX and the Canadian Astronomy Data Centre as part of
the Canada-France-Hawaii Telescope Legacy Survey, a collaborative
project of NRC and CNRS. The VIPERS web site is
http://www.vipers.inaf.it/. } 
}
\subtitle{Never mind the gaps: comparing techniques to restore 
  homogeneous sky coverage}

\titlerunning{Comparing techniques to restore gaps}

   \author{O.~Cucciati\inst{1,2}
           \and
	   B.~R.~Granett\inst{3}
           \and
	   E.~Branchini\inst{4,5,6}
	   \and F.~Marulli\inst{1,7,2}
\and A.~Iovino\inst{3}
\and L.~Moscardini\inst{1,7,2}
\and J.~Bel\inst{3}
\and A.~Cappi\inst{2}
\and J.~A.~Peacock\inst{8}
\and S.~de la Torre\inst{8}
\and M.~Bolzonella\inst{2}           
\and L.~Guzzo\inst{3,9}
\and M.~Polletta\inst{10}
\and A.~Fritz\inst{10}
\and C.~Adami\inst{11}
\and D.~Bottini\inst{10}
\and J.~Coupon\inst{12}
\and I.~Davidzon\inst{2,1}
\and P.~Franzetti\inst{10}
\and M.~Fumana\inst{10}
\and B.~Garilli\inst{10,11}     
\and J.~Krywult\inst{13}
\and K.~Ma{\l}ek\inst{14}
\and L.~Paioro\inst{10}
\and A.~Pollo\inst{15,16}
\and M.~Scodeggio\inst{10} 
\and L.~A.~M.~Tasca\inst{11}
\and D.~Vergani\inst{17}
\and A.~Zanichelli\inst{18}
\and C.~Di Porto\inst{2}
\and G.~Zamorani\inst{2}
}

   \offprints{Olga Cucciati (olga.cucciati@oabo.inaf.it)}

   \institute{Dipartimento di Fisica e Astronomia - Universit\`{a} di Bologna, viale Berti Pichat 6/2, I-40127 Bologna, Italy 
\and INAF - Osservatorio Astronomico di Bologna, via Ranzani 1, I-40127, Bologna, Italy 
\and INAF - Osservatorio Astronomico di Brera, Via Brera 28, 20122 Milano, via E. Bianchi 46, 23807 Merate, Italy 
\and Dipartimento di Matematica e Fisica, Universit\`{a} degli Studi Roma Tre, via della Vasca Navale 84, 00146 Roma, Italy 
\and INFN, Sezione di Roma Tre, via della Vasca Navale 84, I-00146 Roma, Italy 
\and INAF - Osservatorio Astronomico di Roma, via Frascati 33, I-00040 Monte Porzio Catone (RM), Italy 
\and INFN, Sezione di Bologna, viale Berti Pichat 6/2, I-40127 Bologna, Italy 
\and SUPA, Institute for Astronomy, University of Edinburgh, Royal Observatory, Blackford Hill, Edinburgh EH9 3HJ, UK 
\and Dipartimento di Fisica, Universit\`a di Milano-Bicocca, P.zza della Scienza 3, I-20126 Milano, Italy 
\and INAF - Istituto di Astrofisica Spaziale e Fisica Cosmica Milano, via Bassini 15, 20133 Milano, Italy
\and Aix Marseille Universit\'e, CNRS, LAM (Laboratoire d'Astrophysique de Marseille) UMR 7326, 13388, Marseille, France  
\and Astronomical Observatory of the University of Geneva, ch. d'Ecogia 16, 1290 Versoix, Switzerland
\and Institute of Physics, Jan Kochanowski University, ul. Swietokrzyska 15, 25-406 Kielce, Poland 
\and Department of Particle and Astrophysical Science, Nagoya University, Furo-cho, Chikusa-ku, 464-8602 Nagoya, Japan 
\and Astronomical Observatory of the Jagiellonian University, Orla 171, 30-001 Cracow, Poland 
\and National Centre for Nuclear Research, ul. Hoza 69, 00-681 Warszawa, Poland 
\and INAF - Istituto di Astrofisica Spaziale e Fisica Cosmica Bologna, via Gobetti 101, I-40129 Bologna, Italy 
\and INAF - Istituto di Radioastronomia, via Gobetti 101, I-40129, Bologna, Italy 
}

 
  \abstract
   {}
   {Non-uniform sampling and gaps in sky coverage are common in galaxy
   redshift surveys, but these effects can degrade galaxy
   counts-in-cells measurements and density estimates.  We carry out a
   comparative study of methods that aim to fill the gaps to correct
   for the systematic effects.  Our study is motivated by the analysis
   of the VIMOS Public Extragalactic Redshift Survey (VIPERS), a
   flux-limited survey at $i_{AB}<22.5$ consisting of single-pass
   observations with the VLT Visible Multi-Object Spectrograph (VIMOS)
   with gaps representing 25\% of the surveyed area and an average
   sampling rate of 35\%.  However, our findings are generally
   applicable to other redshift surveys with similar observing strategies.}
   {We applied two algorithms that use photometric redshift information
   and assign redshifts to galaxies based upon the spectroscopic
   redshifts of the nearest neighbours.  We compared these methods with
   two Bayesian methods, the Wiener filter and the Poisson-Lognormal
   filter. Using galaxy mock catalogues we quantified the accuracy and
   precision of the counts-in-cells measurements on scales of $R=5\mpcoh$
    and $8\mpcoh$ after applying each of
   these methods.  We further investigated how these methods perform to
   account for other sources of uncertainty typical of spectroscopic
   surveys, such as the spectroscopic redshift error and the sparse,
   inhomogeneous sampling rate. We analysed each of these sources
   separately, then all together in a mock catalogue that mimicks the
   full observational strategy of a VIPERS-like survey.}
   {In a survey such as VIPERS, the errors in counts-in-cells
   measurements on $R<10\mpcoh$ scales are dominated by the
   sparseness of the sample due to the single-pass observing
   strategy. All methods under-predict  the counts in high-density 
   regions by 20-35\%, depending on the cell size, method, and underlying
   overdensity. This systematic bias is similar to random
   errors.  No method outperforms the others: differences are not
   large, and methods with the smallest random errors can be more
   affected by systematic errors than others. Random and systematic
   errors decrease with the increasing size of the cell. All methods
   can effectively separate under-dense from over-dense regions by
   considering cells in the $1^{st}$ and $5^{th}$ quintiles of the
   probability distribution of the observed counts.}
   {We show that despite systematic uncertainties, it is possible to
   reconstruct the lowest and highest density environments on scales
   of $5\mpcoh$ at moderate redshifts $0.5\lesssim z
   \lesssim 1.1$, over a large volume such as the one covered by the
   VIPERS survey. This is vital for characterising of cosmic
   variance and rare populations (e.g, brightest galaxies) in
   environmental studies at these redshifts.}


   \maketitle


\section{Introduction}\label{intro}

Large-volume spectroscopic redshift surveys have emerged as the best
tool for investigating the large-scale structure of the Universe and,
eventually, for constraining cosmological models. Measuring spectroscopic redshifts and
angular positions allows us to trace the ${\rm 3D}$ distribution
of galaxies and, assuming that they trace the underlying density
field, of the matter.  Effective constraints on the cosmological model
can be obtained by comparing the statistical properties of the galaxy
distribution with theoretical predictions.  Moreover, the apparent
distortions in galaxy clustering induced by peculiar velocities
provide a unique observational test for non-standard gravity models
as alternatives to dark energy to account for the accelerated expansion
of the Universe (e.g. \citealp{guzzo08}).  Indeed, setting these types of
constraints is one of the main scientific drivers of the VIMOS Public
Extragalactic Redshift Survey
(VIPERS\footnote{http://vipers.inaf.it}), an ongoing spectroscopic
survey of about 100~000 galaxies at $z\simeq 0.8$
\citep{guzzo2013_vipers}. VIPERS has already achieved this goal (see
e.g. \citealp{delaTorre2013_clustering}) using part of the dataset now
made available with the first Public Data
Release\footnote{http://vipers.inaf.it/rel-pdr1.html}
\citep{garilli2013_VIPERS}.

The effect of peculiar velocities is just one of the reasons that prevent us from
observing the full 3D distribution of objects. Other effects,
either intrinsic (e.g. Galactic absorption of extragalactic light) or
induced by the observational strategy (selecting objects above
some flux threshold, measuring spectra only for a subset of
potential targets, etc.) and instrumental setup, effectively modulate
the spatial distribution of objects.  All these effects are potential
sources of systematic errors that need to be accurately quantified and
accounted for.

The impact of these effects and their correction depend on the
characteristics of the dataset, on the kind of systematic effects that
need to be corrected for, and on the type of analysis one wishes to
perform.  One of these observational biases is incomplete sky
coverage. Correction for this effect is quite trivial when measuring
clustering statistics in configuration space. In the specific case of
the VIPERS survey, this has been efficiently done through the extensive
use of random samples mimicking the observational biases to estimate
the two-point correlation function of different types of galaxies
\citep{marulli2013_clustering, delaTorre2013_clustering}. 

The types of analysis that are most sensitive to inhomogeneous sky coverage
include the study of galaxy properties and their relation to the local
environment.  Clearly, the presence of unobserved areas with size
comparable to the physical scale that one wishes to investigate can
have a serious impact on the analysis. In this case a large-scale
statistical correction is not sufficient.  Instead, a more local and
deterministic recovery of the missing information is mandatory 
(e.g. \citealp{cucciati2006}).

The best known example of 3D reconstruction is that of the
extragalactic objects behind the Galactic plane where observations are
hampered by strong photon absorption.  This is a long-lasting issue
triggered in the '90s by the search for the `Great Attractor', a
putative large-scale structure responsible for the coherent large-scale 
flows in our cosmic neighbourhood.  The need to fill the
so-called Zone of Avoidance not only has triggered a long-term
observational programme (see e.g. \citealp{kk05}) but also the
development of techniques able to fill the unobserved regions while
preserving the coherence of the large-scale structure.
Among these techniques, those more relevant for our work are the
cloning, or randomised cloning, of the 3D distribution of objects into
unobserved areas (e.g. \citealp{Y91,B99}) and the application of the
Wiener Filtering technique (e.g. \citealp{L94}).

Here we build on these and other more recently developed and
sophisticated techniques, such as ZADE \citep{kovac2010_density} and
Poisson-Lognormal Filtering \citep{Kitaura2010}, to tackle the problem
of reconstructing the 3D distribution of galaxies in the unobserved
regions of a much deeper redshift survey. As anticipated, this
analysis is targeted to a specific dataset, VIPERS in this
case.  We can nevertheless draw some general conclusions from this
exercise, since some of the problems we address are indeed of
interest for future potential surveys at similar or higher redshift
that aim at maximising both volume and sampling.

From the point of view of the angular coverage, VIPERS can be considered as
a typical example of a survey in which unobserved areas constitute a 
sizeable fraction ($\sim 25$\%) of the total and are characterised by their regular
pattern that reflects the footprint of the spectrograph. 
Some of the systematic effects, such as 
the violation of the local Poisson hypothesis in cell counting statistics
\citep{bel13},  can be amplified by the presence of regular gaps, given the 
significant fraction of unobserved sky.

We do not limit our analysis to the effect of inhomogeneous
sky coverage.  All other effects, ranging from sparse, inhomogeneous
and clustering-dependent galaxy sampling, radial selection induced by
the flux threshold, redshift measurement errors, as well as
incompleteness induced by selection criteria, are also folded into our
analysis. Again, some of these effects are specific to VIPERS but the
dominant ones (sparse sampling and flux limit cut) are quite common to
general-purpose surveys that aim at both cosmological and galaxy
evolution studies.

The paper's layout is as follows. In Sect.~\ref{data_mocks} we describe
VIPERS data and the mock galaxy catalogues we use in the analysis. We
also list the sources of uncertainties for counts in cells.  In
Sect.~\ref{methods} we discuss the methods that we use to fill the
gaps, and in Sect.~\ref{countsincells} we describe how we perform our
analysis (the considered sources of uncertainty included in the
different mock catalogues, the kinds of comparison we carry on, and
the samples and redshift range we consider). Our results are presented
in Sect.~\ref{results}, and in Sect.~\ref{discussion} we summarise and
discuss them. The Appendix gives more details about some specific results. 
In this work, we use the same cosmology assumed in the
dark matter N-body simulation on which our mock galaxy catalogues are
based (see Sect.~\ref{mocks}), i.e. a flat $\Lambda$CDM cosmology with
$\Omega_m=0.27$, $\Omega_{\Lambda}=0.73$, and $H_0=70\kms {\rm
Mpc}^{-1}$. Magnitudes are expressed in the AB system \citep{oke74,fukugita96}.


\section{Data and mock samples}\label{data_mocks}

\subsection{Data}\label{data}

The VIMOS Public Extragalactic Redshift Survey
(VIPERS) is an ongoing Large Programme
aimed at measuring redshifts for $\sim 10^5$ galaxies at redshift $0.5
< z \lesssim 1.2$. The main scientific drivers of this survey are a
robust and accurate measurement of galaxy clustering and of the growth
of structure through redshift-space distortions and the study of
galaxy properties at an epoch when the Universe was about half its
current age.

\begin{figure*} \centering
\includegraphics[width=18cm]{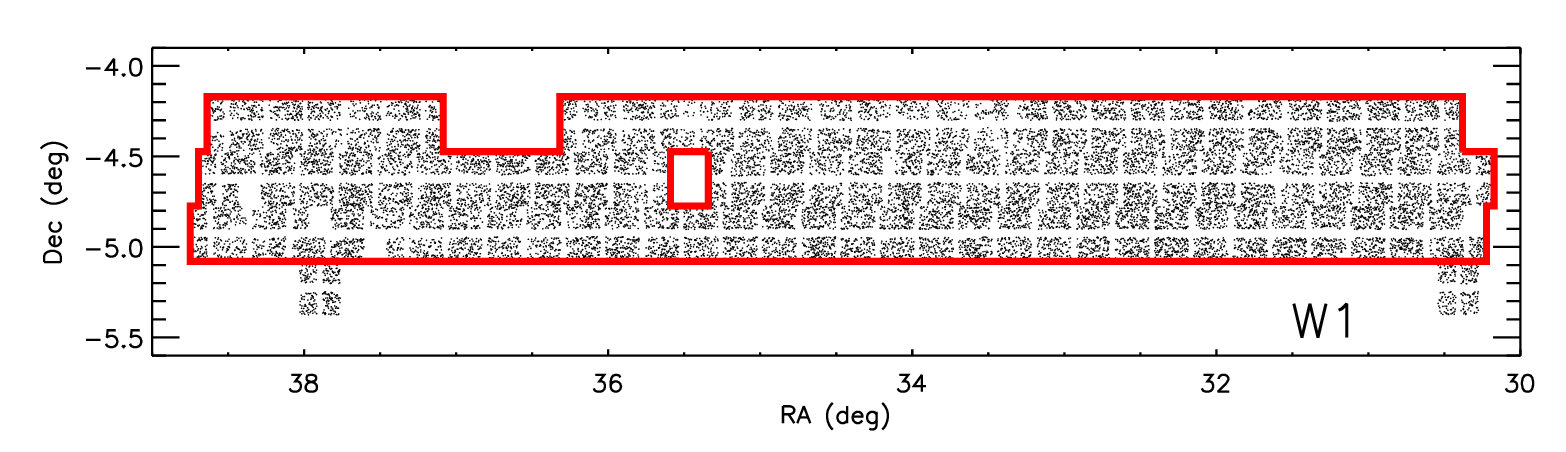}
\includegraphics[width=18cm]{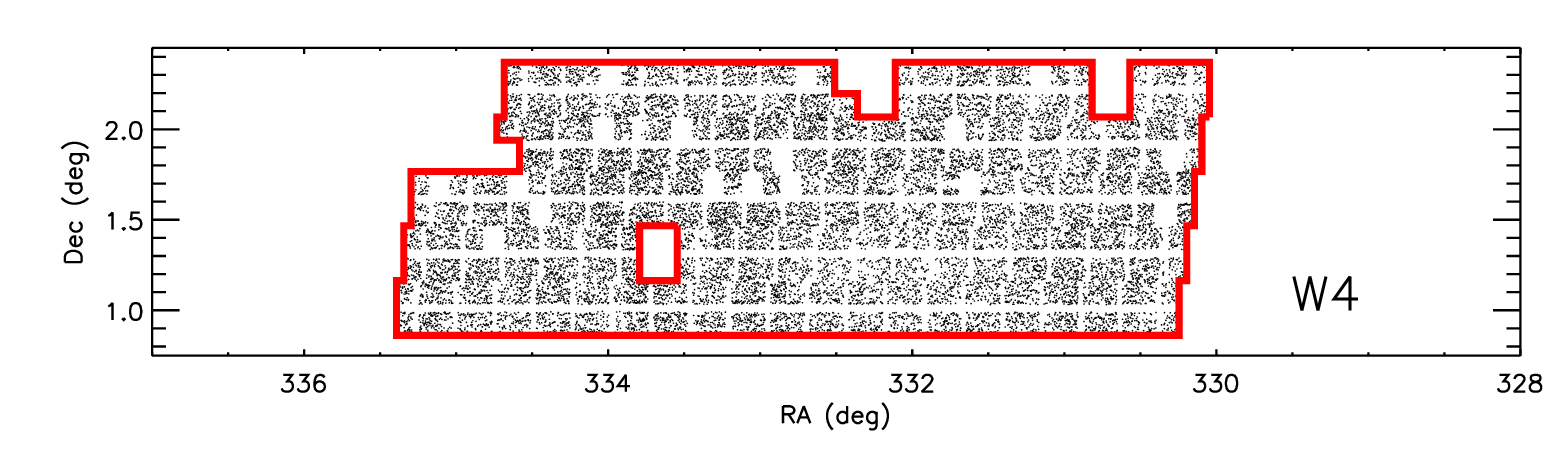}
\caption{$RA$-$Dec$ distribution of galaxies with reliable redshift (see
text for details) in the VIPERS Public Release PDR-1 in W1 (top) and
W4 (below) fields. The thick red line in each panel is the `field boundary'
that we consider in this work. The  VIMOS footprint with four quadrants
 is visible from the two single pointings at $Dec\lesssim-5.08$ deg in the W1 field
at $RA=37.9$ and $RA=30.4$.}  \label{fields_fig}
\end{figure*}

At completion, VIPERS will cover $\sim24$ deg$^2$ on the sky, divided
into two areas within the W1 and W4 CFHTLS fields. The parent
photometric catalogue from which VIPERS targets are selected is the
Canada-France-Hawaii Telescope Legacy Survey Wide (CFHTLS-Wide)
optical photometric catalogues \citep{CFHTLS}. Galaxies are selected
to a limit of $i_{AB}<22.5$, and a colour pre-selection in $(g-r)$ vs
$(r-i)$ is also applied to efficiently remove galaxies at $z<0.5$. In
combination with an optimised observing strategy
\citep{scodeggio2009_VIMOS}, this allows us to double the galaxy
sampling rate in the redshift range of interest with respect to a
purely flux-limited sample ($\sim 40\%$).  The final surveyed volume
will be $5 \times 10^{7}$ h$^{-3}$ Mpc$^{3}$, similar to that of
the 2dFGRS at $z\sim0.1$.  VIPERS spectroscopic observations are
carried out with the VLT Visible Multi-Object Spectrograph \citep[VIMOS,][]{lefevre2002,lefevre2003}, 
using the LR Red grism (resolution
$R=210$, wavelength coverage of 5500-9500$\AA$). The typical
radial velocity error is of $\sim140\kms$.

A discussion of the survey data reduction and management
infrastructure is presented in \citet{garilli2012}, and the complete
description of the survey is given by \cite{guzzo2013_vipers}.  The
data set used in this paper represents the VIPERS Public Data Release
1 (PDR-1) catalogue, which was made publicly available in Fall 2013
\citep{garilli2013_VIPERS}. It consists of $\sim 47000$ galaxies and
AGN with reliable spectroscopic redshift. We consider reliable
redshifts to be those with spectroscopic quality flag equal to 2, 3, 4, 9 (see
\citealp{garilli2013_VIPERS} for a detailed description).

The VIMOS instrument is composed of four CCDs (`quadrants') with a field
of view of $8' \times 7'$ each, for a total field of view of $218$
arcmin$^2$. The four CCDs are placed on a $2\times2$ grid with a $2'$ 
separation. The VIMOS pointing footprint is thus characterised by a 
cross with no data (see below).

As mentioned above, the `parent' photometric catalogue from which
targets are selected has a flux limit of $i_{AB}<22.5$ and a colour
cut to remove galaxies at $z<0.5$. The consequent cut in redshift is
not sharp, but has a smooth transition from $z\sim0.4$ to
$z\sim0.6$. This has an effect on the radial selection function of the
survey. We quantify this effect by the colour sampling rate (CSR from now
on). The CSR depends on redshift, and it is equal to 1 for $z\geq
0.6$.  Moreover, the VIPERS observational strategy consists
in targeting for observations $\sim40$\% of the galaxies in the parent
photometric catalogue, and in addition not all the targeted galaxies
yield a reliable redshift measurement. All these effects need to be
taken into account when deriving the VIPERS selection function. We
define the target sampling rate (TSR) as the fraction of galaxies in
the parent photometric catalogue that have been targeted, and the
spectroscopic success rate (SSR) the fraction of targeted galaxies for
which a reliable redshift has been measured. The average VIPERS
sampling rate, considering the TSR and SSR together, is $\sim35$\%.
The VIPERS sampling rate is not uniform on the surveyed area.  Both
TSR and SSR depend on VIMOS quadrant.  The TSR is higher when the
surveyed sky region has a lower target surface density, while the SSR
can vary quadrant per quadrant because of different observational
conditions. In each quadrant, the number of slits is maximised using
the SPOC algorithm \citep{bottini2005}. We refer the reader to
\cite{delaTorre2013_clustering} and \cite{fritz14_red} for more
details on the VIPERS selection function.

Figure \ref{fields_fig} shows the $RA$-$Dec$ distribution of the
galaxies and AGN with reliable redshift (see above) in the PDR-1 in
the fields W1 and W4. The cross-like pattern of void regions is
evident, together with larger empty regions corresponding to quadrants
or pointings that have been discarded owing to technical problems or
poor observational conditions.

Photometric redshifts ($z_p$) were computed for all the galaxies
in the photometric catalogue, as described in \cite{coupon2009} but
using CFHTLS T0005 photometry. At $i_{AB}<22.5$, the photometric
redshift error is $\sigma_{zp} = 0.035(1+z)$, with an outlier fraction of
3-4\%. From now on we call `spectroscopic galaxies' the galaxies with 
a reliable spectroscopic redshift, and `photometric galaxies' all the 
other galaxies in the parent photometric catalogue having a 
photometric redshift.

Absolute magnitudes were obtained via spectral energy distribution
(SED) fitting technique, using the algorithm {\it Hyperzmass}, an
updated version of {\it Hyperz}
\citep{bolzonella2000,bolzonella2010}. We used a template library
from \cite{BC03}, with solar and sub-solar ($Z=0.2\,Z_\odot$) metallicity, exponentially
declining star formation histories and a model with constant star
formation.  The extinction laws of the Small Magellanic Cloud (SMC,
\citealp{prevot84,bouchet85}) and of \cite{calzetti00} have been applied to the
SEDs, with $A_V$ ranging from 0 to 3 magnitudes.  The observed filters
used to compute the SED fitting are the T0005 CFHTLS $u*g'r'i'z'$
filters plus ancillary photometry from UV to IR. For more details we
refer the reader to \cite{davidzon2013} and \cite{fritz14_red}.

\subsection{Mock samples}\label{mocks}

We use 26 independent mock galaxy catalogues constructed using the halo 
occupation distribution (HOD) method as
detailed in \cite{delaTorre2013_clustering}. These mock
catalogues were obtained by assigning galaxies to the DM haloes of the
MultiDark simulation, a large N-body run based on $\Lambda$CDM
cosmology \citep{prada2012_multidark}.  The mass resolution limit of
this simulation ($10^{11.5} h^{-1} M_{\odot}$) is too high to
include the less massive galaxies observed in VIPERS. To
simulate the entire mass and luminosity range covered by VIPERS, the
MultiDark simulation has been repopulated with haloes of mass below
the resolution limit. We refer the reader to \cite{delaTorre2012_HOD}
for details. We note that, although these HOD catalogues are based
on $\Lambda$CDM, ideally one would like to use a range of different
cosmological models and study the desired statistics in all cases, 
but this goes beyond the aim of this paper.

These HOD mock catalogues contain all the information we need, including, for each galaxy,
right ascension, declination, redshift (cosmological redshift with
peculiar velocity added), $i$-band observed magnitude, and $B$-band absolute
magnitude. Moreover, we simulated the spectroscopic and photometric
redshift, adding to the redshift an error extracted randomly from a
Gaussian distribution with standard deviation equal to the spectroscopic and
photometric redshift errors, respectively.

Applying the cut $i_{AB}\leq 22.5$ to the HOD mock catalogues, we
obtain the `total photometric mock catalogues.' We extract from these
catalogues the `parent photometric mock catalogues,' applying a radial
selection corresponding to the VIPERS CSR (i.e., we deplete the mock
catalogues at $z<0.6$ according to the CSR). Next, we apply to the parent photometric
catalogues the same slit positioning tool (VMMPS/SPOC,
\citealp{bottini2005}) as was used to prepare the VIPERS observations.
In this way we have mock catalogues with the same footprint on the sky as VIPERS, 
and we further deplete such catalogues to mimic the effects of the SSR to
obtain galaxy mock catalogues that fully reproduce the effects of the
VIPERS observational strategy. We call these mocks VIPERS-like mock catalogues.


\subsection{Sources of uncertainty for counts in cells}\label{err_sources}

The observational strategy of the VIPERS survey implies some specific
observational biases, some due to the instrumentation (and so common
to all surveys that observe with the same configuration), and some
specific to VIPERS.  We point out that we work in redshift space, but see
Sect.~\ref{testA_sec} for a brief discussion of counts in real space.

Here we describe in details the VIPERS observational biases.

\begin{itemize}

\item[A.] {\bf Redshift measurement error.} As mentioned in
Sect.~\ref{data}, the typical spectroscopic redshift error is
$140\kms$, corresponding to $\Delta z_s \sim 0.0005(1+z)$, while
the photometric redshift error is $\Delta z \sim 0.035(1+z)$. The first
is due to the combination of the resolution of adopted grism (which 
is the main source of uncertainty), the flux limit of our
sources and the exposure time.  The second is mainly due 
to the number of photometric bands
available in the surveyed fields. In the present work, the effects of
the spectroscopic redshift error will be accounted for in Test A (see
Sect.~\ref{levels}), while the photometric redshift error will be used
only in the gaps-filling methods that make use of photometric
redshifts.

\item[B.] {\bf Gaps and field boundaries.} The total area covered by a
VIMOS pointing (the 4 CCDs plus the space between them) is about 290
arcmin$^2$. The effective area covered by the four CCDs is about 218
arcmin$^2$. This means that the gaps between the quadrants represent
$\sim25$\% of the VIPERS field. The distance between the CCDs (2')
corresponds to $\sim0.7\mpcoh$ and $\sim1.5\mpcoh$ (comoving) at $z=0.5$ and
$z=1.1$, respectively. From Fig.~\ref{fields_fig} it is evident that
there are also other unobserved regions, such as missing quadrants or
even pointings. 

Figure \ref{fields_fig} shows what we call the
`field boundaries', i.e. the borders of the total surveyed area,
disregarding the presence of gaps and missing quadrants.  Fully
missing pointings, however, are considered to be outside the survey
boundaries. Moreover, in W1 we exclude the two observed pointings at
$Dec<-5.08$.

In this work, we make use of the VIPERS galaxy sample enclosed in such
field boundaries. The total areas enclosed in these regions are
7.35 and 7.19 deg$^2$ in the W1 and W4 fields, respectively. If we exclude gaps
and missing quadrants, we have an effective area of 5.37 deg$^2$ in W1
and 5.11 deg$^2$ in W4. This means that the sky area to be `filled'
for the counts in cell is about 27\% in W1 and 29\% in W4 
(see Test B in Sect.~\ref{levels}). 

We note that, in our analysis of counts in cells, we  only
consider cells that are fully contained within the survey boundaries
(i.e. that do not overlap with the red edges in Figure~\ref{fields_fig}),
which do not require any statistical correction for edge-induced
incompleteness. It has already been shown that, in a spectroscopic
survey with a sampling rate of 25-35\%, boundary effects can be
corrected by computing the fraction of the volume of each cell falling
outside the surveyed field (see
e.g. \citealp{cucciati2006}).

\item[C.] {\bf Sampling rate and effect of slit positioning.} The VIPERS
selection function (see Sect.~\ref{data}) is given by the product of
CSR, TSR, and SSR and depends on observed magnitude, redshift, and
quadrant. The net effect is that the overall sampling rate in VIPERS,
with respect to a full photometric catalogue limited at $i_{AB}=22.5$,
is well below 100\%. This increases the shot noise, making it more
difficult to properly recover the tails of the counts-in-cell
distribution.  Moreover, the slit positioning system
(SPOC, \citealp{bottini2005}) induces scale-dependent sampling of the
objects within each quadrant.  We notice that such inhomogeneities are
produced on much smaller scales ($<1\mpcoh$)  than the ones we will
explore in this work (see \citealp{delaTorre2013_clustering}).

The overall sparseness of the sample will be accounted for in Test C1,
while Test C2 will also consider i) the fact that the
sampling rate depends on quadrant and ii) the effects induced by the
slit-positioning software (see Sect.~\ref{levels}). 

\end{itemize}

In the real VIPERS sample, all these effects are present, and their
overall effect will be tested in Test D (see Sect.~\ref{levels}).


\section{Filling the gaps}\label{methods}

In this section we discuss the methods we tested to
fill/correct the gaps. With the aim of reliably reproducing the
counts in cell in a complete (100\% sampling) galaxy catalogue, we
necessarily have to deal also with the other observational biases (low
and not homogeneous sampling, redshift error). The effects of all
these biases are also studied (see Sect.~\ref{levels}), and their
impact on the gap-filling accuracy is described in Sect.~\ref{results}.

\subsection{ZADE}\label{methods_ZADE}

This method is a modified version of the ZADE approach described in
\cite{kovac2010_density}. It can be briefly described as follows. We
take all galaxies in gaps. For each of these galaxies, we keep its
angular position ({\it RA} and {\it Dec}), but we spread its
photometric redshift ($z_{p,i}$) over several probability peaks along
its line of sight (l.o.s). We assign a weight ($w_{ZADE}$) to each of
these peaks according to their relative height, normalised so that the
the total probability corresponding to the sum of the weights is
unity. For a given galaxy $i$ in the gaps, the weights $w_{ZADE}$ and
the positions along the l.o.s. of the peaks are computed as
follows. First, we consider the measured photometric redshift of the
$i$-th galaxy, $z_{p,i}$, and set the probability of $z_{p,i}$,
$P(z_{p,i})$, equal to a Gaussian centred on $z_{p,i}$ with standard
deviation equal to the 1 $\sigma$ error in the photometric redshift,
$\sigma_{zp}$=0.035(1+$z$). Then, we select all galaxies in the
spectroscopic sample that are within a cylinder centred on the
position of the galaxy ({\it RA}$_i$, {\it Dec}$_i$, $z_{p,i}$) with
radius $R_{ZADE}$ (see below) and half-length equal to $3\sigma_{pz}$ and compute
their redshift distribution, $n(z_{s,i})$. Finally, we form the
probability function associated with the $i$-th galaxy as
$P(z_i)=AP(z_{p,i})n(z_{s,i})$, where $A$ is a normalisation factor.
This function, which represents the probability of the galaxy along
the l.o.s., is characterised by several peaks ($20-25$, depending on
the sparseness of the spectroscopic galaxies, given by the kind of
mock catalogue and by the luminosity limit, see Sects.~\ref{levels} and 
\ref{used_samples}). The value of $P(z_i)$ at each redshift peak
corresponds to the weight $w_{ZADE}$ at that given redshift. We note
that, with the ZADE approach, the resulting distribution of `redshift
peaks', for a given value of $z_p$, is unbiased. As a comparison, we
refer the reader to \cite{francis10} for the discussion of a different
$z_p$ recovery method.

ZADE exploits the correlation properties of the spatial distribution
of galaxies. Therefore it is natural to choose a value for $R_{ZADE}$ 
close to that of the correlation length of VIPERS galaxies
\citep{marulli2013_clustering}. Based on this consideration, the value
of $R_{ZADE}$ is a compromise between the need to maximise the number
of galaxies with spectroscopic redshift within the cylinder, to reduce
shot noise, and to minimise the size of the cylinder, to probe the
smallest possible scales. Here we use $R_{ZADE}=5\mpcoh$ as a
reference case, but we have systematically checked the robustness of our
results by varying $R_{ZADE}$ between 3 and 10 $\mpcoh$.

Our tests show that the performance of ZADE degrades significantly
when one considers only a few prominent peaks of the probability
distribution $P(z_i)$. Therefore in our implementation of ZADE we
decided to use the full probability distribution with all probability
peaks.

We used this method to fill gaps and missing quadrants. Given the size
of $R_{ZADE}$, we cannot use ZADE to fill areas as big as the missing
pointings, for which we would lack (a reasonable number of) galaxies
with $z_s$ within $R_{ZADE}$. The ZADE method also allows us to
increase the sampling rate within the observed quadrants, because we
can apply ZADE to all galaxies having a photometric redshift.

\begin{figure} \centering
\includegraphics[width=9cm]{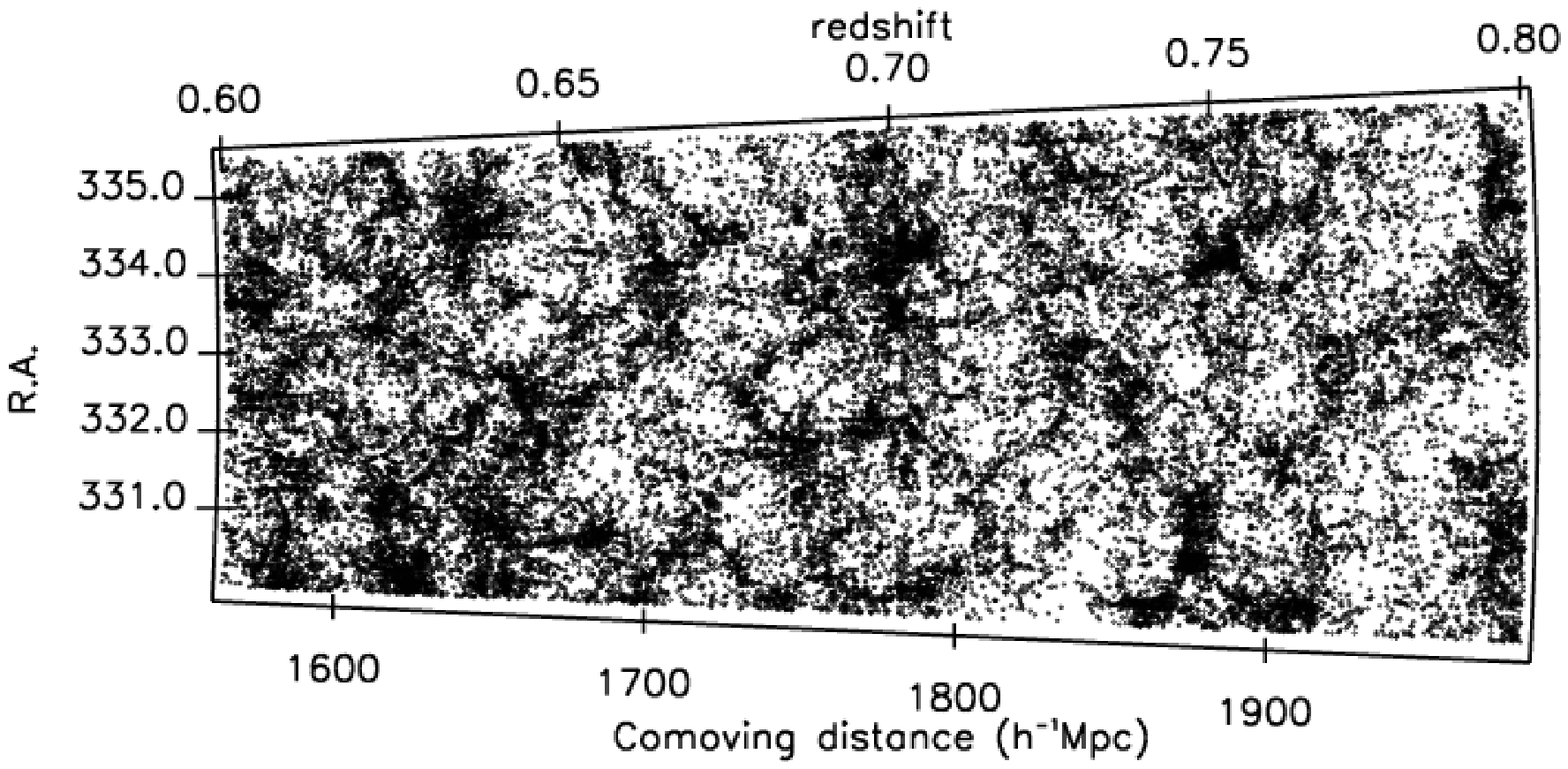}
\includegraphics[width=9cm]{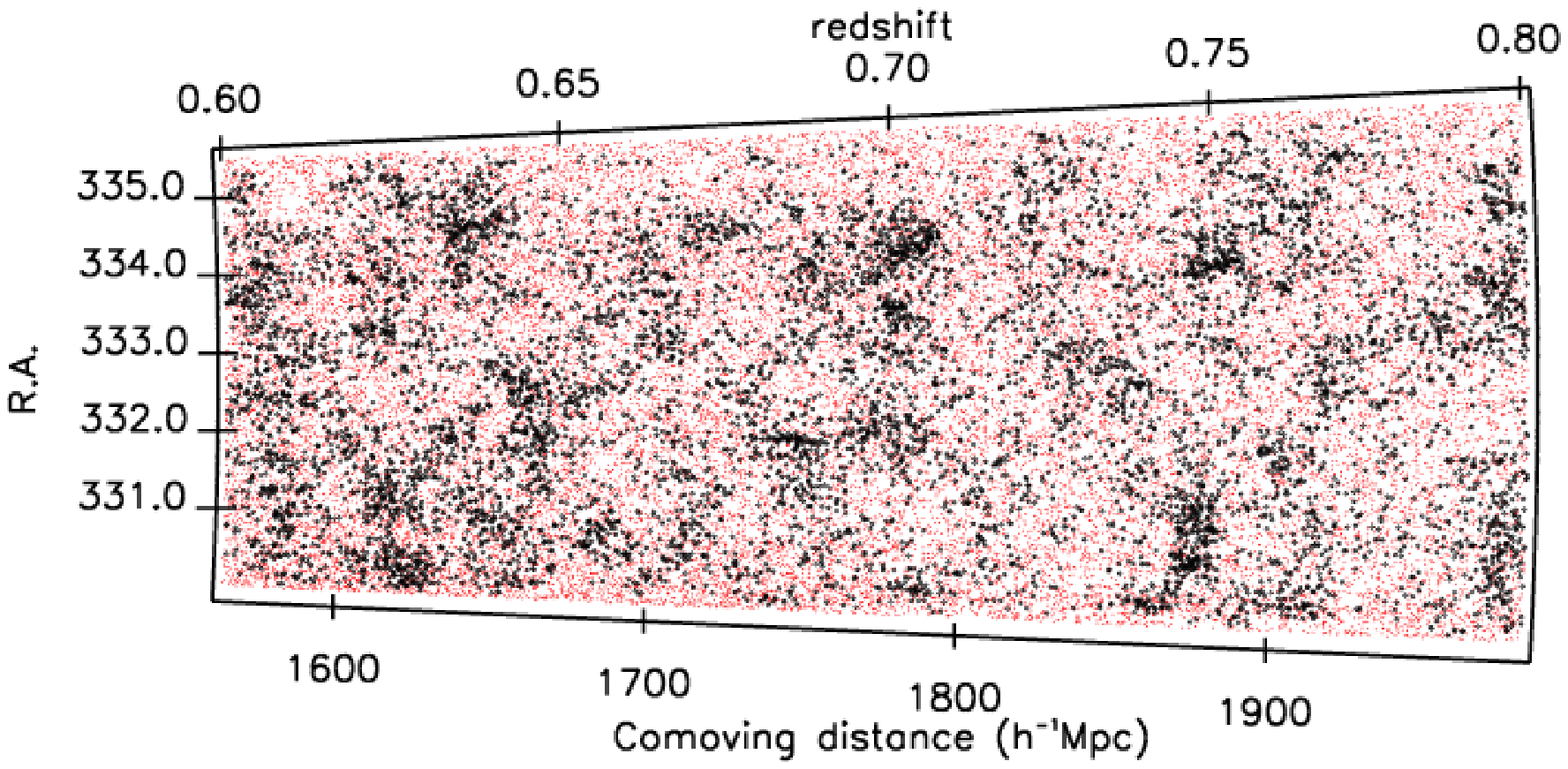}
\includegraphics[width=9cm]{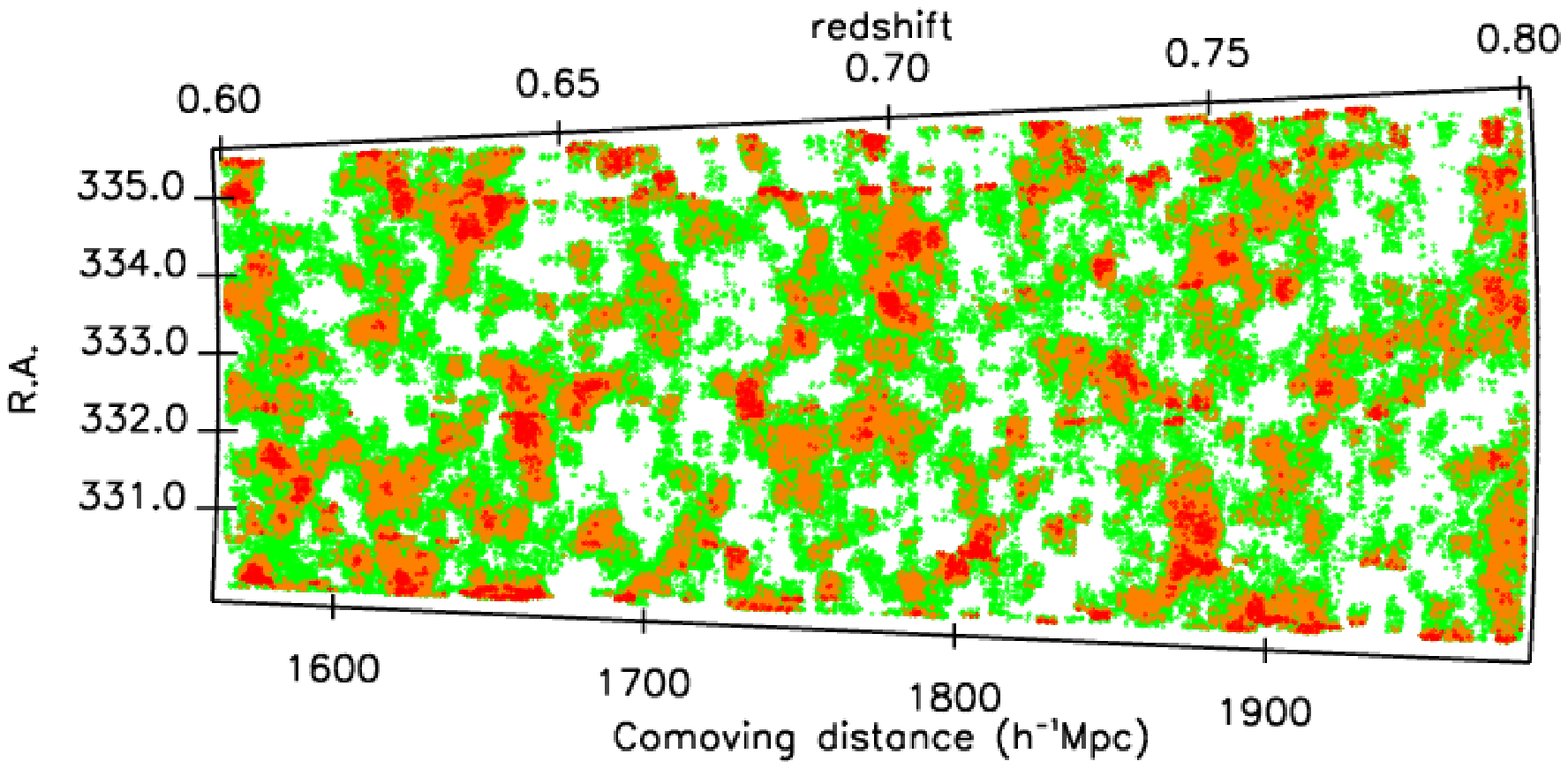}
\caption{2D distribution of galaxies (where the
projected $Dec$ covers a range of $1.5^\circ$) in one of the W4 field mock catalogues used
in this work.  {\it Top:} all galaxies in the reference mock catalogue
(100\% sampling rate, the redshift includes cosmological redshift and
peculiar velocity). {\it Middle:} galaxies in the Test D mock catalogue
(see Sect.~\ref{levels} for more
details). The catalogue includes $\sim35$\% of galaxies in quadrants
(black points) with spectroscopic redshift, while the remaining 65\%
in quadrants and all the galaxies in gaps have a photometric redshift
(red dots).  {\it Bottom:} Equal probability contours for the ZADE probability 
function $P(z_i)$ (see Sect.~\ref{methods_ZADE} for details). Colour code is set according to 
the statistical weight $w_{ZADE}$. Red: $w_{ZADE} \geq 0.2$, Orange: $0.1 \leq w_{ZADE} < 0.2$.
Green: $0.05 \leq w_{ZADE} < 0.1$.
} \label{ZADE_fig} \end{figure}

Figure ~\ref{ZADE_fig} shows how well ZADE, when applied to a VIPERS-like mock
catalogue, is able to recover the position of photometric galaxies
along the l.o.s. The top panel shows how galaxies are
distributed in a mock catalogue with 100\% sampling rate and no error
on the galaxy redshift. Comparing the top and the mid panels, it
is clear that galaxy structures are blurred when introducing a
photometric redshift error. The bottom panel shows how the weights $w_{ZADE}$
assigned by ZADE are distributed: it is evident that weights are
greater in correspondence with the prominent concentrations of 
objects in the spectroscopic catalogue, as expected (see top panel of Fig.~\ref{ZADE_fig}).

\subsection{Cloning}\label{cloning}

This method replicates the spatial position of spectroscopic galaxies 
near the edges of the surveyed areas into the unobserved regions 
(gaps and missing quadrants) to preserve the coherence of the large-scale 
structure without necessarily reconstructing the actual position 
of the missing objects.

We proceed as follows. A spectroscopic galaxy in a
quadrant with 3D coordinates $ra_0$, $dec_0$ and $z_0$ (where $z_0$ 
is the spectroscopic redshift) is cloned in a gap 
with new coordinates $ra_g$, ~$dec_g$, ~$z_g$ assigned in two steps:

{\bf - STEP 1.} We define $z_g$ = $z_0$ plus an error extracted 
from a Gaussian distribution with $\sigma$=0.0005(1+$z_0$), i.e. the typical 
spectroscopic redshift error; $ra_g$ and $dec_g$ are cloned from $ra_0$ and 
$dec_0$, by adding or subtracting an angular offset to either the 
$ra_0$ or $dec_0$ equal to the angular distance between the real 
galaxy and the edge of gap multiplied by two.

{\bf - STEP 2.} Once $ra_g$, $dec_g$, and $z_g$ are assigned according
to STEP 1, $ra_g$ and $dec_g$ are set equal to $RA$ and $Dec$ of the
closest photometric galaxy in the gaps. We compute an adimensional
distance between the position of the cloned galaxy ($ra_g$, ~$dec_g$,
~$z_g$) and the position of the photometric galaxies ($ra_p$, ~$dec_p$,
~$z_p$) using the formula 
\begin{equation} \displaystyle 
dist^2 = \frac{(z_g-z_p)^2}{\Delta_z^2} + \alpha  \frac{(ra_g-ra_p)^2+(dec_g-dec_p)^2}{\Delta_{ang}^2}, 
\label{cloning_distance} 
\end{equation}
\noindent where $\Delta_z$ and $\Delta_{ang}$ are two normalisations, namely
$\Delta_z=0.035(1+z_g)$ (i.e. the photometric redshift error) and
$\Delta_{ang}=1$ arcmin; $RA$ and $Dec$ are expressed in arcmin;
$\alpha$ is an ad-hoc chosen factor used to transform angular distances
into redshift distances. A photometric galaxy can only be assigned to one single cloned galaxy.

\begin{figure} \centering
\includegraphics[width=9cm]{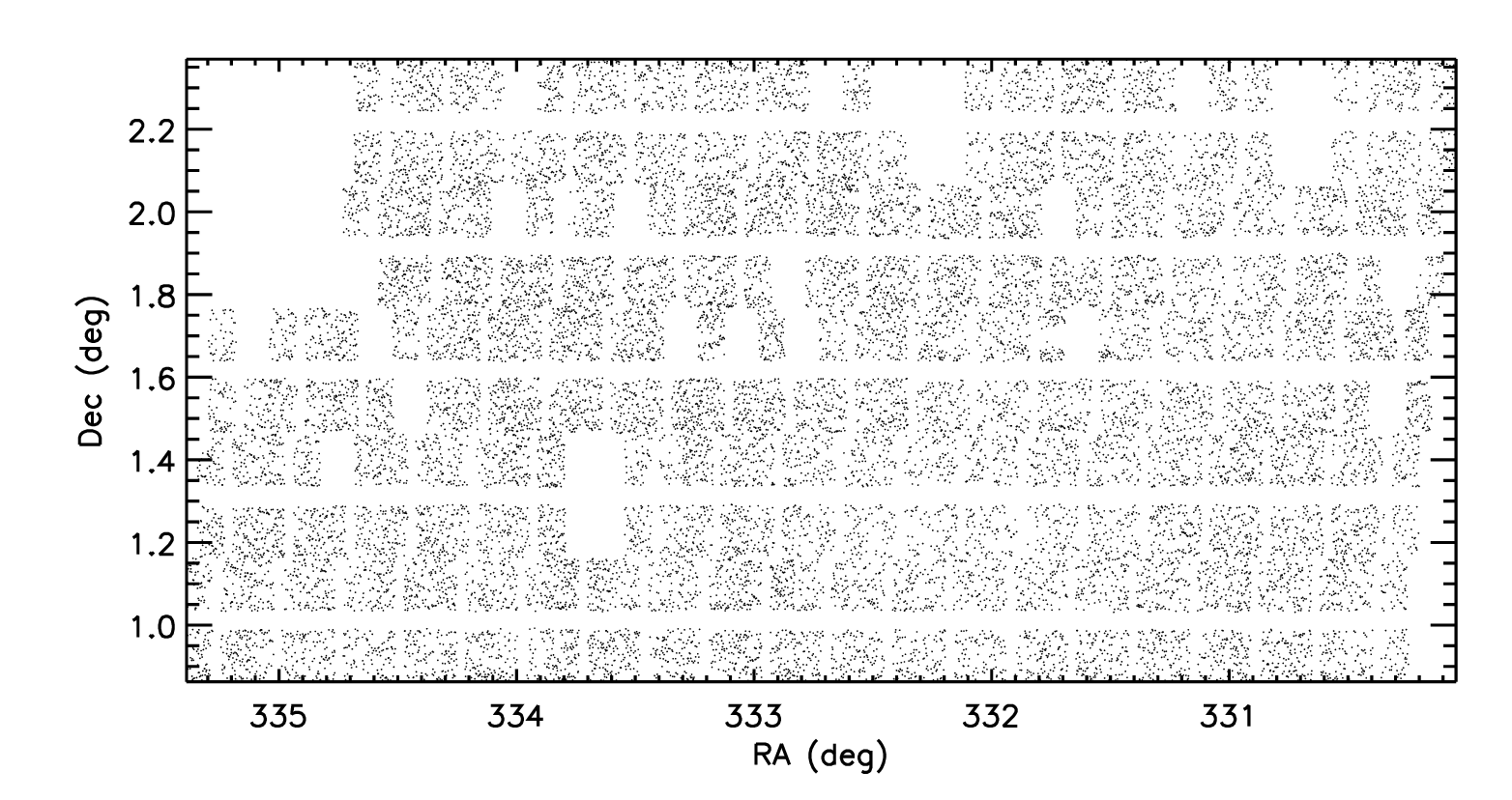}
\includegraphics[width=9cm]{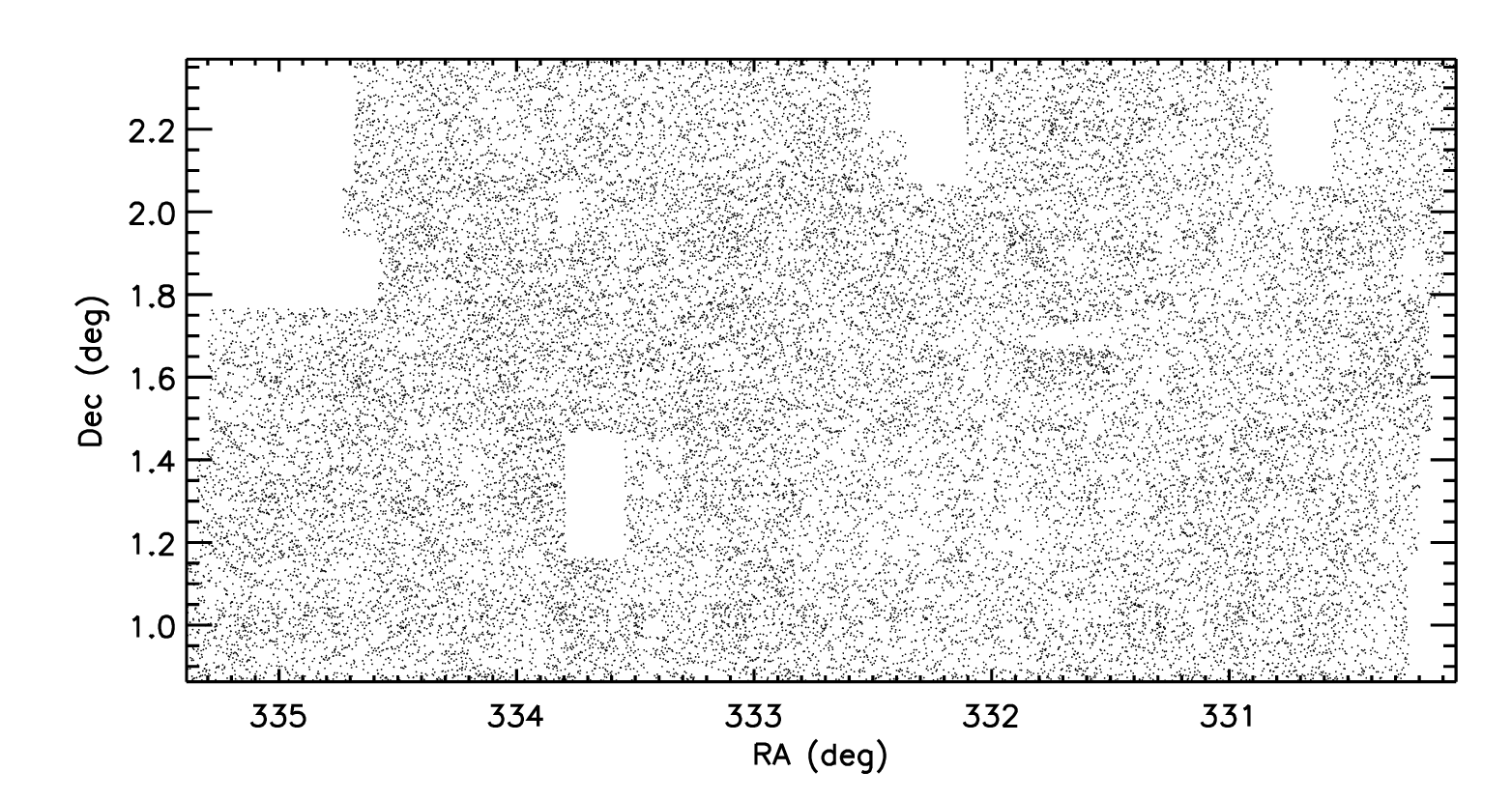}
\caption{{\it Top:} the real $RA-Dec$ distribution of VIPERS spectroscopic 
galaxies in the W4 field. {\it Bottom:} as in the top panel, but now 
gaps  are filled with cloned galaxies. }
\label{cloning_fig} 
\end{figure}

Like ZADE, we also used the cloning method to fill single missing
quadrants. The result of cloning applied to the VIPERS field W4 is
shown in Fig.~\ref{cloning_fig}. In principle, we could also fill 
regions bigger than a quadrant (i.e. missing pointings), but in this way
galaxies would be cloned from more and more distant regions, not
preserving the large-scale coherence of galaxy clustering on the
scales of interest.

We did not attempt to use cloning to correct for a low sampling rate.
We find some studies that used cloning to correct
small-scale incompleteness due to fibre collisions (see
e.g. \citealp{blanton05,lavaux11}). Given the way we implemented the
cloning (i.e., moving the cloned galaxy to the $RA-Dec$ position of a
real photometric galaxy), its use on all scales to correct for
sampling rate would simply result in retaining 
all the photometric galaxies in the new catalogue, and move their $z_p$ to the $z_s$ of the
closest spectroscopic galaxy. This is very similar to implementing
ZADE and retaining only the highest peak (with weight equal to 1) of
the $z$ distribution obtained by multiplying the $z_p$ PDF by 
$n(z_s)$. We tested this ZADE configuration (see the previous
section), and we found that it does not perform as well as full ZADE
in recovering the counts in cells.

The cloning method allows us to fill the gaps with a sampling rate
similar to the one in the nearby cloned areas. On such a galaxy
sample, it is possible to apply a non-parametric method to compute the
3D local density. As an example, we applied a 3D Voronoi-Delaunay
tessellation to the cloned real VIPERS samples. The Voronoi diagram
\citep{voronoi1908} consists in a partition of 3D space in polyhedra,
where each polyhedron encloses a galaxy and defines the unique volume
containing all the points that are closer to that galaxy than to any
other in the sample.  The Delaunay complex
\citep{delaunay1934} defines the tetrahedra whose vertices are
galaxies that have the property that the unique sphere that
circumscribes them does not contain any other galaxy. The centre of
the sphere is a vertex of a Voronoi polyhedron, and each face of a
Voronoi polyhedron is the bisector plane of one of the segments that
link galaxies according to the Delaunay complex. 

\begin{figure*} \centering
\includegraphics[width=6.cm]{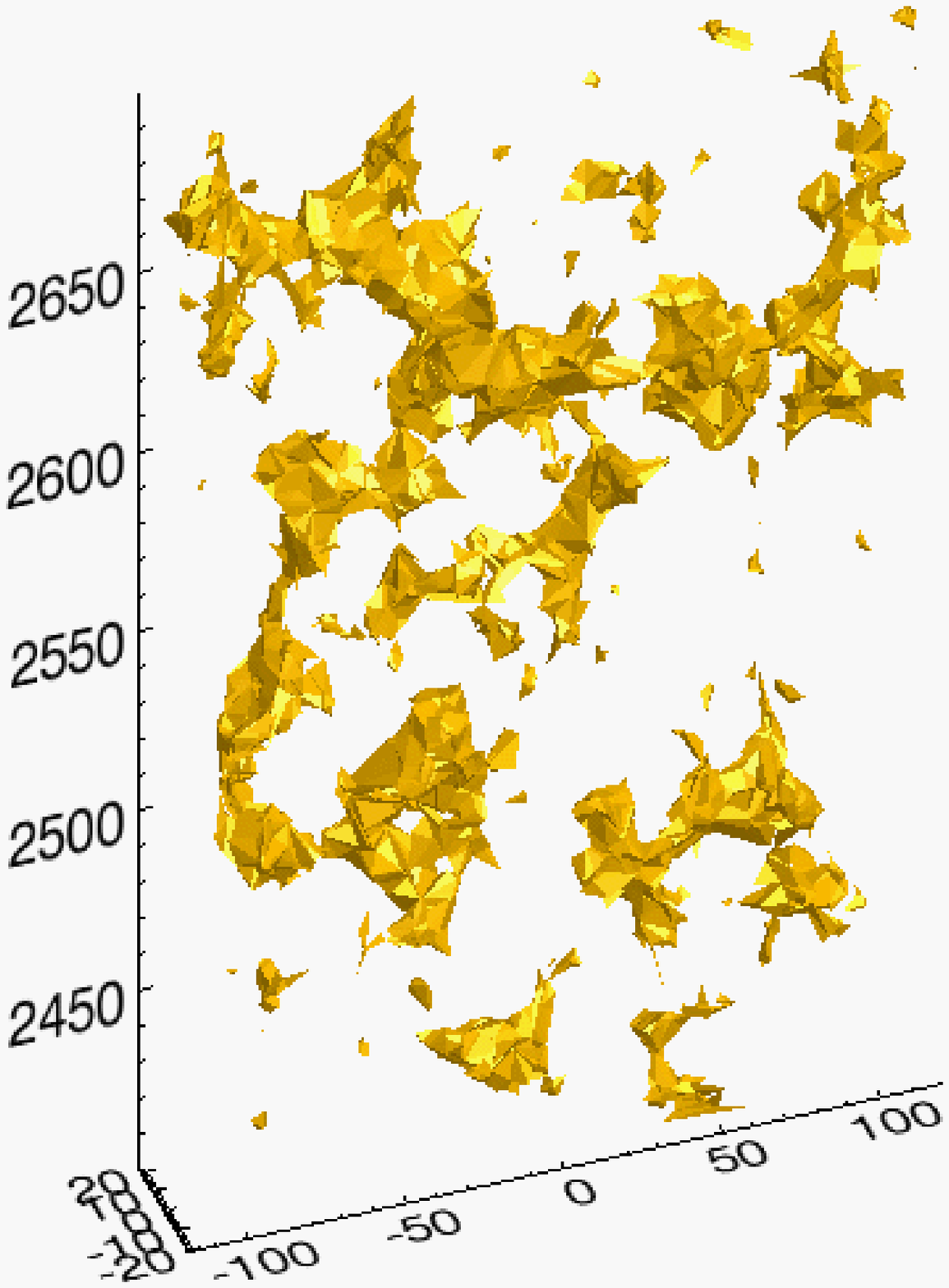}
\includegraphics[width=6.cm]{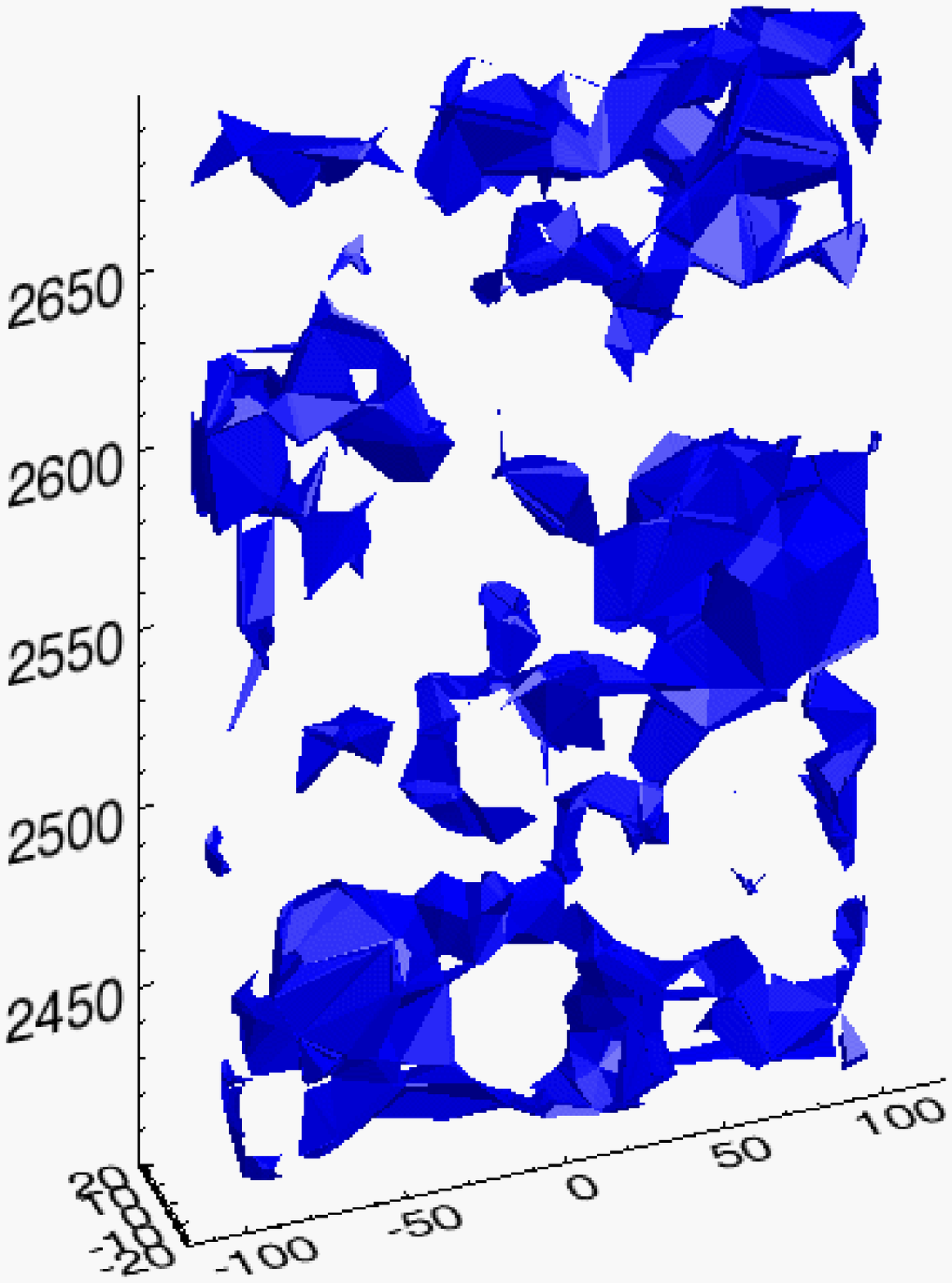}
\includegraphics[width=6.cm]{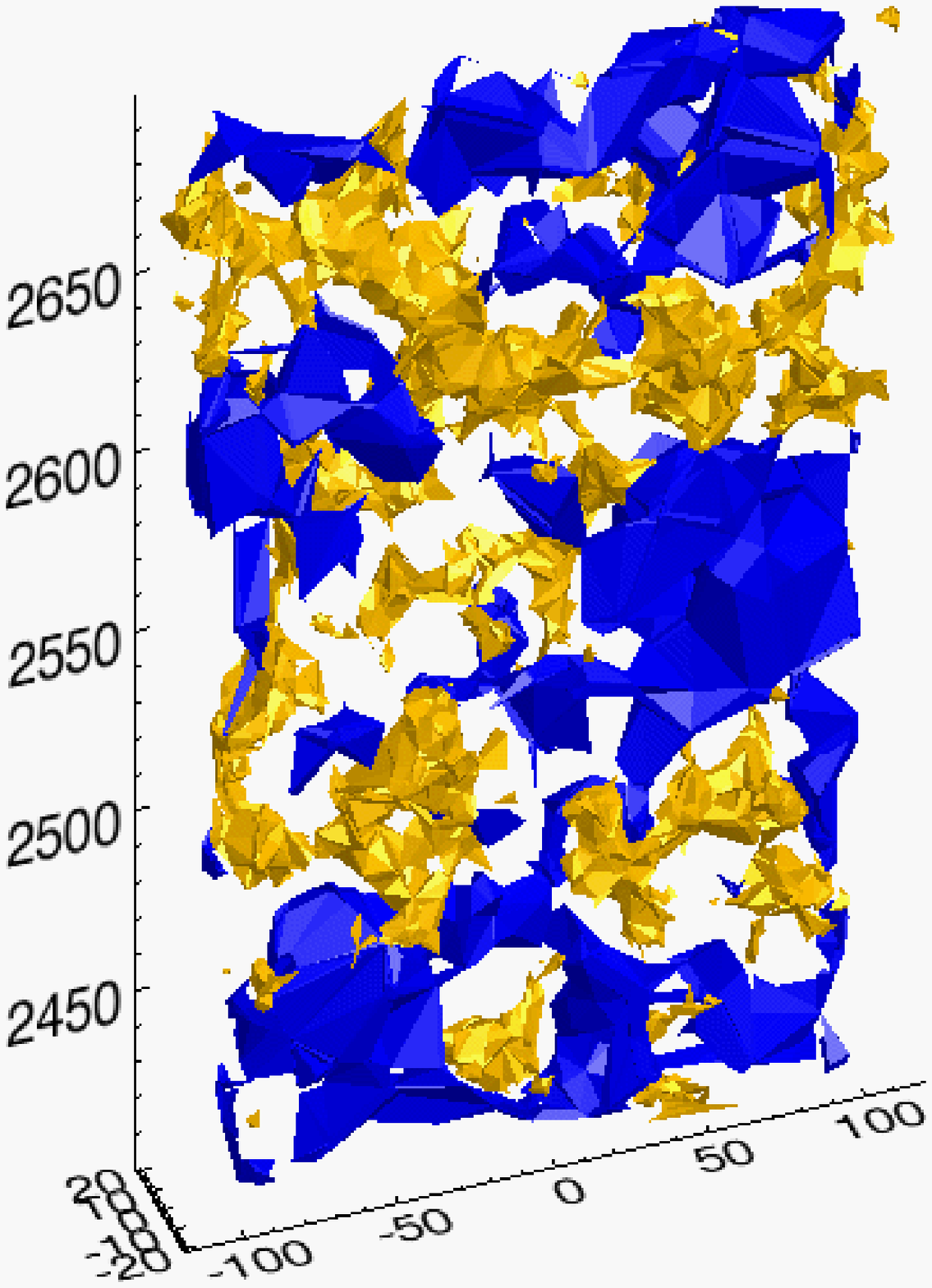}
\caption{The 3D VIPERS density field obtained by filling gaps with the cloning method
and computing the density within Voronoi volumes. {\it Left:} the isosurface enclosing the highest
densities. {\it Centre:} the isosurface enclosing the lowest densities. 
{\it Right:}
the isosurfaces of lowest and highest densities overplotted. One can see
 clearly that the highest densities form filamentary structures, while
the lowest densities enclose more spherical regions. The plot shows the W1
field with RA$<$35.3 deg to avoid the missing pointing at
RA$\sim35.5$. Axes are in comoving coordinates (expressed in $\mpcoh$): $x$ and $y$ are
arbitrarily centred on the $RA-Dec$ centre of the considered field, and
the $z$-axis corresponds to the redshift range $0.65\lesssim z
\lesssim 0.75$. }
\label{3D_fig} 
\end{figure*}

A 3D Voronoi-Delaunay
tessellation has already been used successfully for cluster
identification in optical spectroscopic surveys \citep{marinoni2002,
gerke2005_groups, knobel09, cucciati10}, its power residing in
exploiting the natural clustering of galaxies without any scale length
chosen a priori. We applied a 3D Voronoi-Delaunay tessellation to the
cloned PDR-1, and we used the inverse of the Voronoi volumes as a
proxy for local density.  Figure ~\ref{3D_fig} shows 3D maps of the
isosurfaces enclosing the regions with measured densities in the highest 
and lowest tails of the density distribution (i.e. densities above or 
below given thresholds). One can see very
clearly that the highest densities form filamentary structures, while
lowest densities enclose more spherical regions. The study of the
topology of such regions goes beyond the aim of this paper, but 
Fig.~\ref{3D_fig} illustrates the potential of using a large and deep
spectroscopic survey without gaps and roughly homogeneous coverage.

\subsection{Wiener filter} 

The Wiener filter differs from the previous methods in that it aims at
reconstructing the continuous density field rather than the position of
specific galaxies.  It is a Bayesian method based upon statistical
assumptions on the density field, namely that both the distribution of
the over-density field $p(\delta)$ and the likelihood of observing $N$
galaxies given $\delta$, $p(N|\delta)$, are Gaussian distributions.
The filter may be derived by maximising the posterior distribution
given by Bayes formula:
\begin{equation}
p(\delta | N) = p(N | \delta) p(\delta) / p(N).
\end{equation}
Although the prior and likelihood are modelled as Gaussian, the
resulting estimated density field may be strongly non-Gaussian
according to the constraints given by the observations.

To estimate the continuous density field, we bin the survey volume
with cubic cells with size $2\mpcoh$.  Galaxies are assigned to
the nearest cell.  For a survey sub-volume indexed by $i$, the
expected galaxy count is modelled as
\begin{equation}
\langle n_i \rangle = w_i \bar{n} (1+\delta_i),
\end{equation}
\noindent where $w_i$ is the selection function and $\bar{n}$  the
mean number of galaxies per cell.  The selection function is
determined by the product of the TSR, SSR, and CSR and also accounts
for the angular geometry of the survey.  The mean density $\bar{n}$
may be inferred from the observations by
\begin{equation} \bar{n} = \frac{\sum_i
n_i}{\sum_i w_i}, \end{equation} 
\noindent while the observed galaxy
over-density, weighted by the selection function, is
\begin{equation} 
\label{eq:delta}
\delta_{obs,i} \equiv w_i \delta_i = \frac{n_i}{\bar{n}} - w_i.  
\end{equation}
The Wiener filter depends on the covariance between cells given by a
model correlation function, $S_{ij} \equiv \langle \delta_i \delta_j
\rangle$, which we write in matrix notation.  The computation is
carried out in Fourier space and so, rather than the correlation
function, we use the power spectrum computed with CAMB \citep{Lewis:1999bs,Takahashi:2012em} for the fiducial
cosmology.  We also express the
selection function (in configuration space) as a diagonal matrix
$W_{ii}=w_i$.  Following the derivation by
\citet{Kitaura2009,Kitaura2010}, we may find the Wiener estimate for
the over-density $\hat{\delta}$ by solving
\begin{equation}
\label{eq:wf}
\sum_{j}\left(\left[\bS^{-1}\right]_{ij} + W_{ij} \bar{n} \right)\hat{\delta_j} = \bar{n} \delta_{obs,i}.
\end{equation}
To solve this equation we use the iterative linear conjugate gradient
solver included in the scientific python library {\tt
SciPy}\footnote{http://www.scipy.org}. From now on, we refer to this method simply as `WF'.

\subsection{Poisson-Lognormal Filter}

We further investigate Bayesian estimation methods by adopting a
lognormal form for $p(\delta)$.  The lognormal distribution gives a
more accurate description of the density field particularly in low and
high-density environments \citep{Coles1991}.  We take a Poisson model
for the likelihood $p(N|\delta)$.  Following \citet{Kitaura2010}, we
define $s$ as the logarithmic transform of the over-density, $\bs
\equiv \ln(1+\delta)$.

Empirical tests show that the power spectrum of $\bs$,
$P_{\ln(1+\delta)}(k)$, follows the shape of the linear power
spectrum: $P_{\ln(1+\delta)}(k) \approx a P_{linear}(k)$
\citep{Neyrinck2009}.  The amplitude $a$ depends on the higher order
moments of the field $\delta$ and is sensitive to the adopted
smoothing scale.  In the present analysis, we define the density over a
grid with resolution $2\mpcoh$ and set $a=0.2$ on the basis of numerical
tests. Expressing the covariance in Fourier space, we take (in
redshift space) $S_{L,ii} = a P_{linear}(k_i)$.

Following the Gaussian case discussed above, we may derive the
following equation that may be solved for $\hat{\delta}$
\citep{Kitaura2010}
\begin{equation}
\label{eq:lp}
n_i-w_i\bar{n}(1+\hat{\delta}_i) - \sum_j\left[\bSL^{-1}\right]_{ij} \left(\ln(1+\hat{\delta}_j)-\lambda_j\right) -1 = 0.
\end{equation}
We solve Eq.~\ref{eq:lp} for $\hat{\delta}$ with a nonlinear conjugate gradient solver
using a Newton-Raphson solver with a Polak-Ribiere step. From now on, we refer to this method simply as `LNP'.


\section{Counts-in-cells reconstruction}\label{countsincells}

The aim of this paper is to find the best way to fill the gaps between
quadrants and the areas where there are missing quadrants, 
in presence of all observational biases listed in Sect.~\ref{err_sources}.
 To gauge the success of a filling method
we assess its ability to reconstruct the counts on a cell-by-cell basis
and to separate high- from low-density regions from the 
probability distribution of the counts.

In Sect.~\ref{levels} we list the mock galaxy catalogues used to study
the effects of the different VIPERS observational biases to
counts in cells. In Sect.~\ref{used_samples} we describe the redshift bins
and samples used in the analysis. In Sect.~\ref{comp_layout} we describe the tests
carried out to estimate the robustness of the counts in cells given
the biases.

\subsection{Test levels}\label{levels} 

To estimate the contribution of each observational bias to the total
error in counts in cells reconstruction, we mimic their effect
separately in the mocks. Each source of error is investigated by means
of a specific `test'.

For each light cone, our reference catalogue is the parent
photometric mock catalogue (see Sect.~\ref{mocks}), which has flux limit
$i_{AB}=22.5$ and 100\% sampling rate. We work in redshift space, so
galaxies in these mocks have redshifts obtained combining cosmological
redshift and peculiar velocities.

All mock catalogues used to assess the impact of the individual
sources of error are drawn from the reference catalogue, and are
called `test catalogues'. They are listed below, specifying the source
of error they were designed to include.

{\bf Test A:} The impact of the spectroscopic redshift error. We mimic
this effect by adding the VIPERS spectroscopic
redshift error to the reference catalogues. We do this by adding to the redshift $z_i$ of each galaxy
a random value extracted from a Gaussian with
$\sigma=0.0005(1+z_i)$. In these mocks there are no gaps, and the
sampling rate is 100\%.

{\bf Test B:} The impact of the performances of the gap-filling method. We
use mock catalogues obtained from the reference catalogue, by removing
galaxies in gaps and missing quadrants. Sampling rate and redshift of
the galaxies falling in observed quadrants are unaltered. Since some
of the methods used to fill the gaps (ZADE and cloning) make use of
the photometric redshift of the galaxies falling in the gaps, we
actually re-insert such galaxies into the catalogues, when needed, but
we added to their redshift a photometric redshift error. We do this by
adding to the redshift $z_i$ of each galaxy a random value extracted
from a Gaussian with $\sigma=0.035(1+z_i)$.

{\bf Test C1:} The impact of a low sampling rate. We use mock catalogues
obtained from the reference catalogue by removing randomly 65\% of
the galaxies to reach an overall sampling rate of 35\% as in
VIPERS. In these mocks there are no gaps, and the redshift of the
retained galaxies is unaltered. As in Test B, we re-inserted the
unretained galaxies mimicking for them a photometric redshift error,
because the ZADE method uses them to correct for the sampling rate.

{\bf Test C2:} The impact of low {\it and} inhomogeneous sampling
rate. Here we want to test not only an average sampling rate of 35\%,
but also its modulation quadrant by quadrant, as given by the tool
SPOC, used to choose VIPERS targets (see Sect.~\ref{data}). We build
the required mock catalogue as follows. First we apply SPOC to the
reference catalogue, obtaining a catalogue with gaps and a sampling
rate varying quadrant by quadrant. This is likely to mimic the VIPERS
TSR, which is $\sim40$\%, so we further depopulate this
catalogue randomly to reach an average sampling rate of 35\% (to account for
the SSR). This way we have a VIPERS-like catalogue (varying sampling
rate and empty gaps). Since we do not want to test the effects of gaps in
this test, we add the galaxies again in the gaps, with a homogeneous
sampling rate of 35\%, so at the end in these mock catalogues there
are no gaps, and the redshift of the retained galaxies (also the 35\%
in gaps) is unaltered with respect to the reference catalogue.  Again,
the remaining $\sim65$\% of galaxies have been assigned a photometric
redshift error to make them available if needed (see the ZADE method).

{\bf Test D:} The impact of all of the above effects together, i.e. using
VIPERS-like mock catalogues. These catalogues have been prepared as in
Test C2, with the exceptions that i) galaxies retained in quadrants
($\sim35$\%) have a spectroscopic redshift error as in Test A, and ii)
all of the other galaxies (100\% of galaxies in gaps or in missing
quadrants, and the remaining $\sim65$\% in quadrants) have a photometric
redshift error.

\subsection{Galaxy samples}\label{used_samples}

VIPERS covers a wide redshift range and, because of its flux-limited
selection, the survey samples only the more luminous galaxies at
higher redshift.  As a result, the mean number density of objects
decreases at higher redshift. In this work, we want to use samples
with a constant number density as a function of redshift, to ease the
interpretation of our results and to better compare them with similar
choices in the literature.  We have divided the redshift range in
three shells and applied a luminosity cut (in $B$-band absolute
magnitude $M_B$) to obtain a set of volume-limited, luminosity
complete subsamples, with constant number density in the given redshift
bin. The three samples adopted in this work are

\begin{enumerate}[I]
\item - $0.5<z<0.7$, with cut at $M_B - 5\log_{10}(h) = -18.9 -z$,
\item - $0.7<z<0.9$, with cut at $M_B - 5\log_{10}(h) = -19.4 -z$,
\item - $0.9<z<1.1$, with cut at $M_B - 5\log_{10}(h) = -19.9 -z$,
\end{enumerate}

\noindent where the redshift dependence of the luminosity thresholds is designed
to account for evolutionary effects, since it roughly follows the same
dependence on redshift as the $M^{*}$ of the galaxy luminosity
function (see e.g. \citealp{kovac2010_density}). In the reference 
mock catalogues, i.e. those with flux limit
$i_{AB}=22.5$ and 100\% sampling rate, the galaxy number 
densities (averaged over all the mock catalogues) 
in the three samples are $1.1\times10^{-2}$, $4.3\times10^{-3}$, and 
$1.7\times10^{-3}$ galaxies per $(\mpcoh)^3$. The 
variance of these values among the 26 catalogues is $\sim10$\% 
in the sample at lowest redshift and $\sim5$\% in the other two.

We show results only for the central redshift bin, to minimise the 
number of plots in the paper,  but we will discuss  the results obtained 
in all three redshift bins (see Sect.~\ref{testD_redshift} and Fig.~\ref{testD_allz}).

\subsection{Counts-in-cells comparison}\label{comp_layout}

In each kind of mock catalogue, we perform counts in cells on
spherical cells with radius $R = 5$ and $8\mpcoh$ comoving ($R_5$ and
$R_8$ from now on), distributed randomly in the field. Galaxy
over-densities are obtained by counts in cells as $\delta_N =
N/\langle N\rangle -1 $, where $N$ represents the number of objects in
the cell, and $\langle N\rangle$ is the mean galaxy count in cell in
each of the considered redshift bins.

Our first test, described in Sect.~\ref{dens_dens_results} consists of
a cell-by-cell comparison of $\delta_N$ in the reference mock
catalogue ($\delta_N^R$) and in the different test catalogues
($\delta_N^A$, $\delta_N^B$, $\delta_N^{C1}$, $\delta_N^{C2}$,
$\delta_N^D$). The results of this comparison are described in
Sect.~\ref{dens_dens_results}. 
This is indeed a very demanding test. More demanding than, for example,
the recovery of the one-point probability of galaxy counts, $P(N)$.
This latter issue will be addressed in detail
elsewhere in the  more general framework of the recovery of the 
probability distribution function of the underlying galaxy density field
(Bel {\it et al.} in prep.) and of the galaxy bias  (Di Porto {\it et al.} in prep.). 
Here we  also consider the $P(N)$ of the reconstructed counts,  
but we use it only as a tool to separate high- from low- density regions.
Assessing the ability to effectively separate these environments is the goal of 
our second test.

Our second test, explained in Sect.~\ref{tails} aims at finding which
gaps-filling method allows us to best disentangle the lowest and
highest $\delta_N^R$ when selecting the extremes of the
$P(\delta_N^D)$ distribution.

We note that the WF and LNP methods give an estimate of the density
field on a grid, and so the counts in cells measurement cannot be made
in the same way as for the ZADE and cloning methods.  To make the
comparison, we compute $\delta$ in a given spherical cell by averaging
the enclosed grid cell values.

\section{Results}\label{results}

\subsection{Density-density comparison}\label{dens_dens_results}

The results of Tests A, B, C1, and C2 are extensively presented in
Appendix \ref{app_testABC}.  In this section we summarise them, and
discuss Test D in detail, which includes all the sources of uncertainty
of the other tests.

\begin{figure*} \centering
\includegraphics[width=4.4cm]{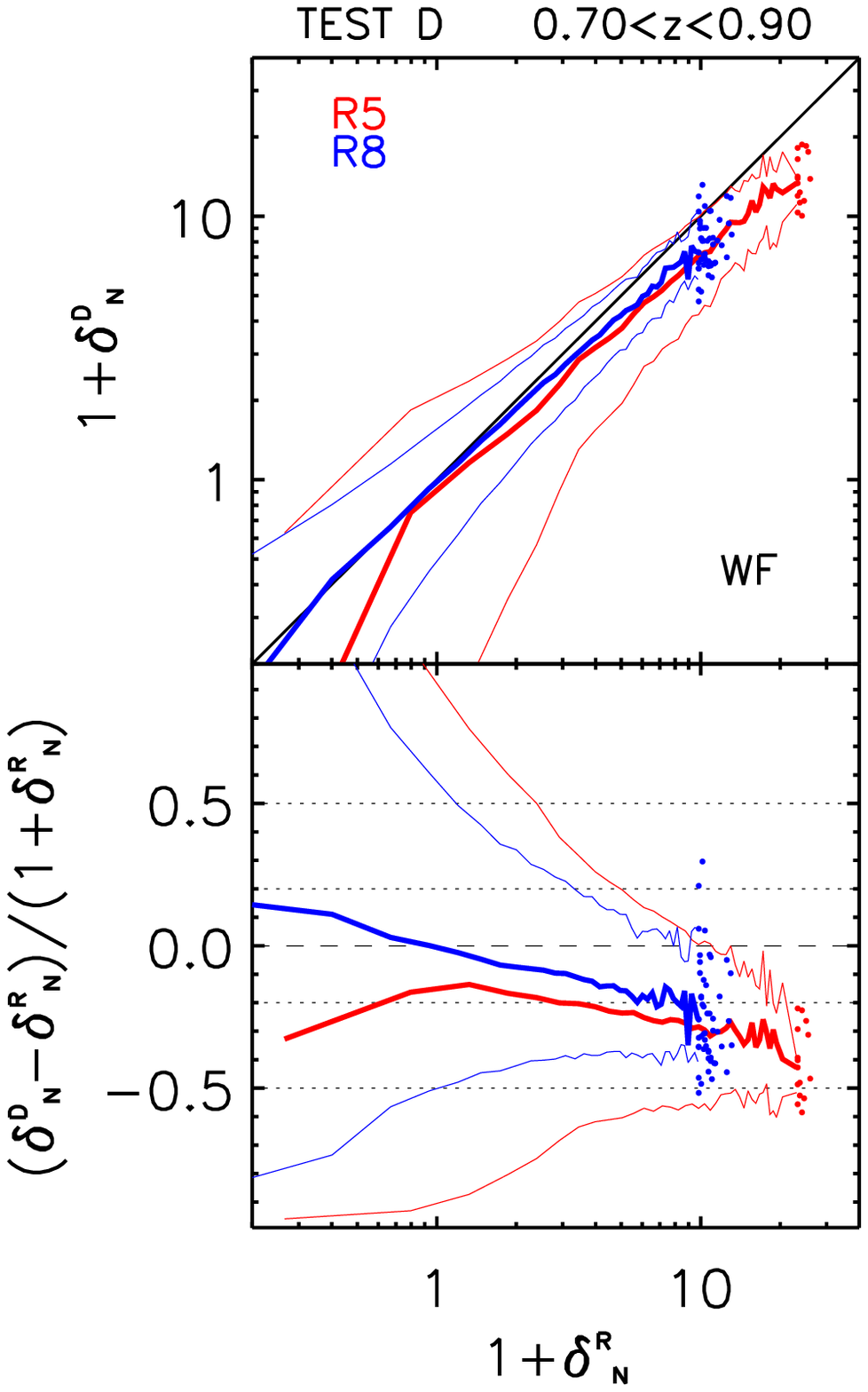}
\includegraphics[width=4.4cm]{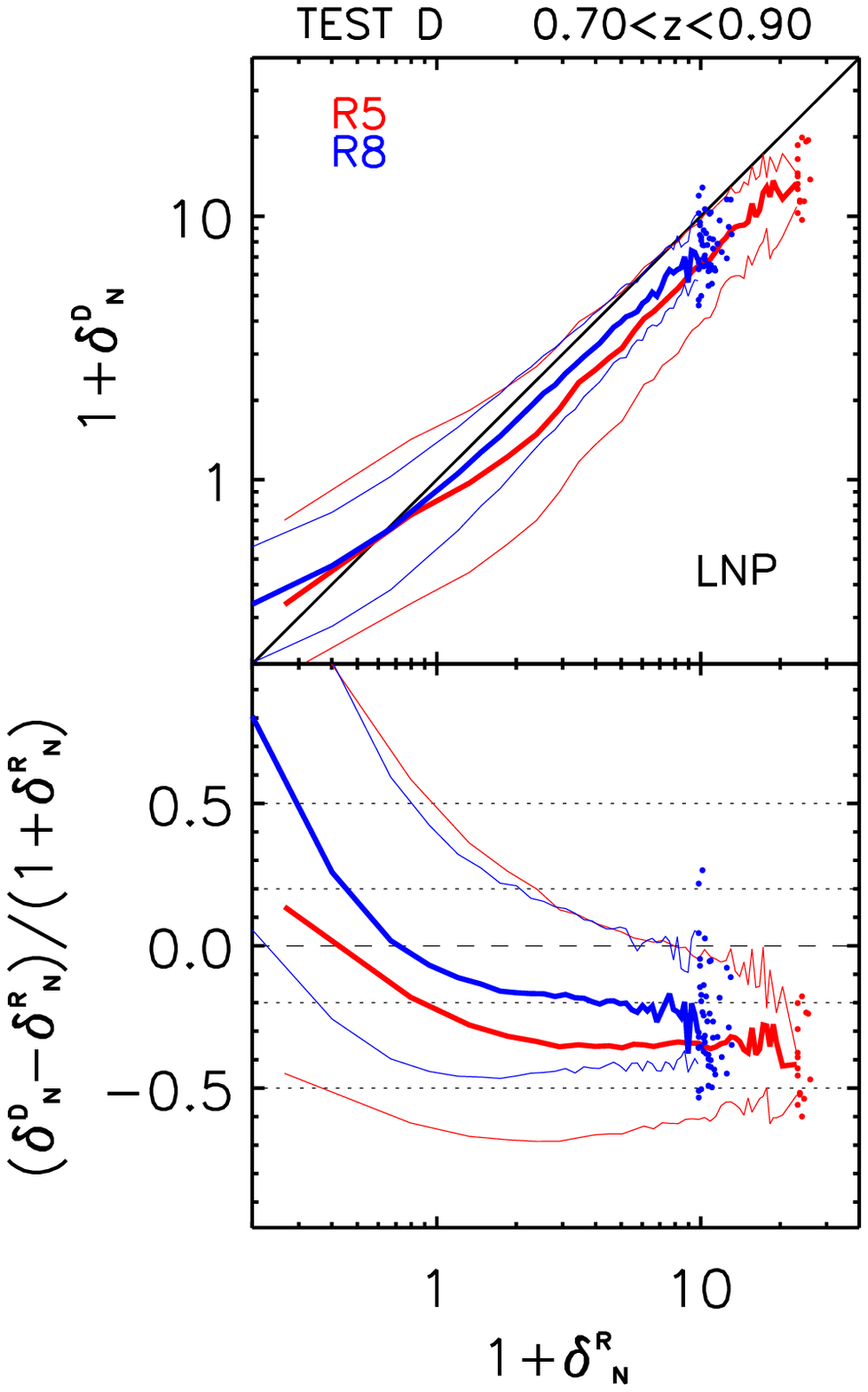}
\includegraphics[width=4.4cm]{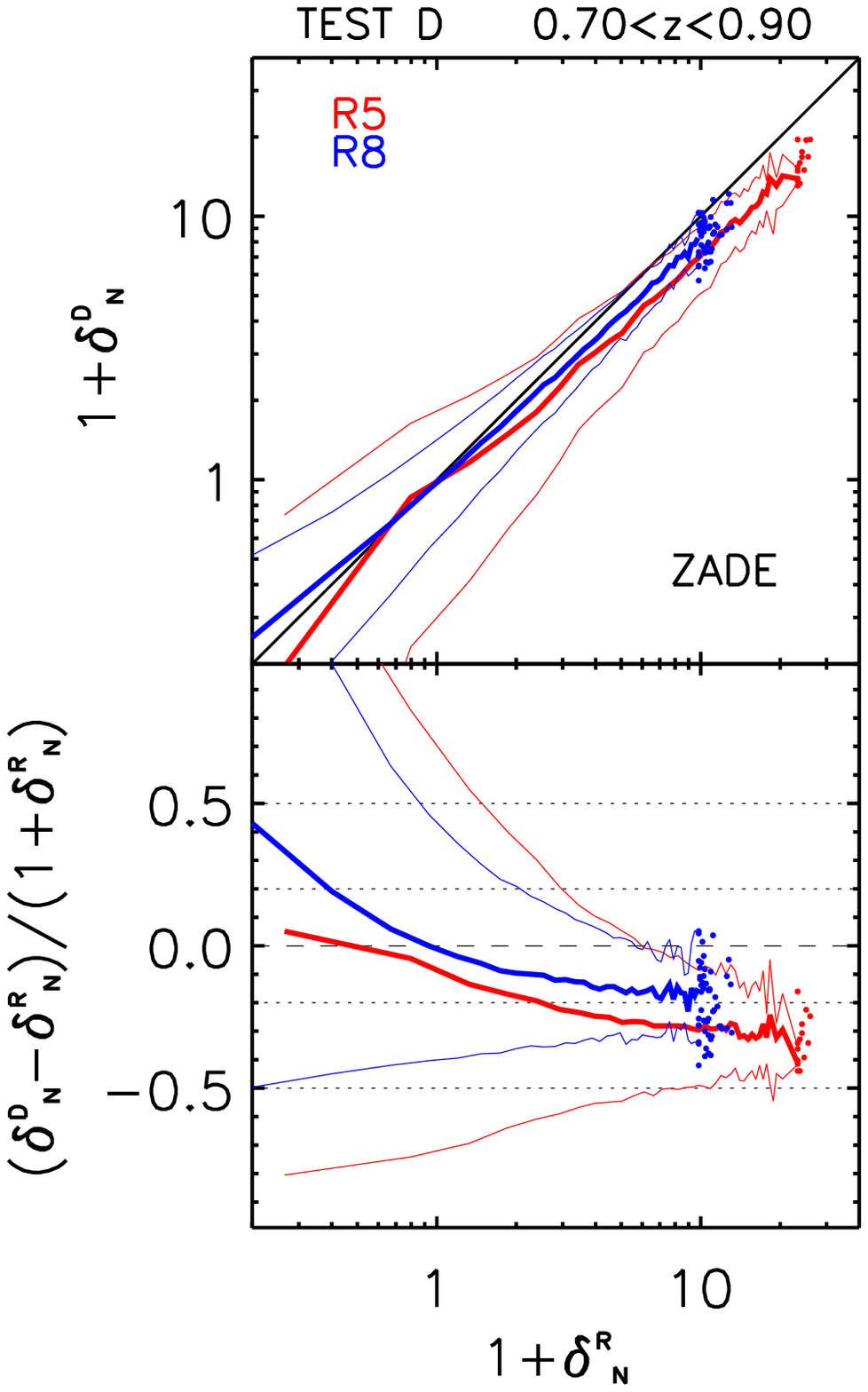}
\includegraphics[width=4.4cm]{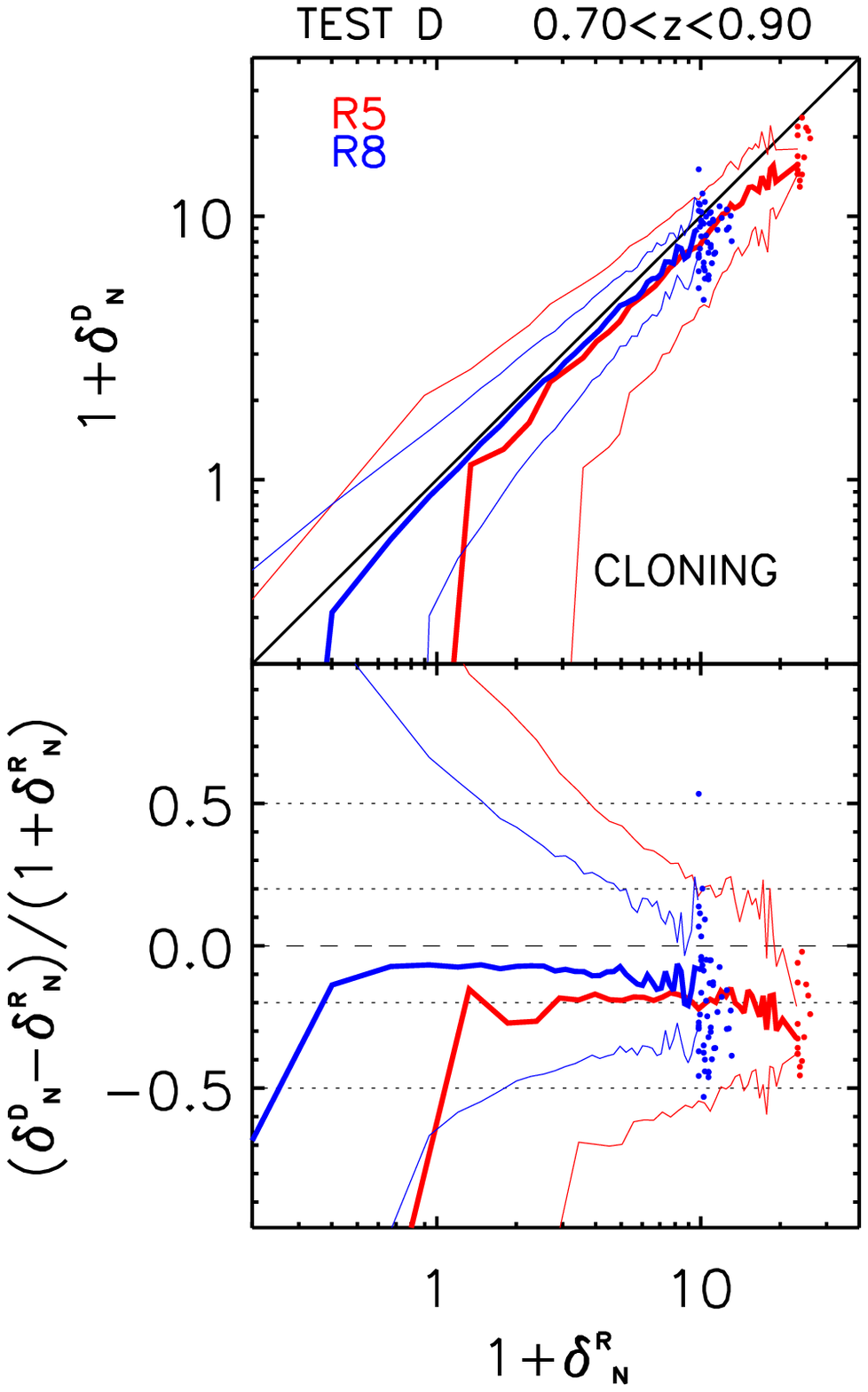}
\caption{Comparison of overdensities for Test D, for WF, LNP, ZADE, and cloning (from left to right),
for the redshift bin $0.7<z<0.9$ and tracers with $M_B - \log_{10}(h) \leq
-19.4 -z $. x-axis: overdensity in the reference catalogue ($1+\delta^R_N$);
y-axis, top panels: overdensity in the test catalogue ($1+\delta^B_N$);
y-axis, bottom panels: relative error ($(\delta^B_N-\delta^R_N)/(1+\delta^R_N)$). The thick lines are the median value of the
quantity displayed on the y-axis in each given x-axis bin. Thin lines
represent the 16th and 84th percentiles of its distribution. Points are single cells, when cells per bin are $<20$ (in
which case we do not compute a median and percentiles). Red and blue
lines/symbols are for spheres with radius $R_5$ and $R_8$,
respectively.  See text for details about tests.}
\label{testD_1940} 
\end{figure*}

The plots of Figures~\ref{testA_1940}, ~\ref{testB_1940},
\ref{testC1_1940}, \ref{testC2_1940}, and \ref{testD_1940} show the
comparison between the density contrast in the reference catalogue
($\delta_N^R$, on the x-axis) and in the test catalogues
($\delta_N^A$, $\delta_N^B$, $\delta_N^{C1}$, $\delta_N^{C2}$,
$\delta_N^D$, on the y-axis). A quantitative comparison of all the
tests shown in these figures is reported in Table \ref{fit_tab}. The
reference catalogue has a 100\% sampling rate down to $i$=22.5, and no
gaps. Test catalogues are listed in Sect.~\ref{levels} and are
different in each set of plots.

We notice that, for $\delta  \rightarrow -1$, the lines in 
the top panels tend to diverge because of the logarithmic 
scale of the axes  (which emphasises the low- and intermediate-density 
regimes). Moreover, for $\delta \rightarrow -1$, the denominator 
of the normalised residuals (the variable in $y$-axis in the bottom panels)
approaches zero and residuals rapidly increase.  This is an artefact
related to our definition of residuals.

The results of Tests A, B, C1, and C2 can be summarised as follows.

{\bf Test A.} The effects of the spectroscopic redshift error on the
counts in cell is to induce a small systematic underestimate at high
densities (for $1+\delta^R_N \gtrsim 5$). For both radii, the
systematic error is comparable to the scatter for intermediate or high
densities. Applying the WF or LNP method to recover the counts in the
reference catalogue does not improve the reconstruction.

{\bf Test B.} For all methods the scatter is larger than found in Test
A, while the systematic error is comparable. The ZADE method shows the
smallest scatter with low systematic error for both cases $R_5$ and
$R_8$. The accuracy of the reconstruction is better for $R_8$ than
$R_5$.

{\bf Test C1 and C2.} In Test C1, for all three methods, and for
both $R_5$ and $R_8$, the scatter is larger or comparable to the
systematic error, with possibly the exception of the highest
densities. Moreover, the systematic error and the scatter due to low
sampling rate are always greater than those due to gaps, and much
more than those due to the spectroscopic redshift error. The results
for Test C2 are only slightly worse than those of Test C1.

\subsubsection{Test D}\label{test_D_sec}

In Test D we combine all of the sources of uncertainty of the previous
tests (spectroscopic redshift errors, gaps, low and not homogeneous
sampling rate), thus using mock catalogues that mimic all the VIPERS
characteristics. Results are shown in Fig.~\ref{testD_1940}.

We did not use the cloning method to correct for the
sampling rate (see Sect.~\ref{cloning}), but in Test D we need to
account for it. After applying the cloning, we are left with a sample
of original and copied galaxies with no gaps, but with a sampling rate
varying quadrant by quadrant.  To correct for this, we weighted each
galaxy by the inverse of the relevant sampling rate.

In contrast, ZADE can be used not only for filling the gaps, but also
to correct for the sampling rate in quadrants (see
Sect.~\ref{methods_ZADE}). Alternatively, we can use ZADE to fill the
gaps, but treat the low sampling rate by weighting spectroscopic galaxies in
quadrants by the inverse of the sampling rate.  We verified that this
second method gives a poorer reconstruction of counts in cells than the
one based solely on ZADE (in particular, it gives a random error
15-20\% larger), so here we only show the results with ZADE correcting
for both gaps and sampling rate.

Figure ~\ref{testD_1940}  shows the cumulative effects of all sources of uncertainties
considered in the previous tests: 

\begin{itemize}
\item A general underestimate of the counts, at all densities, 
with the exception of LNP and ZADE over-predicting counts in the very underdense 
regions.

\item At high overdensities the random errors  are smaller than or comparable to 
the systematic ones in all cases except for the cloning method, meaning that
systematic errors are indeed significant.

\item The method less affected by systematic errors is cloning. This is
  evident in the high-density tail.  This, however, is also the
  method affected by the largest random errors.
  Random errors are the smallest when the ZADE method is applied.
\end{itemize}

Table \ref{fit_tab} shows a more quantitative comparison between
$\delta_N^R$ and $\delta_N^D$. We list in the table the values of the
slope and of the intercept of the linear fit performed on the thick
lines in the top panels of Fig.~\ref{testD_1940}. 
Errors on $\delta_N^D$ are set equal to the width of the 
probability of $\delta_N^D$ given $\delta_N^R$, $P(\delta_N^D|\delta_N^R)$
measured at the $16^{th}$ and $84^{th}$ percentiles. The table also
shows the linear correlation coefficient $r$ and the Spearman
coefficient $\rho$, for the same set of $x$ and $y$ values used for
the linear fit. While $r$ tells us how well the two variable are {\it
linearly} correlated, $\rho$ returns the degree of correlation (not
necessarily linear) between $x$ and $y$. This second test is also
important for our analysis: even if $\delta_N^D$ is not linearly
correlated with $\delta_N^R$, if $\rho$ is close to 1 the two variable
can be linked with a monotonic function, allowing us to 
disentangle low and high densities (see Sect.~\ref{tails}). The table
shows that both $r$ and $\rho$ are always very close to unity. 

We obtain a lower Spearman coefficient when we compute it
on the values of $1+\delta_N^R$ and $1+\delta_N^D$ of the single cells
instead of their median value in each density bin. This is shown in
the table as $\rho_{cells}$. This result is due to the fact that the
probability of galaxy overdensity $P(\delta_N)$ is skewed towards
low densities (see Fig.~\ref{distrib_fig} discussed later on): on a cell-by-cell basis,
low densities weight more and, given that the random error at low
densities is larger, the correlation between $\delta_N^R$ and
$\delta_N^D$ is weaker if we use single cells instead of median values
in equally spaced bins of $\delta_N^R$. Nevertheless, both 
$\rho$ and $\rho_{cells}$ are  significatively different from zero, 
since the significance of the Spearman coefficient depends also on the 
number of used points. In our case, the number of spheres is so large 
that $\rho_{cells}$ results more significantly different from zero than $\rho$.

\subsubsection{Redshift dependence}\label{testD_redshift}

\begin{figure} \centering
\includegraphics[width=9cm]{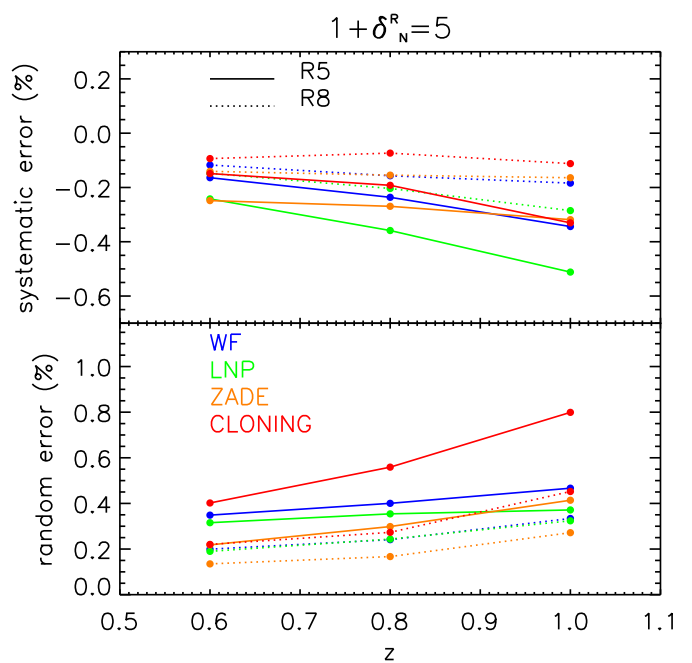}
\caption{Systematic (top) and random (bottom) relative errors in the
reconstruction of $\delta^R_N$ in Test D, in the 3 redshift bins
considered in this study ($x-$axis). We only show results for
$\delta^R_N +1 = 5$, but trends with redshift are similar for other
values.  These values correspond, respectively, to the thick and thin
lines in the bottom panels of Fig.~\ref{testD_1940}, for the given
$\delta^R_N$ value. The value of the scatter is obtained by averaging the
lower and upper values of the scatter (the two thin lines in
Fig.~\ref{testD_1940}). Solid line is for $R_5$ and
dotted line for $R_8$. Different colours correspond to the different
methods (blue: WF; green: LNP; orange: ZADE; red: cloning). }
\label{testD_allz} 
\end{figure}

The results of Test D in the lower ($0.5<z<0.7$) and higher
($0.9<z<1.1$) redshift bins are quite similar to those presented in
Fig.~\ref{testD_1940}. The evolution with redshift of the systematic
and random error in test D for the four methods is shown in
Fig.~\ref{testD_allz}. The figure refers to a given overdensity value
($1+ \delta^R_N =5$), but trends with redshift are qualitatively
similar for other $\delta^R_N$ (the absolute value might be different).

For $R_5$, the systematic error for ZADE does not evolve significantly
with redshift, while its random error is 5-10\% smaller and 5-10\%
larger at $0.5<z<0.7$ and $0.9<z<1.1$, respectively, with respect to
the central redshift bin. For the WF and LNP methods the systematic error
increases with redshift, but the random error does not evolve
significantly. For the cloning method, both random and systematic errors increase
with redshift. For $R_8$, all the trends visible for $R_5$ are much
milder. These results also hold for a intermediate test level (e.g.,
Tests B and C).

We note that we use different luminosity thresholds in each redshift
bin, so we also verified that our results do not change significantly
when keeping the same luminosity threshold and moving to lower redshift.
This means that the different levels of random and systematic errors
found at different redshifts are mainly due to the different density of
the used tracers and not to the evolution with redshift of the typical
density of a given set of tracers.

\begin{table*} 
\caption{Linear fit of the median values in the scatter 
plots (top panels) of Figures~\ref{testA_1940}, \ref{testB_1940}, \ref{testC1_1940},
\ref{testC2_1940}, and  \ref{testD_1940}. For the fit we use $x=1+\delta^R_N$ 
(central value of each $1+\delta^R_N$ bin)  
and $y=median(1+\delta^T_N)$, i.e. the value of the thick line in the plots in 
the given $1+\delta^R_N$ bin. 
Errors of  $\delta^T_N$ are measured at the  $16^{th}$ and $84^{th}$ 
percentiles of the conditional probability function
$P(\delta_N^T|\delta_N^R)$ (thin lines in the 
above-mentioned figures). The table shows the intercept and slope of the linear fit, the 
linear correlation 
coefficient (r), the Spearman coefficient ($\rho$). The columns ``$\rho_{cells}$'' show 
the Spearman coefficient obtained using the density in each single cell instead of 
the median density value per bin.}
\label{fit_tab} 
\centering 
\begin{tabular}{l | r r r r r| r r r r r}
\hline\hline 
 {\bf Method}    &   \multicolumn{5}{c|}{{\bf $R=5\mpcoh $}} & \multicolumn{5}{c}{{\bf $R=8\mpcoh $}} \\    
\hline								   
  &  Intercept  &  Slope  &  $r$ & $\rho$	 &  $\rho_{cells}$ &  Intercept  &  Slope  & $r$ & $\rho$	 &  $\rho_{cells}$\\    
\hline								   
\multicolumn{11}{c}{{\bf TEST A }}\\
\hline	
WF  &   	     -0.03$\pm$ 0.16   &  0.87$\pm$ 0.02  &    0.999  &   0.999 & 0.889 &   0.04$\pm$ 0.07   &  0.93$\pm$ 0.02  &   0.999   &  1.000 & 0.972\\
LNP  &   	     -0.07$\pm$ 0.16   &  0.95$\pm$ 0.02  &    0.999  &   1.000 & 0.866 &   0.05$\pm$ 0.07   &  1.00$\pm$ 0.02  &   0.999   &  0.999 & 0.964\\
Counts &  	     -0.04$\pm$ 0.15   &  0.92$\pm$ 0.02  &    0.999  &   0.999 & 0.907 &   0.02$\pm$ 0.06   &  0.96$\pm$ 0.01  &   0.999   &  1.000 & 0.974\\
\hline								   										      
\multicolumn{11}{c}{{\bf Test B }}\\														      
\hline																		      
WF  &    	    $-0.06\pm 0.19$  & $0.92\pm 0.03$ &   0.998   &  0.997	& 0.868 & $-0.00\pm 0.09$  & $0.98\pm 0.03$ &	0.999	&  0.999 & 0.951\\
LNP  &  	    $-0.03\pm 0.17$  & $0.95\pm 0.03$ &   0.998   &  0.998	& 0.853 & $ 0.03\pm 0.08$  & $1.02\pm 0.03$ &	0.999	&  0.999 & 0.948\\
ZADE  &  	    $ 0.03\pm 0.16$  & $0.94\pm 0.02$ &   0.998   &  0.999	& 0.909 & $ 0.02\pm 0.07$  & $0.97\pm 0.02$ &	0.999	&  1.000 & 0.970\\
Cloning  &    	    $-0.06\pm 0.17$  & $0.93\pm 0.03$ &   0.998   &  0.999	& 0.894 & $-0.02\pm 0.08$  & $0.98\pm 0.03$ &	0.999	&  0.999 & 0.945\\
\hline								   										      
\multicolumn{11}{c}{{\bf TEST C1 }}\\														      
\hline																		      
WF  &   	     -0.16$\pm$ 0.21   &  0.89$\pm$ 0.04  &   0.997   &  0.993	& 0.763 &  -0.02$\pm$ 0.12   &  0.95$\pm$ 0.03  &   0.999   &  0.999 & 0.896\\
LNP  &  	     -0.04$\pm$ 0.19   &  0.86$\pm$ 0.04  &   0.995   &  0.993	& 0.766 &   0.06$\pm$ 0.09   &  0.92$\pm$ 0.03  &   0.998   &  0.999 & 0.896\\
ZADE  &   	      0.00$\pm$ 0.22   &  0.83$\pm$ 0.03  &   0.998   &  0.997	& 0.808 &   0.03$\pm$ 0.10   &  0.93$\pm$ 0.03  &   0.999   &  1.000 & 0.921\\
\hline								   										      
\multicolumn{11}{c}{{\bf Test C2 }}\\														      
\hline																		      
WF  &  	            $-0.06\pm 0.22$  & $0.81\pm 0.04$ &   0.997   &  0.996	& 0.765 &  $0.04\pm 0.12$  & $0.88\pm 0.04$ &	0.998	&  0.997 & 0.896\\
LNP  &   	    $ 0.04\pm 0.19$  & $0.78\pm 0.04$ &   0.997   &  0.995	& 0.768 &  $0.12\pm 0.10$  & $0.84\pm 0.04$ &	0.999	&  0.997 & 0.895\\
ZADE  &   	    $ 0.09\pm 0.23$  & $0.78\pm 0.03$ &   0.998   &  0.999	& 0.811 &  $0.08\pm 0.10$  & $0.88\pm 0.03$ &	0.999	&  0.998 & 0.922\\
\hline								   		 								      
\multicolumn{11}{c}{{\bf Test D }}\\														      
\hline																		      
WF  &  	            $ 0.08\pm 0.27$  & $0.66\pm 0.04$ &   0.993   &  0.994	& 0.685 & $ 0.09\pm 0.15$  & $0.81\pm 0.05$ &   0.995   &  0.992 & 0.837\\
LNP &     	    $ 0.17\pm 0.21$  & $0.63\pm 0.04$ &   0.996   &  0.994	& 0.698 & $ 0.18\pm 0.11$  & $0.74\pm 0.05$ &   0.997   &  0.992 & 0.844\\
ZADE&   	    $ 0.25\pm 0.26$  & $0.65\pm 0.03$ &   0.996   &  0.998	& 0.730 & $ 0.12\pm 0.13$  & $0.82\pm 0.04$ &   0.999   &  0.998 & 0.873\\
Cloning  &          $-0.20\pm 0.01$  & $0.76\pm 0.05$ &   0.995   &  0.992	& 0.650 & $-0.07\pm 0.14$  & $0.90\pm 0.05$ &   0.997   &  0.996 & 0.799\\
\hline 
\end{tabular} 
\end{table*}

\subsection{Distinguishing between low- and high- density environments}\label{tails}

\begin{figure*} \centering
\includegraphics[width=6.0cm]{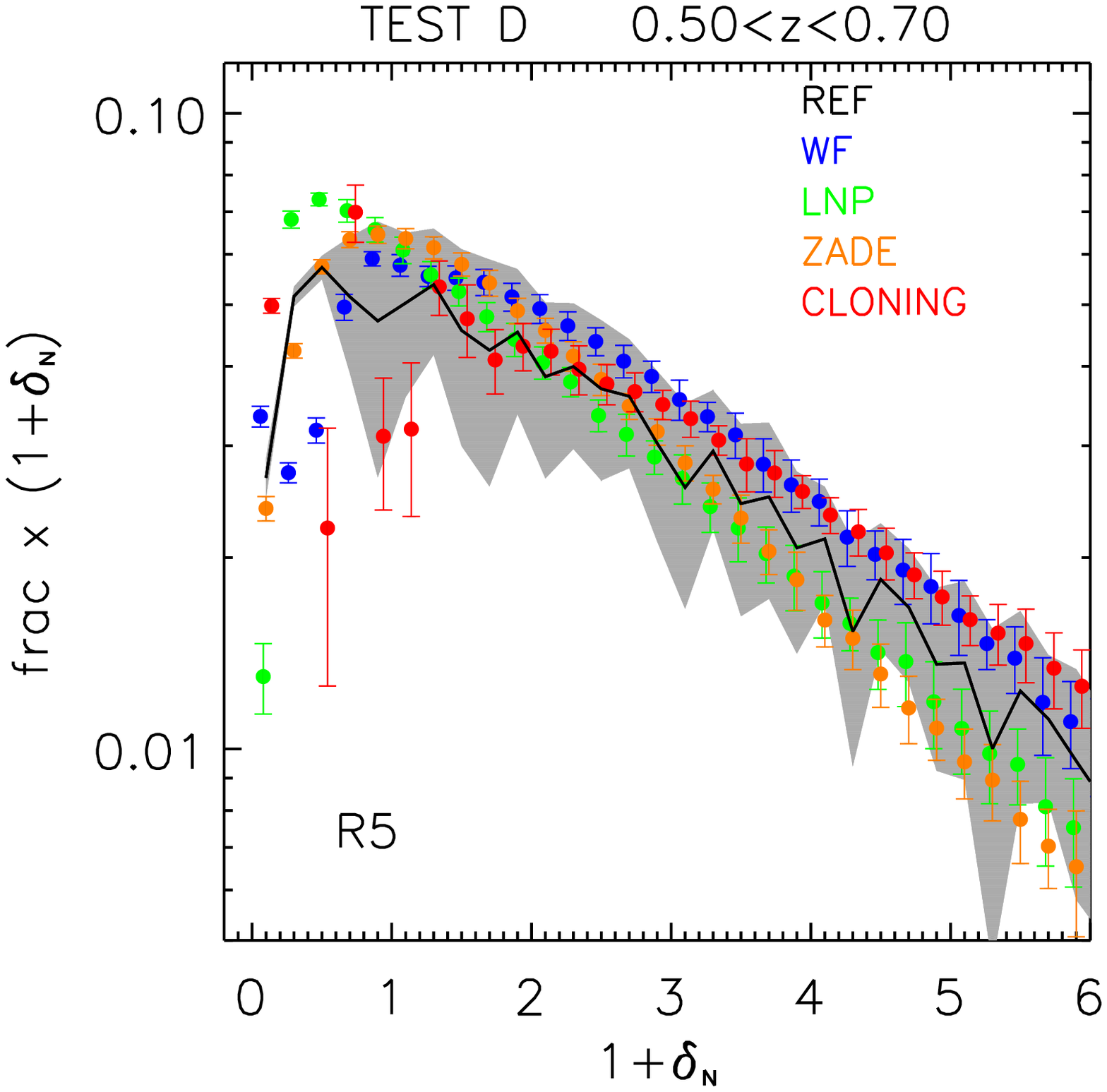}
\includegraphics[width=6.0cm]{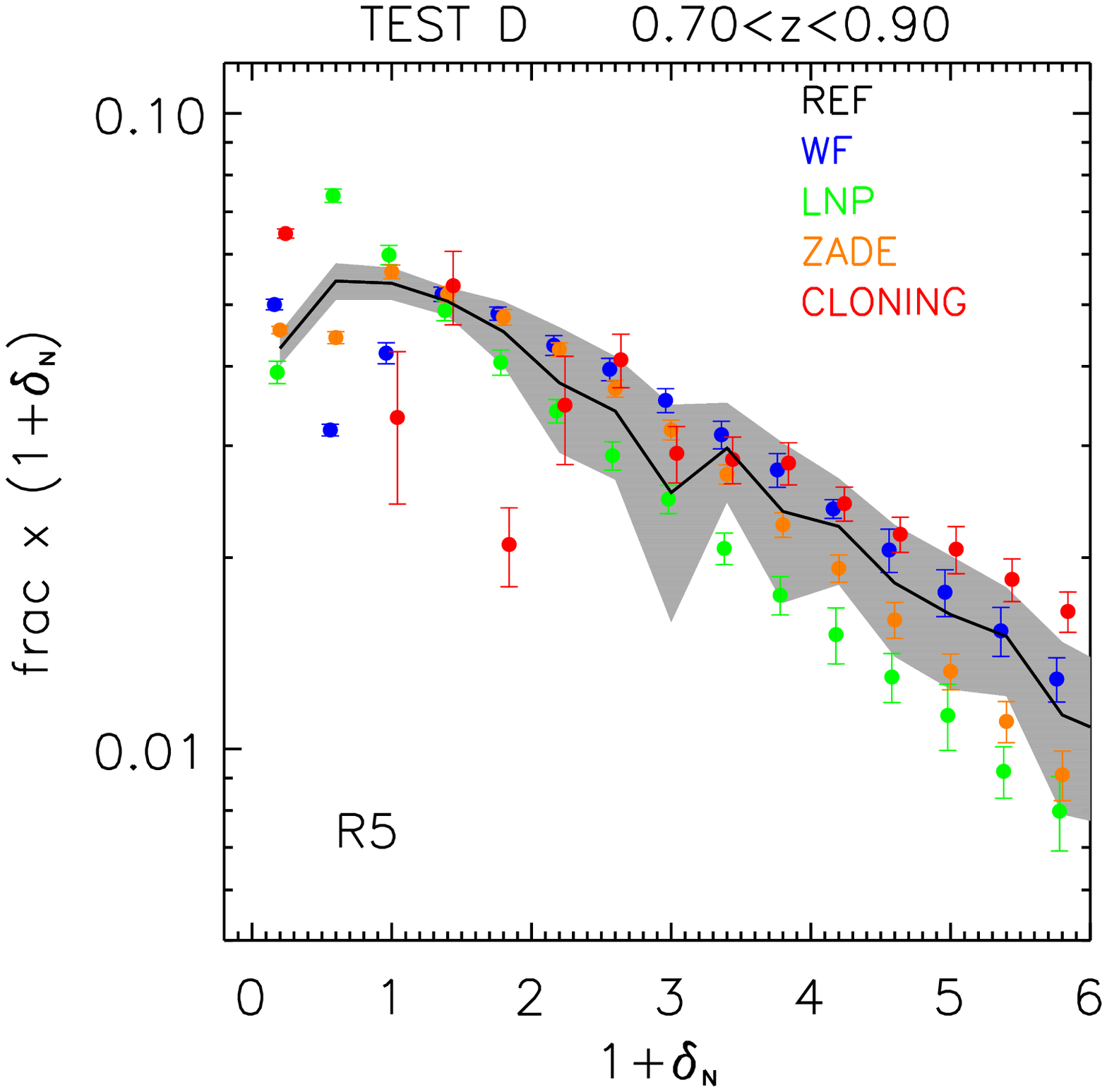}
\includegraphics[width=6.0cm]{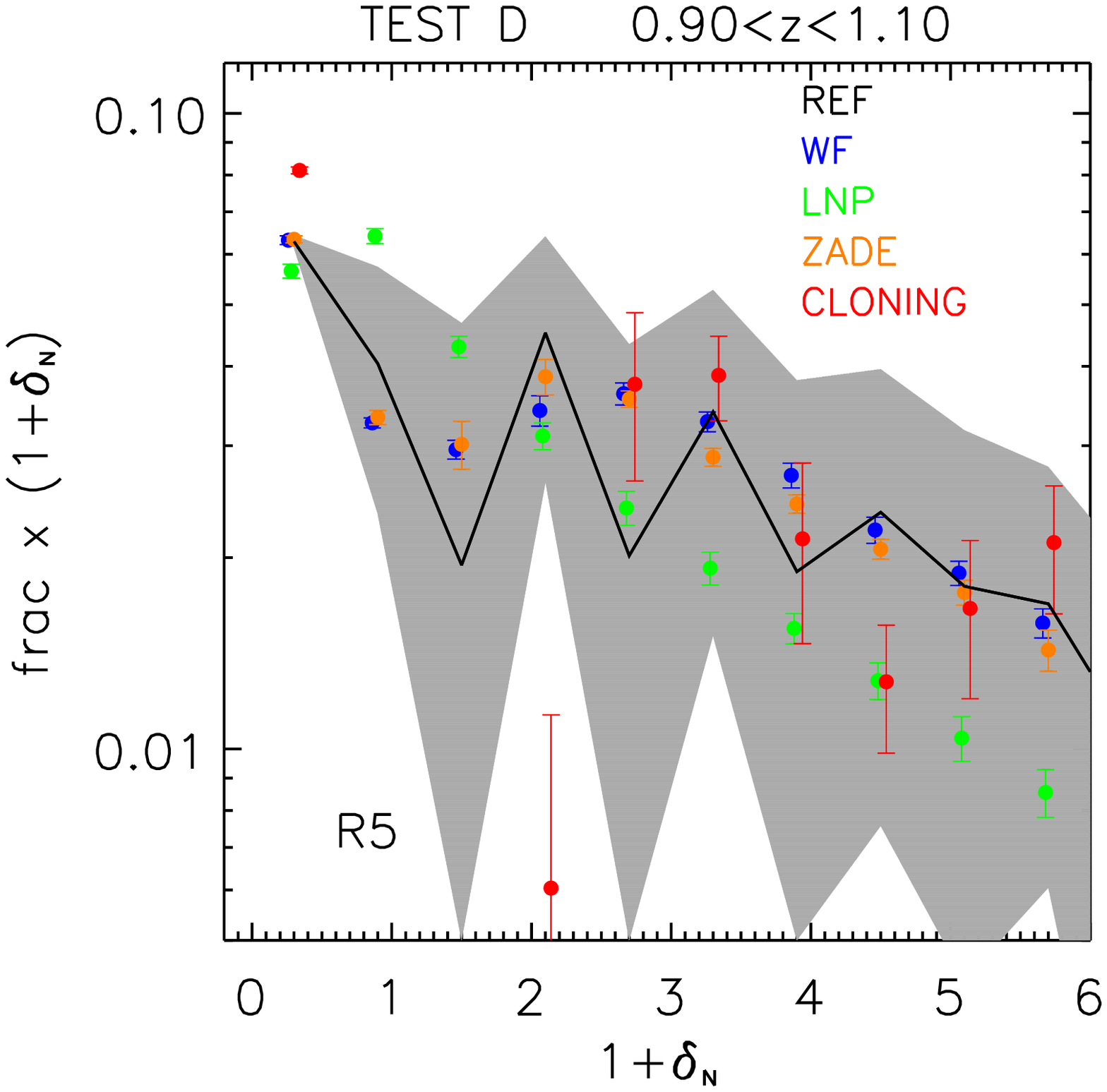}
\includegraphics[width=6.0cm]{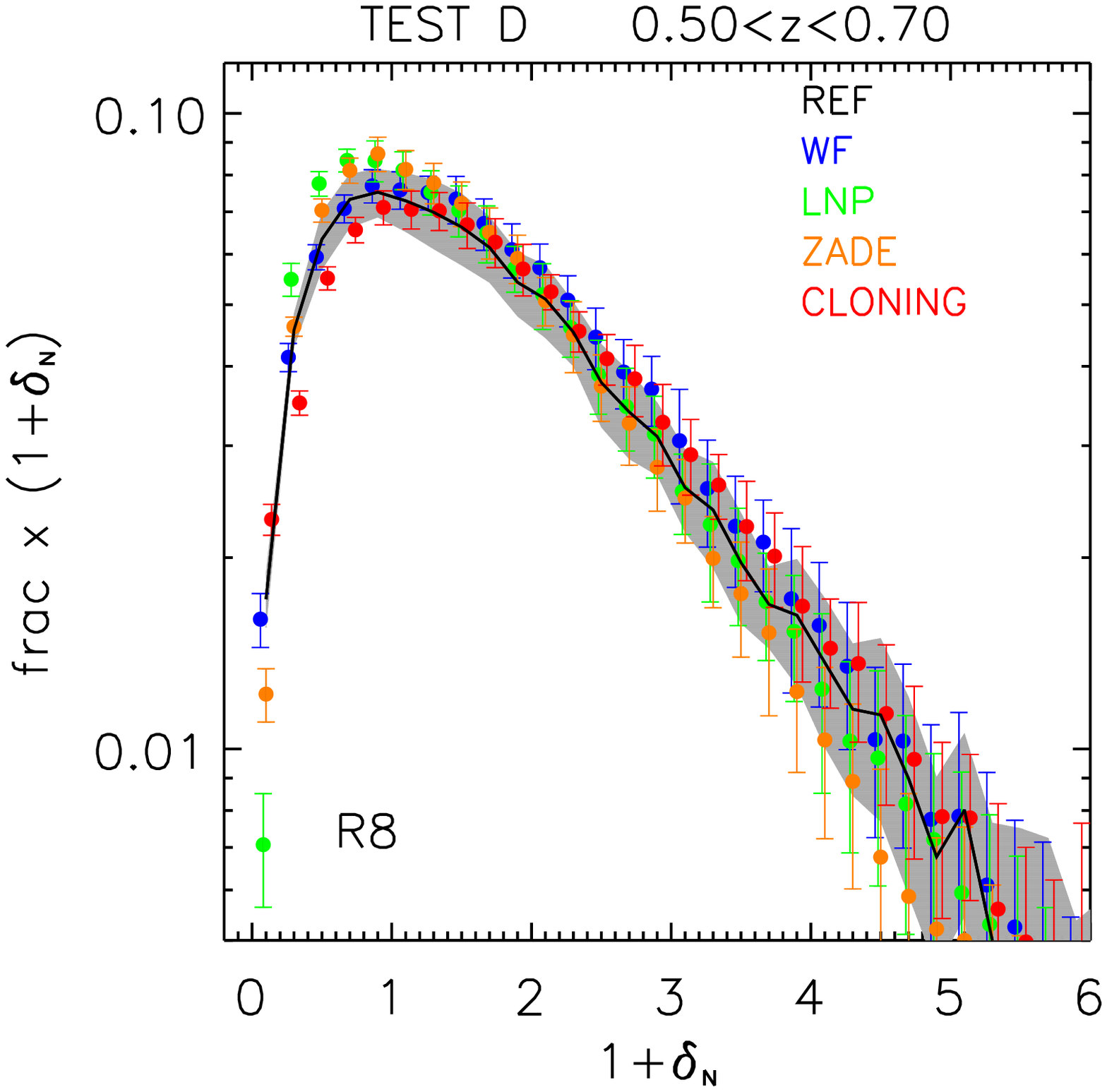}
\includegraphics[width=6.0cm]{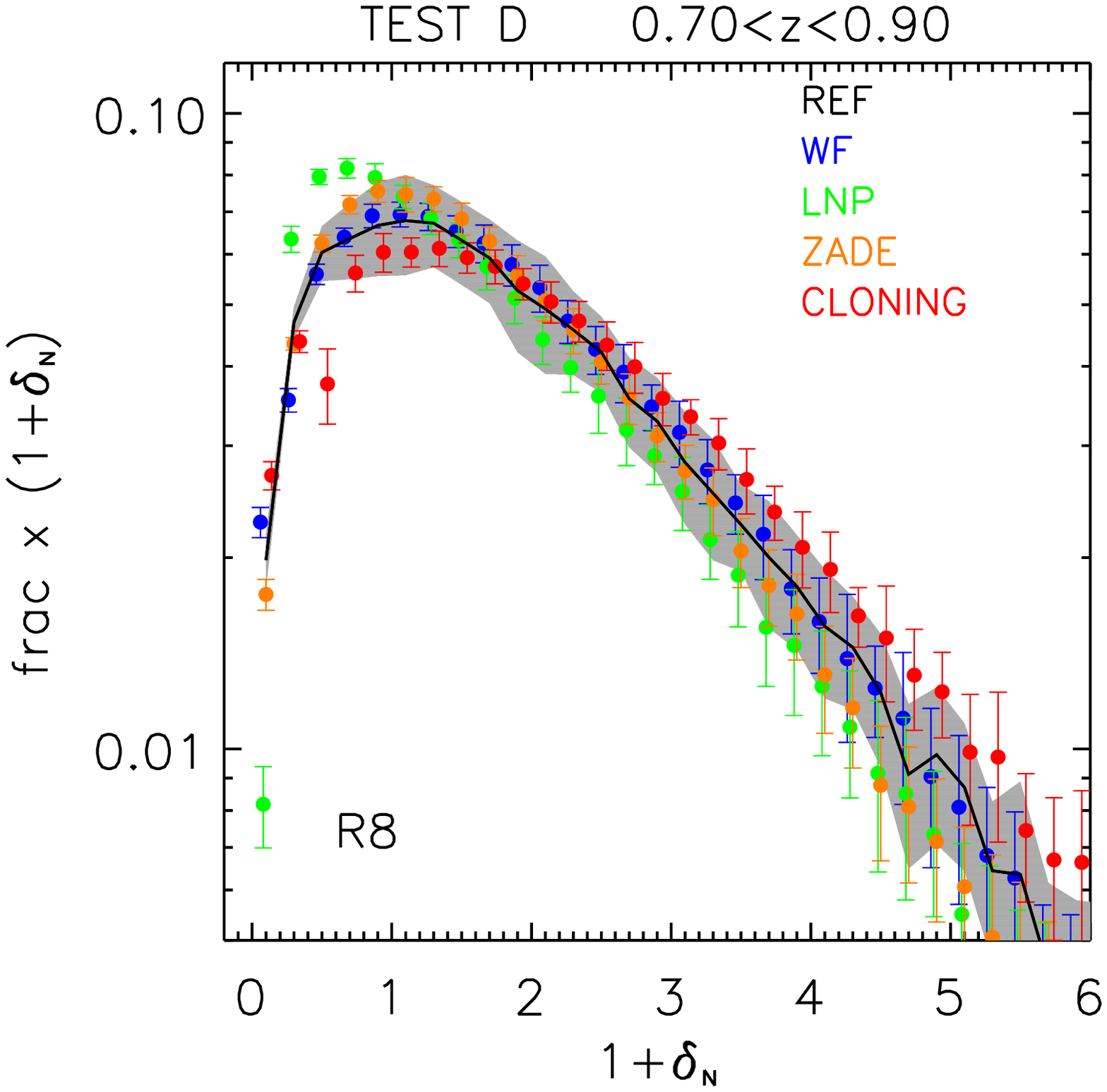}
\includegraphics[width=6.0cm]{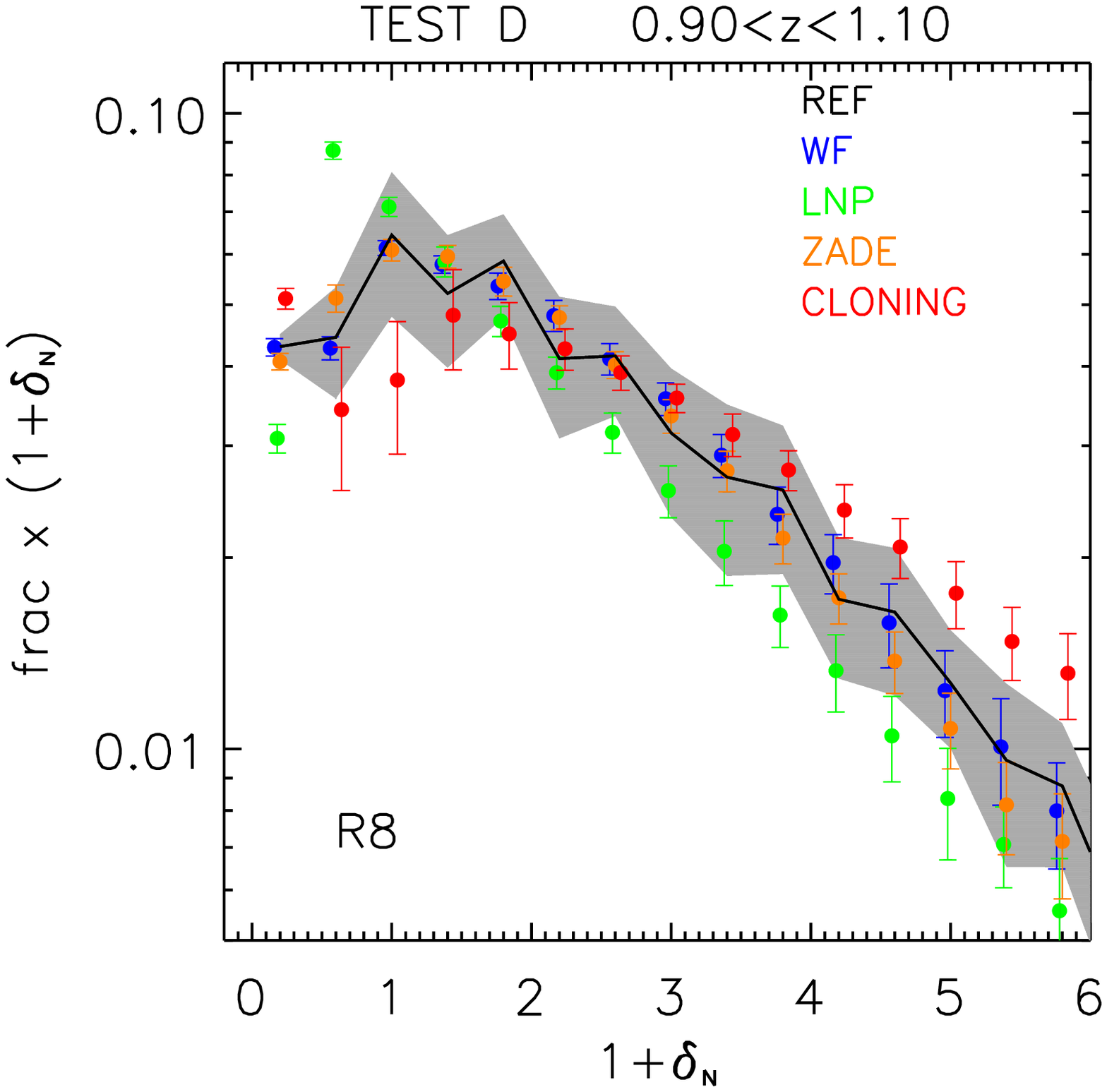}
\caption{Distributions of the reference (black line) and Test D overdensities
(lines) for the three redshift bins $0.5<z<0.7$, $0.7<z<0.9$, and
$0.9<z<1.1$ (from left to right), with tracers with $M_B - \log_{10}(h) \leq
-18.9 -z $, $M_B - \log_{10}(h) \leq -19.4 -z $, and $M_B - \log_{10}(h) \leq -19.9
-z $, respectively. Top row is for cells with $R=5\mpcoh$, bottom row for
cells with $R=8\mpcoh$. Different colours are for different methods, as
specified in the label: WF (blue), LNP (green), ZADE (orange),
cloning (red). Grey-shaded area is the $rms$ around the black line,
computed using the 26 light cones. To better
appreciate the differences among the series of points, the
$1+\delta_N$ distribution on the $y-$axis has been multiplied by 
the corresponding value of $1+\delta_N$ itself. }
\label{distrib_fig} \end{figure*}

In this section we study how well we can disentangle low and high
densities in the reference catalogue using the reconstructed counts in
Test D. Although the recovery of $P(\delta_N)$ is not the main goal of the
paper, but just an intermediate step to separate high-from low-density
environments, it is interesting to compare the $P(\delta_N^D)$
reconstructed with the various methods to the one obtained from the
reference mock catalogues $P(\delta_N^R)$. We make this comparison in
Fig.~\ref{distrib_fig}, in which we show the different probabilities
of counts multiplied by $(1+\delta_N)$ to emphasise differences in the
low- and high-count tails of the distribution.

The
figure shows that all methods recover the reference $P(\delta_N)$ at $\sim
1-2 \sigma$ level, with the largest differences being at the lowest
densities. We remind the reader that the WF and LNP methods produce
filtered counts, while the counts in our reference catalogue are not
filtered. We further verified that, as expected, 
the WF and LNP are in better agreement with the reference $P(\delta_N^R)$ 
when the comparison is made 
using densities obtained from smoothed counts also in the reference catalogue.

To disentangle low and high densities, we take the spheres with 
$\delta_N^D$ falling in the first and those falling in the fifth
quintile of $P(\delta_N^D)$, and we plot the corresponding
$P(\delta_N^R)$ for the selected spheres. As a reference, we do the
same also for the third quintile of $P(\delta_N^D)$. This way we
can verify whether, considering very different environments in the
reconstructed counts in cells, we are also sampling very different
environments in the reference catalogue.

\begin{figure*} \centering
\includegraphics[width=5.5cm]{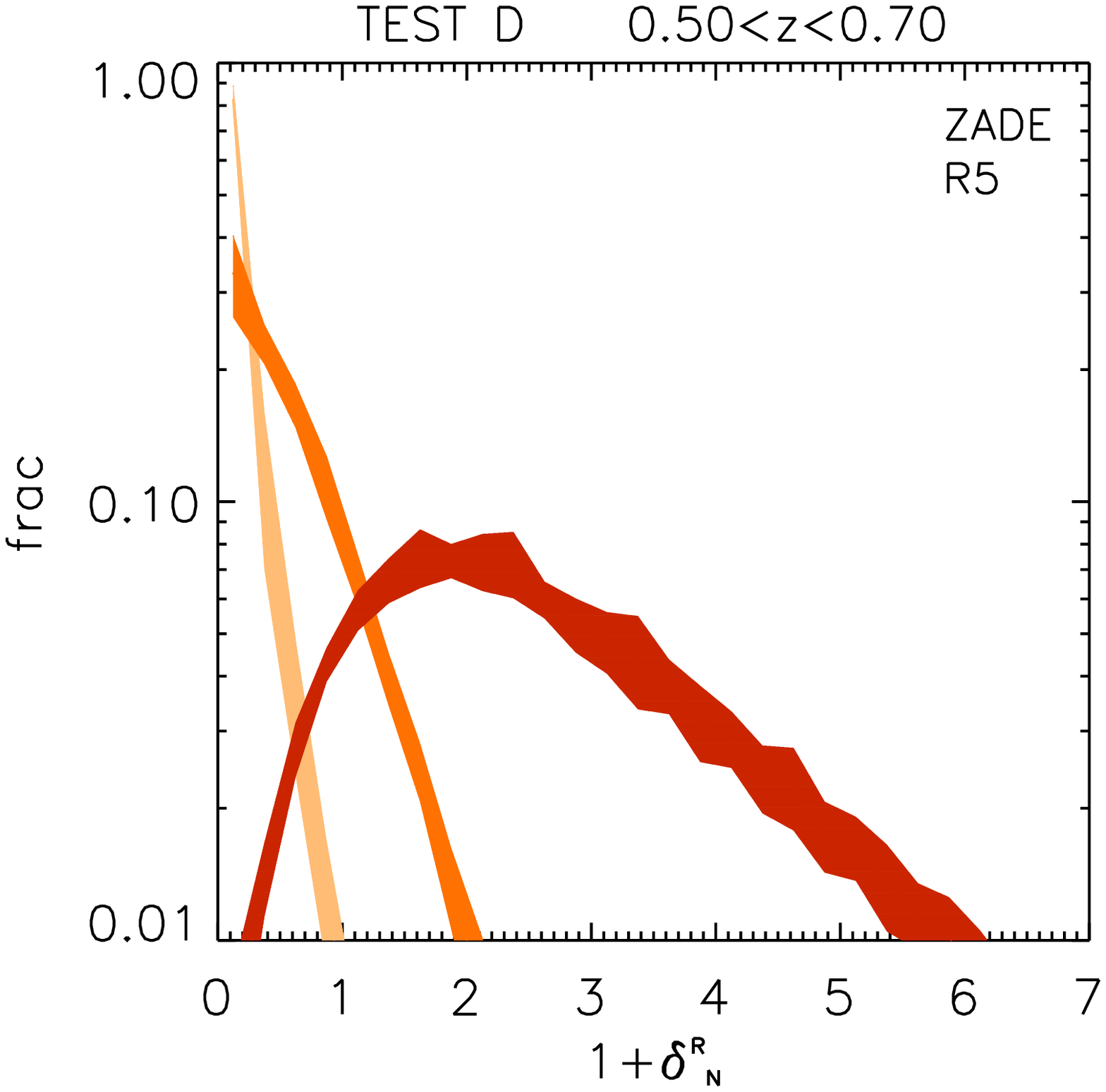}
\includegraphics[width=5.5cm]{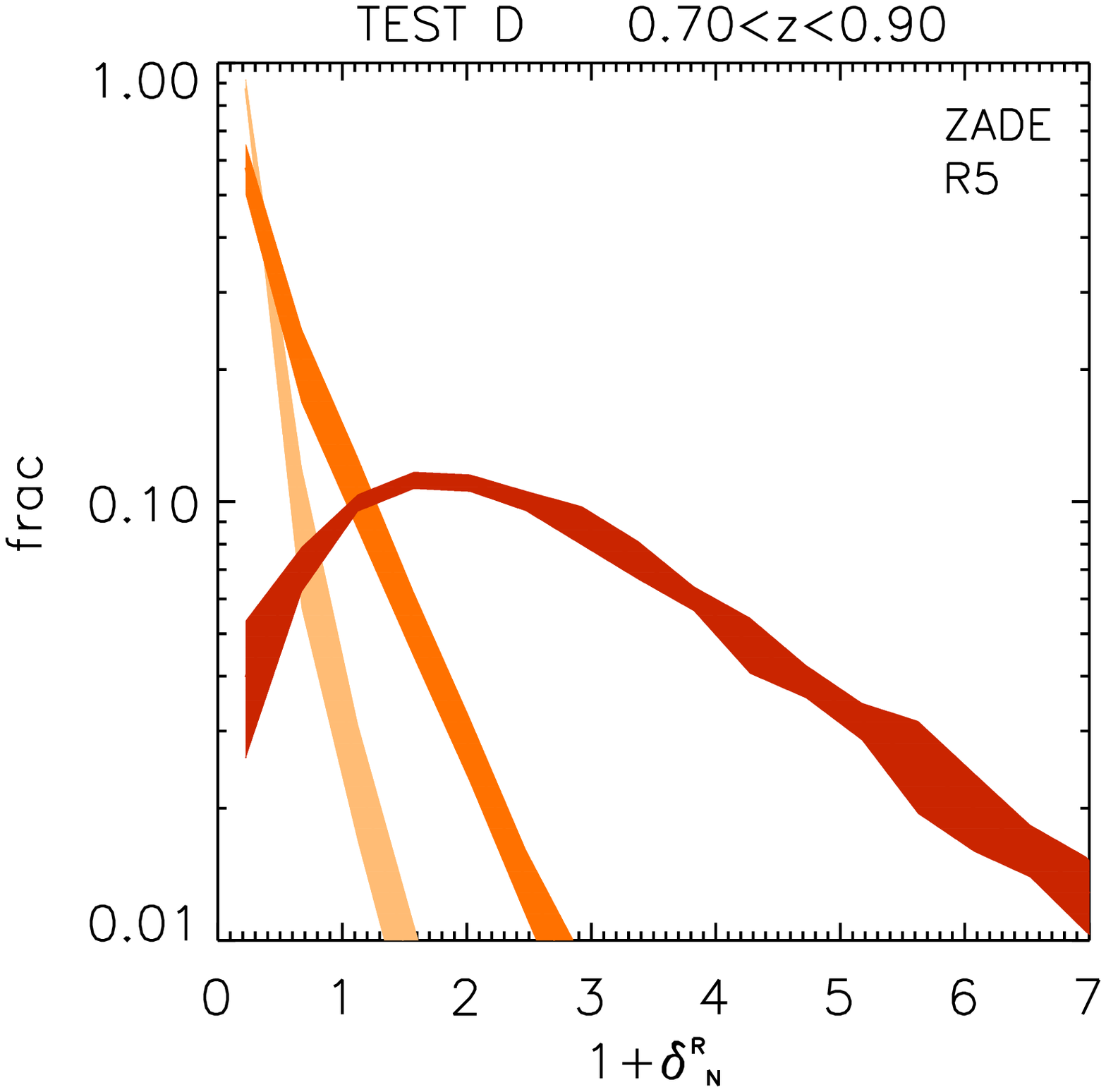}
\includegraphics[width=5.5cm]{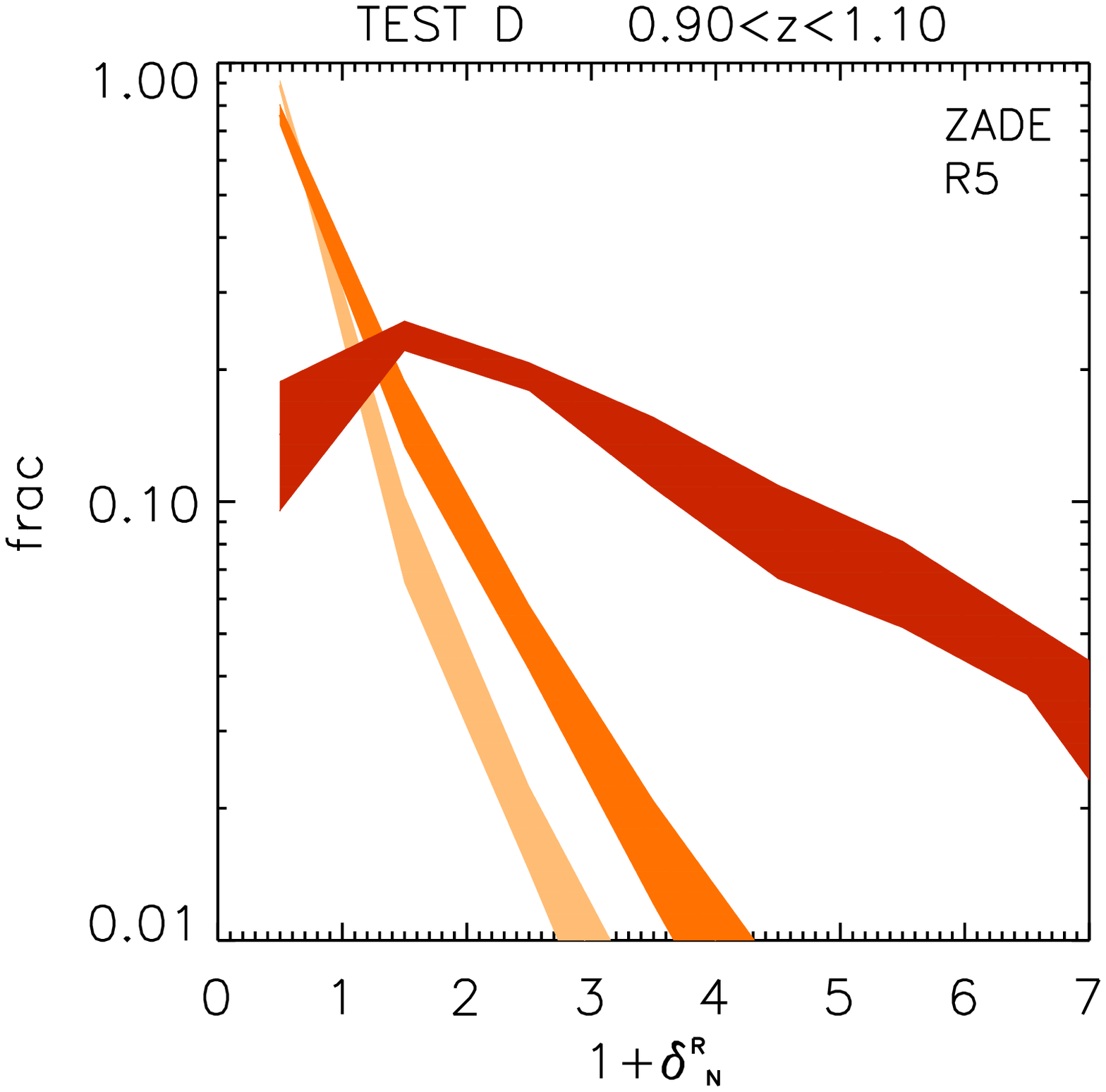}
\includegraphics[width=5.5cm]{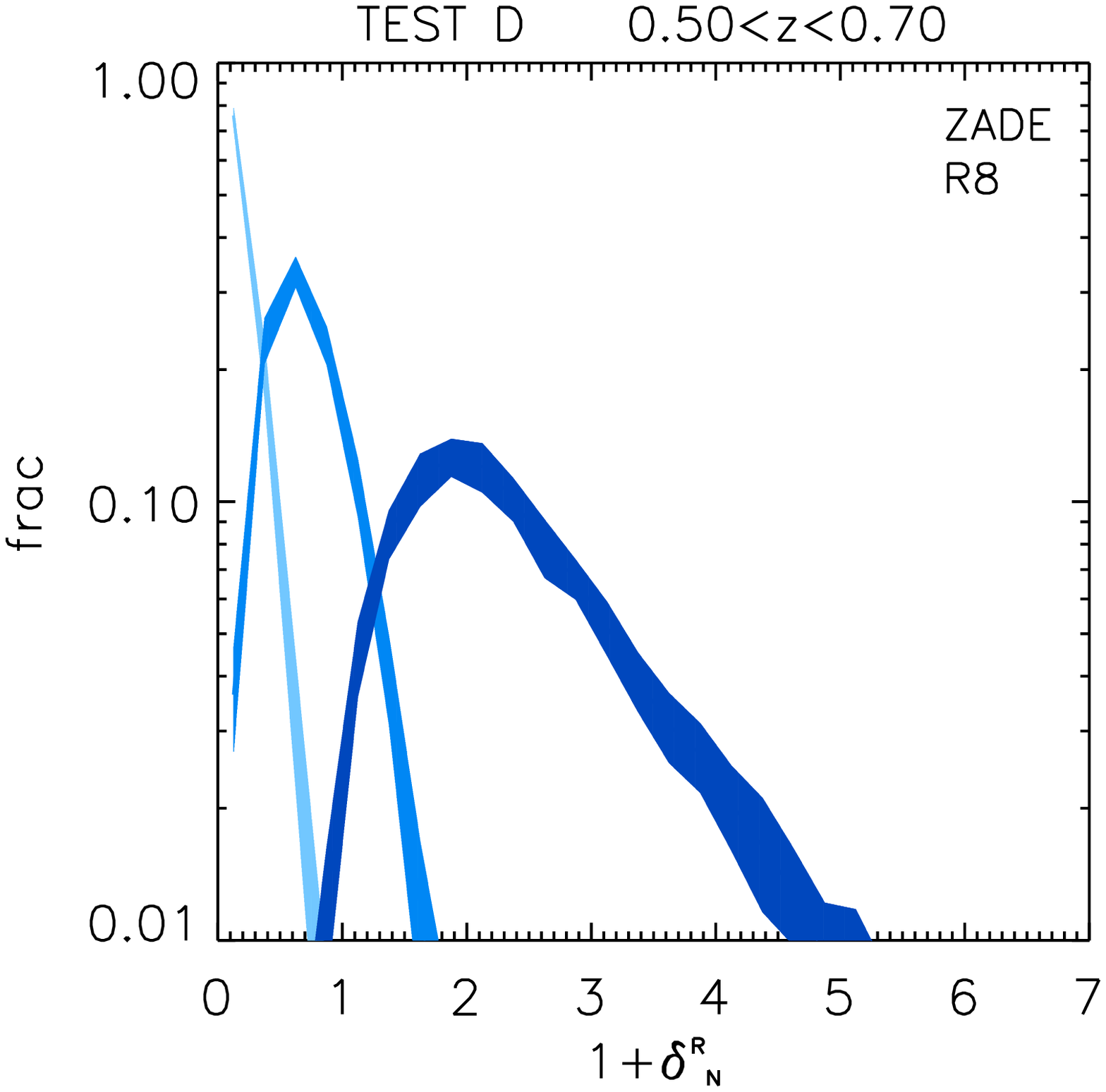}
\includegraphics[width=5.5cm]{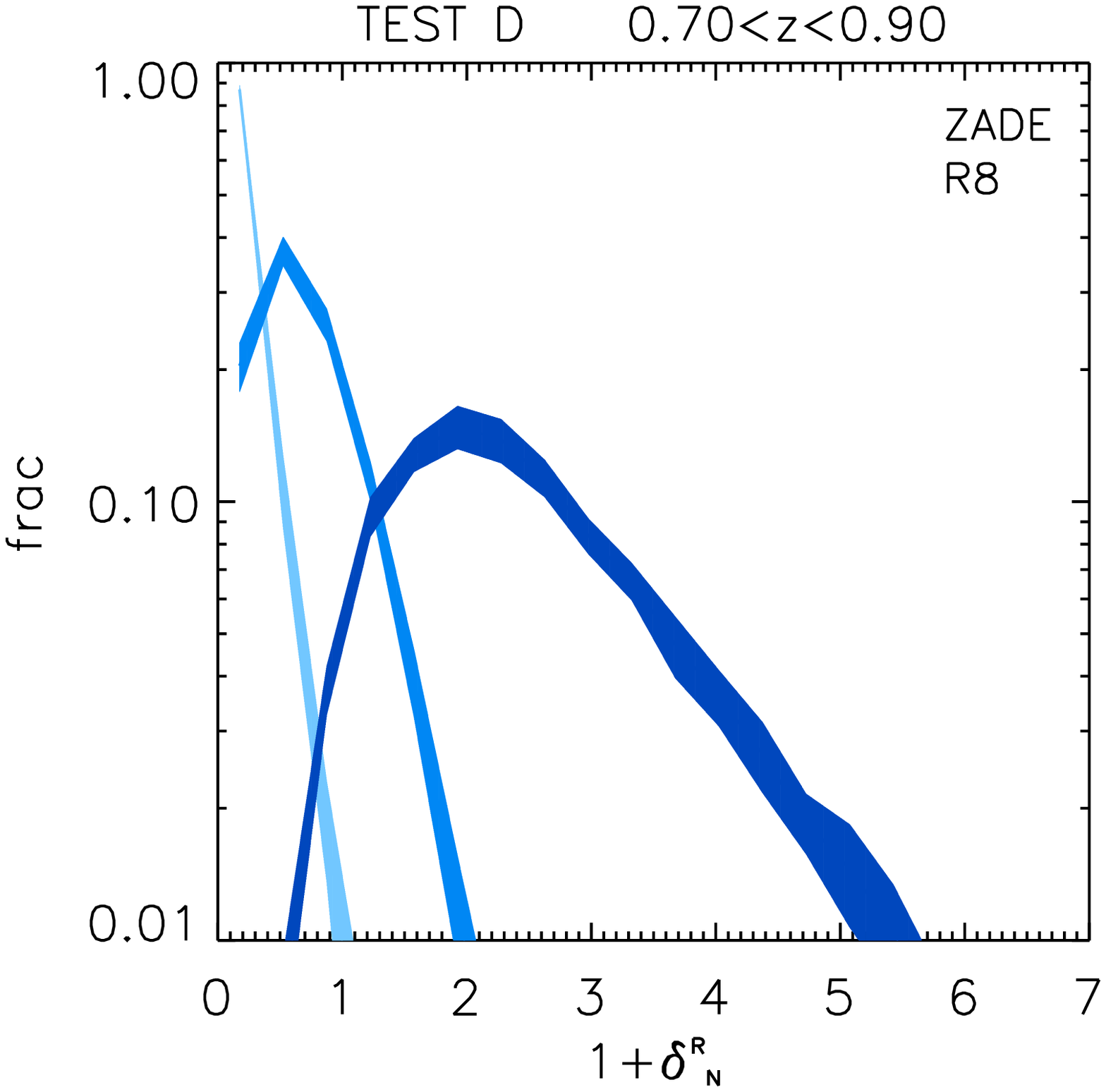}
\includegraphics[width=5.5cm]{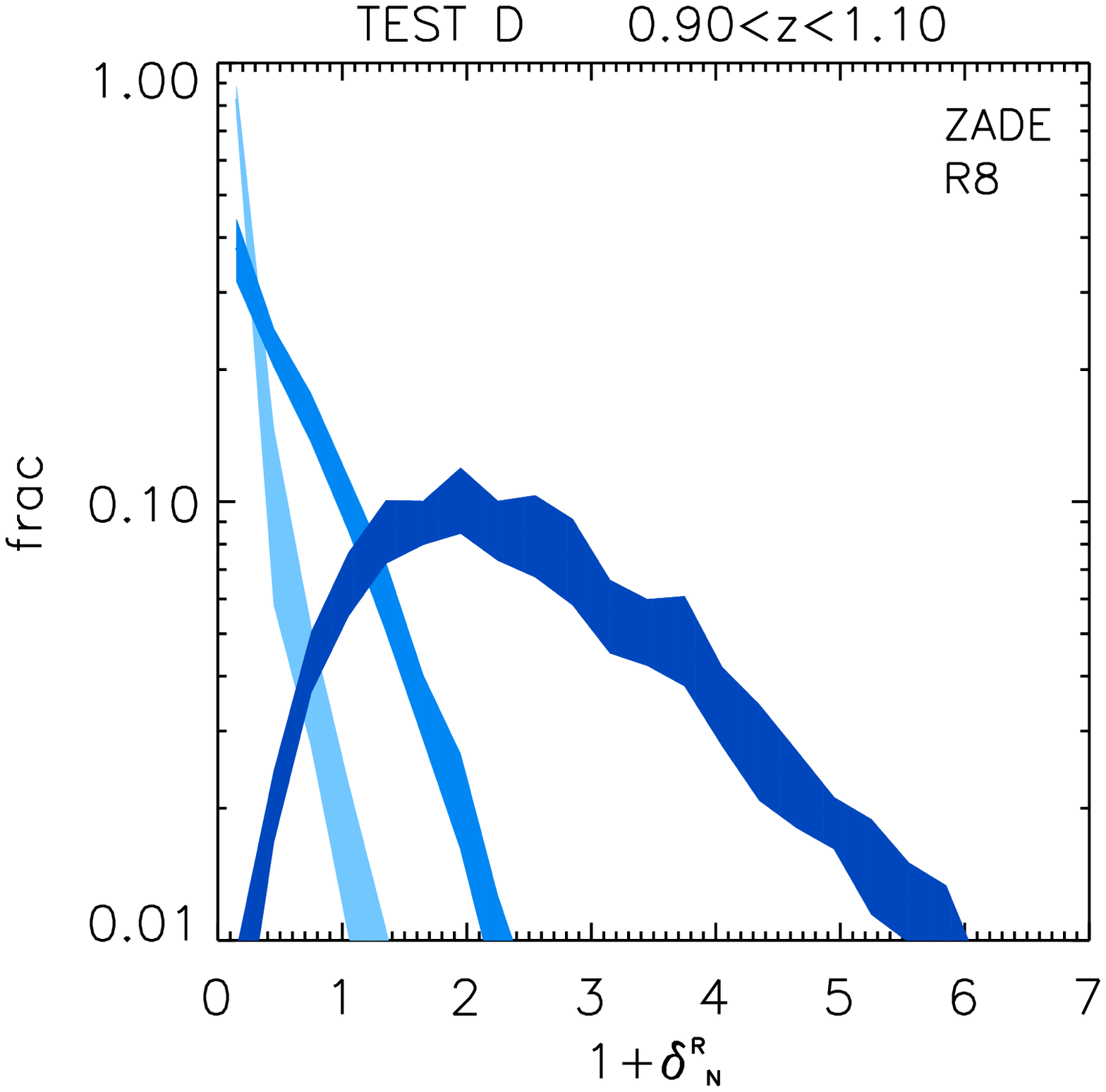}
\caption{Distributions of reference overdensities for the spheres falling in
the first, third and fifth quintiles of the density distribution computed
in Test D, using ZADE. Left, middle and right panels are for the three
redshift bins $0.5<z<0.7$, $0.7<z<0.9$ and $0.9<z<1.1$, with tracers
with $M_B - \log_{10}(h) \leq -18.9 -z $, $M_B - \log_{10}(h) \leq -19.4 -z $ and
$M_B - \log_{10}(h) \leq -19.9 -z$, respectively. Shades of red (top) are for cells
with $R=5\mpcoh$, while shades of blue (bottom) for cells with $R=8\mpcoh$. For
both radii, colours from light to dark are for the first, third and
fifth quintiles of the Test D density distribution. Bins on the x-axis are
different in all panels (for smoothing purposes), so normalisation is not comparable among panels. All distributions
are normalised to unity. Results obtained with the LNP method are very
similar. See also Table \ref{tail_tab}.}
\label{tails_fig} 
\end{figure*}

Results are shown in Fig.~\ref{tails_fig}. The main result is that, in
all cases, the first and fifth quintiles are separated well, for
both $R_5$ and $R_8$.  ZADE, LNP (not shown), and WF (not shown) give
very similar results. This is because none of the methods outperforms
the others, as also seen in Fig.~\ref{testD_1940}.  Table
\ref{tail_tab} lists the average values of the 16\%, 50\%, and 84\%
of all the distributions shown in Fig.~\ref{tails_fig}.  The average
is done on the 26 mock catalogues. These results hold also for the
lower and higher redshift bins. The table shows that the first 
and fifth quintile are always separated at least at $2\sigma$ level.

The intermediate densities (third quintile) are fully separated at
$2\sigma$ level from the highest densities ($5^{th}$ quintile) in almost 
all the cases listed in Table \ref{tail_tab} (the exceptions being the 
highest redshift bin for $R_5$, and the intermediate redshift bin for 
$R_5$ but only for the LNP method). In contrast, it is harder to 
separate the densities in the first and third quartiles for $R_5$, 
irrespective of the redshift bin and of the method used.  This is due 
to the
skewness of the $P(\delta_N)$ and the large relative error in the
reconstruction of low densities.  Even if the Test D
reconstruction is good enough to maintain a good shape of the
$P(\delta_N)$ (see Fig.~\ref{distrib_fig}), the relative error at low
densities is too large to properly distinguish, locally, between low and
intermediate densities. For $R_8$, the lowest and intermediate densities 
are separated at $\sim2\sigma$ level, at least for $z<0.9$.

\begin{table*} 
\caption{Values of the 16\%, 50\% and 84\% of all the distributions of $(1+\delta_N^R)$ shown in 
Fig.~\ref{tails_fig} (ZADE), and the corresponding values obtained with LNP. These values are 
given as mean values among all the used mock catalogues. Their typical $rms$ is of the order of 5-10\% for $R_5$. For $R_8$, 
the typical $rms$  is of the order of 5-10\% for the first and third quintiles, and of the order 
of 2-3\% in the highest density quintile.}
\label{tail_tab} 
\centering 
\begin{tabular}{l l | r r r | r r r | r r r}
\hline\hline 
 {\bf Redshift} & {\bf Method}   &   \multicolumn{3}{c|}{{\bf 1$^{st}$ 20\%}} &  \multicolumn{3}{c|}{{\bf 3$^{rd}$ 20\%}} & \multicolumn{3}{c|}{{\bf 5$^{th}$ 20\%}}  \\    
&   &  16\%  &  50\%  &  84\% & 16\%  &  50\%  &  84\% & 16\%  &  50\%  &  84\% \\    
\hline	\hline

\multicolumn{11}{c|}{{\bf $R=5\mpcoh$}} \\
\hline
{\bf 0.5$<$z$<$0.7}   & LNP  & 0.0 &   0.0 &  0.25 &  0.0 & 0.46 &   1.06 & 1.18 &   2.42  &  4.51	\\
                      & ZADE & 0.0 &   0.0 &  0.25 &  0.1 & 0.46 &   1.01 & 1.31 &   2.50  &  4.53	\\
\hline				   										      
{\bf 0.7$<$z$<$0.9}   & LNP   & 0.0 &	0.0 &  0.44 & 0.0 &  0.43 &	 1.09 &  0.94 &     2.39 &	  4.88 \\
                      & ZADE  & 0.0 &	0.0 &  0.44 & 0.0 &  0.42 &	 0.95 &  1.09 &     2.49 &	  4.89 \\
\hline														      
{\bf 0.9$<$z$<$1.1}   & LNP  & 0.0 &   0.0 &	0.0 &  0.0 &   0.0 &   1.19 &  0.50  &	 2.32 &  5.67  \\
                      & ZADE  &0.0 &   0.0 &    0.0 &  0.0 &   0.0 &   1.16 &  1.10 &	 2.39 &  5.72  \\

\hline	\hline

\multicolumn{11}{c|}{{\bf $R=8\mpcoh$}}\\
\hline

{\bf 0.5$<$z$<$0.7  }  & LNP  & 0.0 &	  0.14 &      0.34 &  0.38 &	  0.66 &     1.05 & 1.44 &	2.20 &       3.50      \\
                       & ZADE & 0.0 &	  0.14 &      0.31 &  0.41 &	  0.67 &     1.01 & 1.54 &	2.24 &       3.53     \\
\hline					
{\bf 0.7$<$z$<$0.9  }  & LNP   & 0.0 &	   0.11  &     0.35 & 0.30 &	  0.62 &       1.12 & 1.36 &	   2.25 &	3.57   \\
                       & ZADE  & 0.0 &	   0.10  &     0.32 & 0.32 &	  0.62 &       1.07 & 1.47 &	   2.29 &       3.61    \\
\hline					
{\bf 0.9$<$z$<$1.1  }  & LNP  &   0.0 &    0.0 &   0.37 &     0.01 &	0.52 &   1.19 & 1.18 &	   2.31 &   4.10  \\ 
                       & ZADE  &  0.0 &    0.0 &   0.30 &     0.13&	0.52 &   1.11 & 1.30 &	   2.37 &   4.14 \\
\hline

\hline				   										      
\end{tabular} 
\end{table*}


\subsection{Application to VIPERS}

We computed the counts in cells in the two VIPERS fields, in the same
spheres used for the mocks catalogues. Gaps and low sampling rate have
been accounted for using the ZADE method. The spectroscopic sample
that we used is the one described in Sect.~\ref{data}. The photometric
sample used to fill the gaps and reach 100\% sampling rate in
quadrants corresponds to the full photometric catalogue in the fields
W1 and W4, complete down to $i_{AB}=22.5$ and from which we removed
the galaxies included in our spectroscopic sample.

\begin{figure*} \centering
\includegraphics[width=6.0cm]{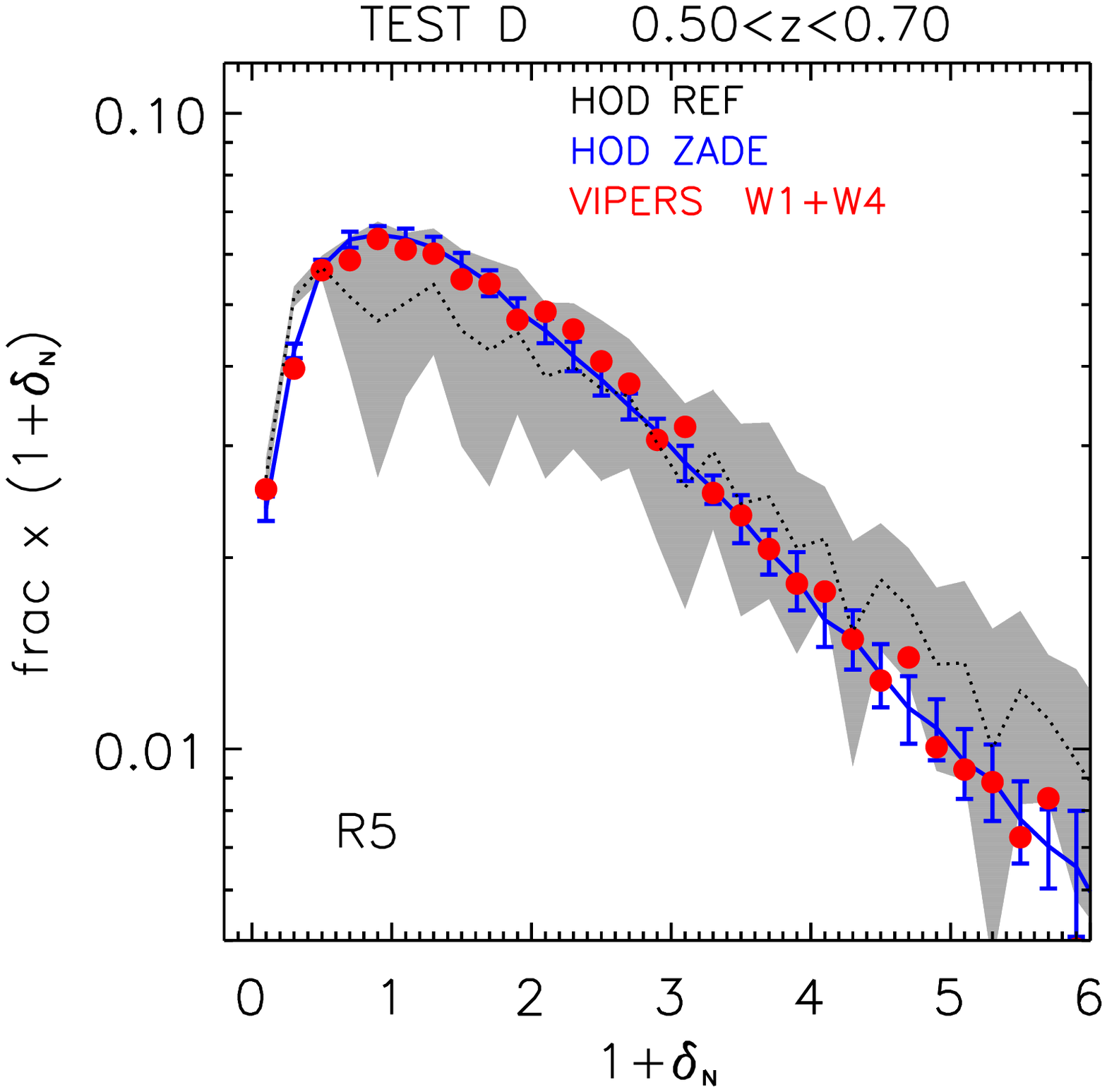}
\includegraphics[width=6.0cm]{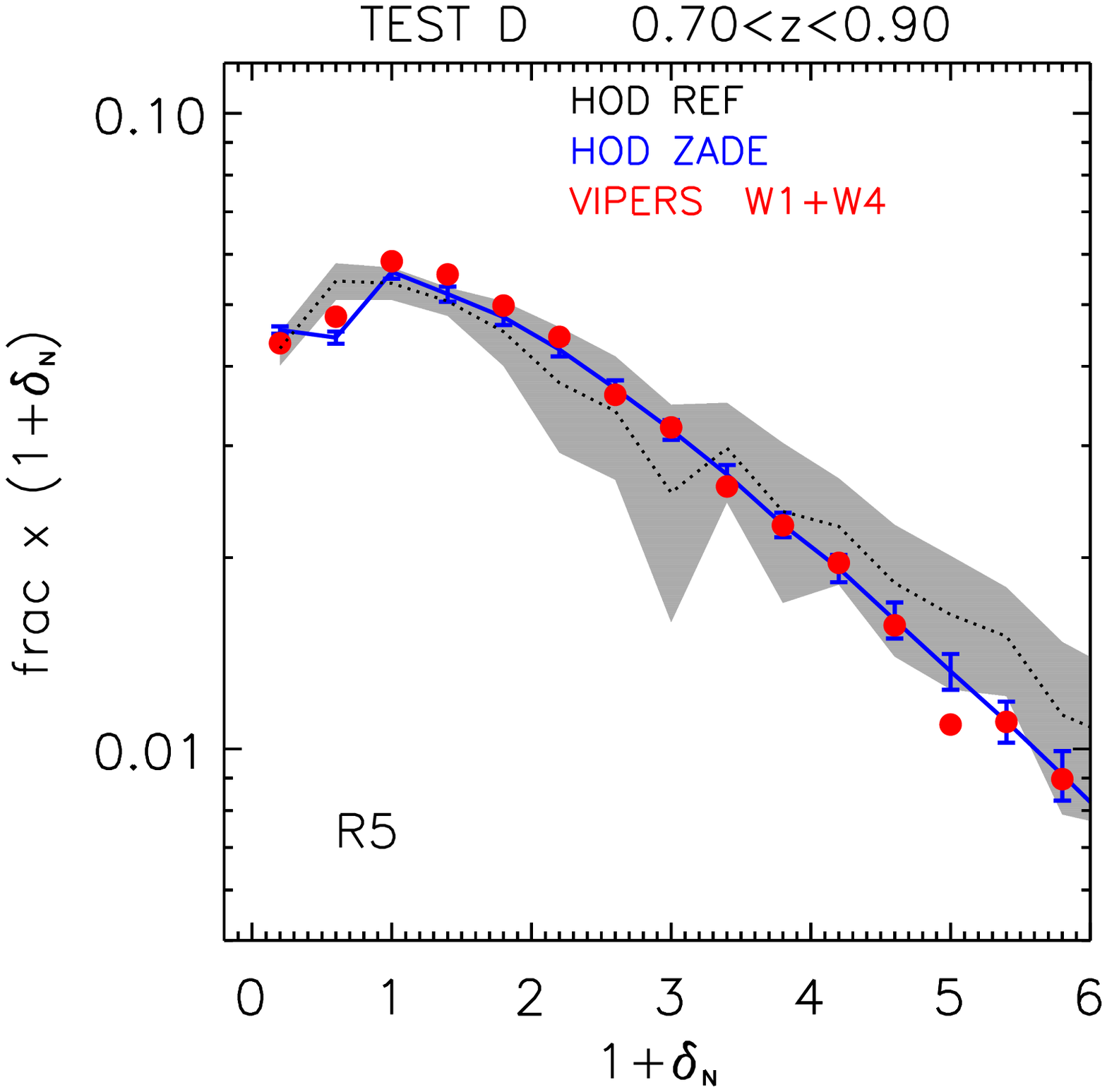}
\includegraphics[width=6.0cm]{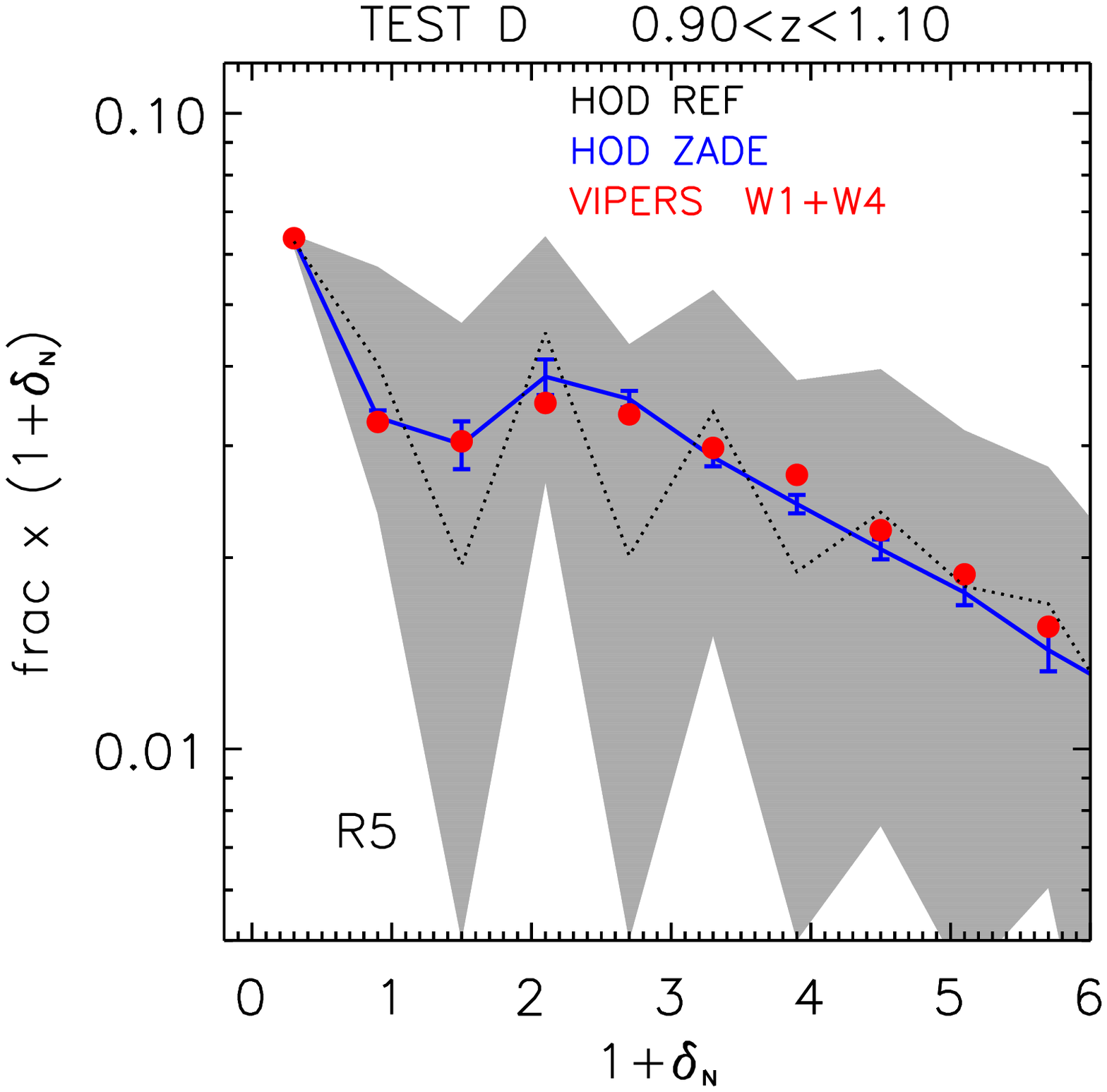}
\includegraphics[width=6.0cm]{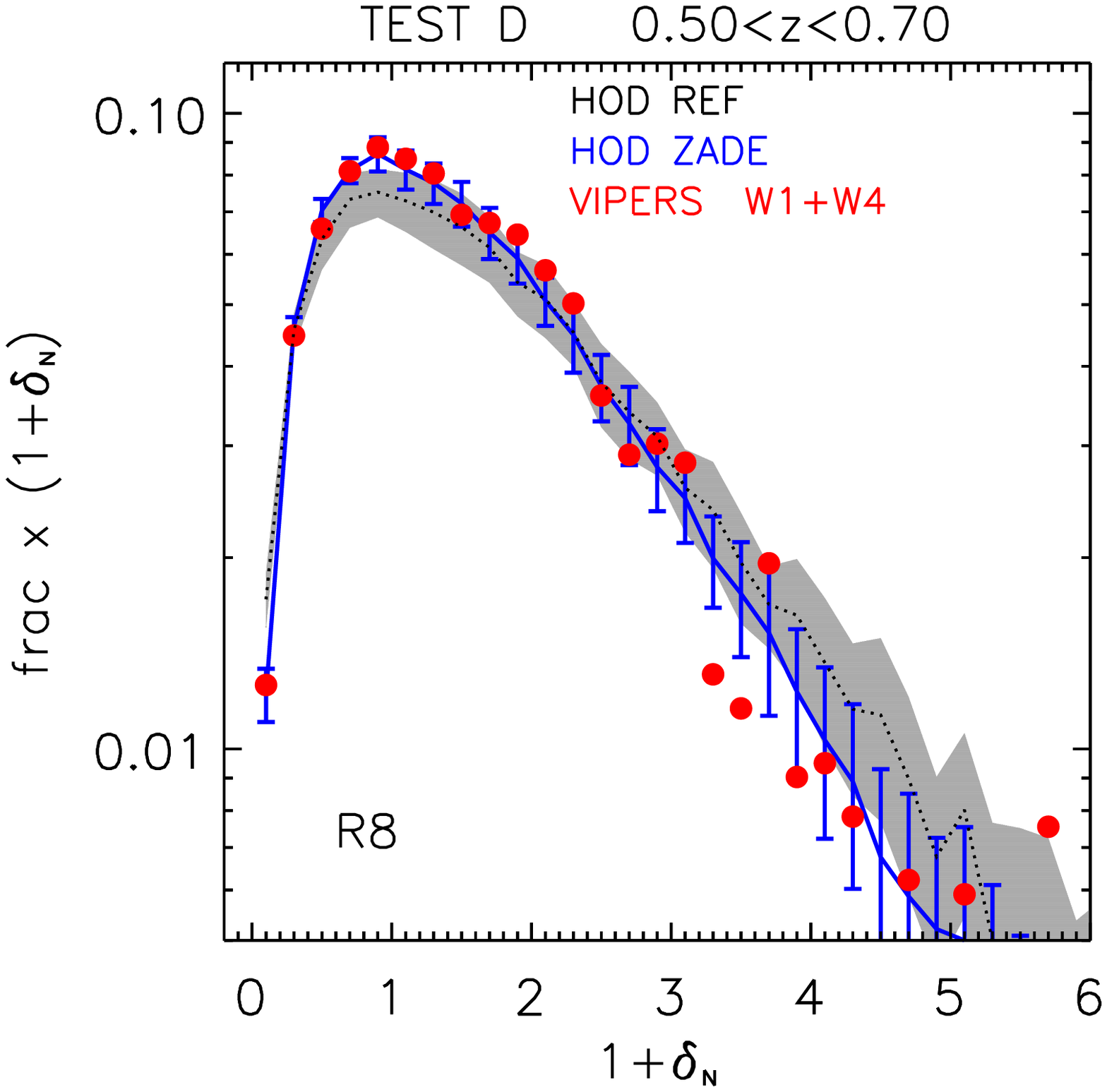}
\includegraphics[width=6.0cm]{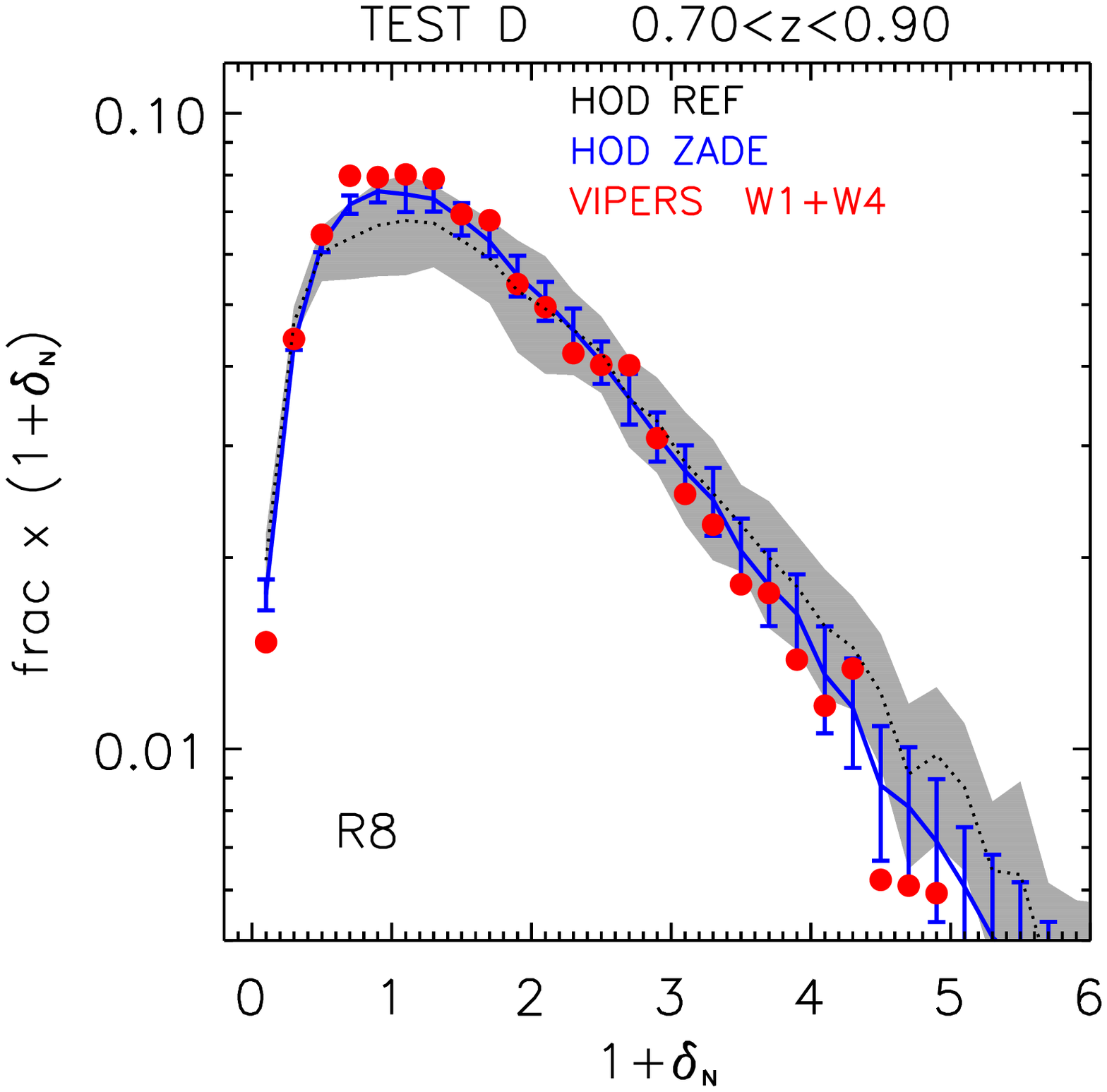}
\includegraphics[width=6.0cm]{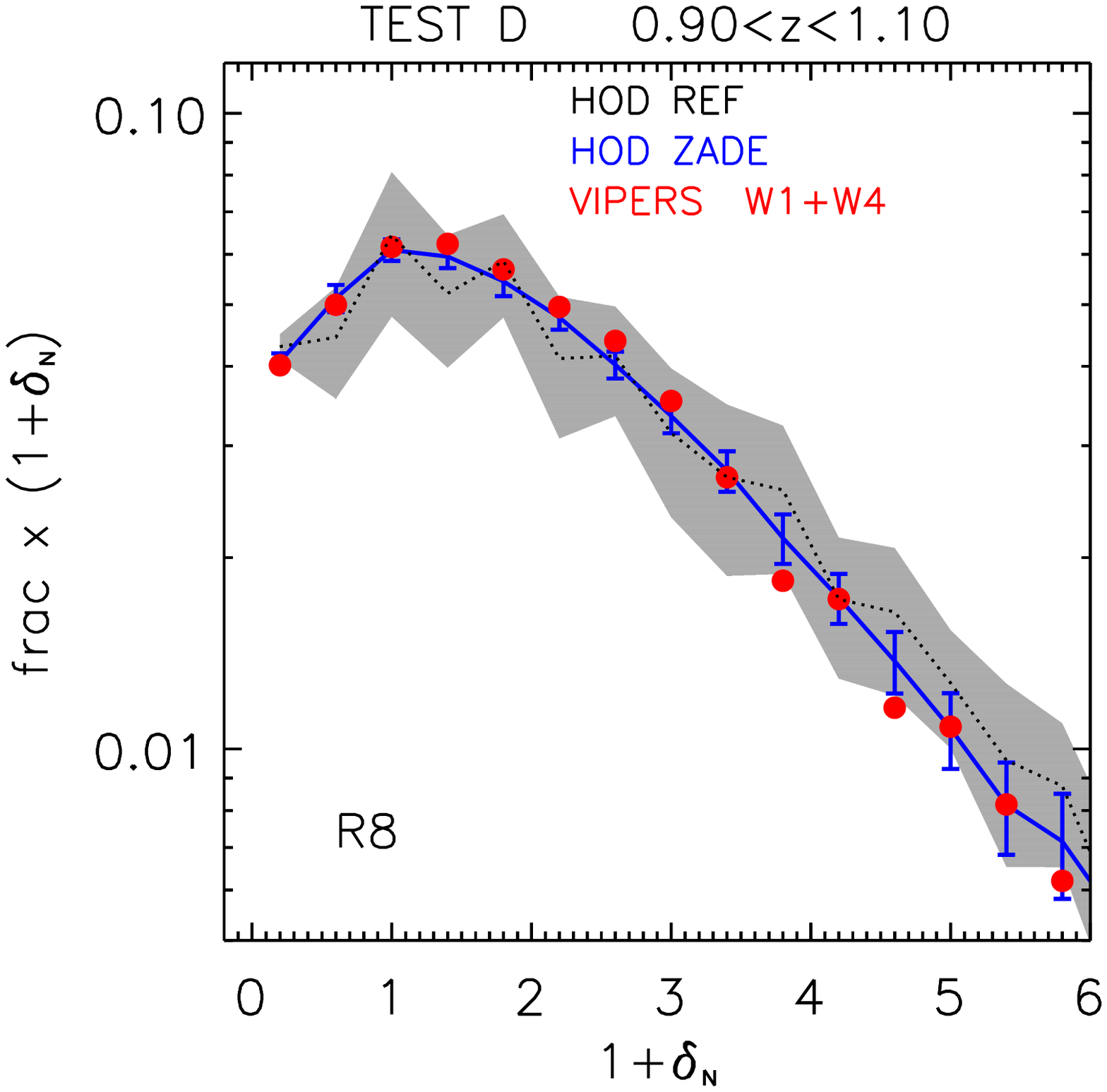}
\caption{Distributions of reference (black line) and real VIPERS
overdensities (points) for the three redshift bins $0.5<z<0.7$,
$0.7<z<0.9$, and $0.9<z<1.1$ (from left to right), with tracers with
$M_B - \log_{10}(h) \leq -18.9 -z$, $M_B - \log_{10}(h) \leq -19.4 -z$, and $M_B -
\log_{10}(h) \leq -19.9 -z $, respectively. Grey-shaded area is the $rms$
around the black line, computed using the 26 light cones. Top row is
for cells with $R=5\mpcoh$, bottom row for cells with $R=8\mpcoh$. Real
VIPERS overdensities are computed with the ZADE method. For
comparison, the blue solid line with vertical error bars is the result
of Test D (ZADE). To better
appreciate the differences among the series of points, the
$(1+\delta_N)$ distribution on the $y-$axis has been multiplied by the corresponding value of $1+\delta_N$ itself.} 
\label{distrib_real_fig} 
\end{figure*}

Figure ~\ref{distrib_real_fig} shows the probability of galaxy
overdensity $P(1+\delta_N)$ for the VIPERS sample (fields W1 and W4
altogether), compared with the $P(1+\delta_N)$ in the reference mock
catalogues and the one reconstructed in test D using ZADE.  It is
evident that the agreement between the VIPERS $P(1+\delta_N)$, and the
one reconstructed in VIPERS-like mocks is very good. For instance, the
$20^{th}$, $50^{th}$, and $80^{th}$ percentiles of the $1+\delta$
distribution for $R_5$ and $0.7<z<0.9$ are 0.09, 0.55, and 1.65 for
VIPERS, $0$, $0.44\pm0.02$, $1.7\pm0.1$ for the reference mock
catalogue, and $0.08\pm0.006$, $0.50\pm0.01$, $1.66\pm0.01$ for the
VIPERS-like mock catalogues (Test D, reconstructed using ZADE). One
can also notice that the real VIPERS density distribution at
$0.7<z<0.9$ around the mean density ($1+\delta_N=1$) is higher than
the one in the mock catalogues, but it is a very weak effect.

We note that this general agreement is not necessarily there by
construction, since the HOD mock catalogues have been tuned to match the
clustering properties (namely, the two point correlation function, see
e.g. \citealp{delaTorre2013_clustering} and
\citealp{marulli2013_clustering}) in VIPERS, and not the over-density
distribution.

\section{Discussion and conclusions}\label{discussion}

The goal of this work has been to find the best way to fill empty areas in
spectroscopic surveys so as to obtain the most reliable counts-in-cell
reconstruction. As a test case, we study how to fill the
gaps across VIMOS quadrants in the VIPERS survey.  To tackle the
problem, we applied four methods to fill the gaps and compared their
performances using mock galaxy catalogues mimicking the sources of
errors we want to correct for (not only gaps, but also sparse sampling
and related systematic effects). For the first time we have performed a systematic comparison
among gap-filling methods using different techniques that
either directly use the observed, discrete galaxy distribution to fill
the gaps or do this within the more general framework of reconstructing
the underlying, continuous mass density field using suitable filtering
techniques.

Our results can be summarised as follows:
\begin{enumerate}
\item On the scales we tested, the error budget is dominated by the
sparseness of the sample.

\item All methods under-predict counts in high-density regions. This bias
is in the range of 20-35\%, depending on the cell size, method, and
overdensity. This systematic bias is similar to random
errors.

\item Random errors have similar amplitude for all methods, except
cloning for which errors are significantly larger.

\item No method largely outperforms the others. LNP is certainly better
than WF, and ZADE is to be preferred to cloning, but differences are
not large, and methods with the smallest random errors (ZADE) can be
more affected by systematic errors than others.

\item Random and systematic errors decrease with the increasing size of the cell,
although in both cases considered ($R_5$ and $R_8$), systematic and random
errors are of similar amplitude.

\item Recovering the correct overdensity in a generic resolution element
is a more demanding test than reconstructing the correct one-point
counts statistic. Our tests show that in a survey such as VIPERS the
galaxy overdensity in a resolution element of 5 to $8\mpcoh$ can be
reconstructed with a typical error of about 25\%, knowing that the
estimate will be biased low by a similar amount.

\item With all methods, the first and fifth quintiles of the
distribution of $(1+\delta_N^D)$ correspond to well-separated regimes of
the $(1+\delta_N^R)$ distribution, at least at a $2\sigma$ level for
both $R_5$ and $R_8$.

\end{enumerate}

These results are quantitatively exact for VIPERS,
because we tested the gap  size and the sampling
rate of this specific survey. Different values of gap  size and/or sampling rate could produce
a different result. Nevertheless, our results are of general interest 
for all surveys utilising a mosaic of non-contiguous fields.

Our results on the reconstruction of counts in cells are the basis 
to understand how well we can parameterise the galaxy local environment 
in a survey like VIPERS. The study of how environment affects galaxy 
properties is crucial to understanding galaxy evolution.

Clearly, for environmental studies, the counts in spherical cells
distributed randomly in the field might not be the best choice,
depending on the specific goal one wants to reach.  For instance,
other cell shapes can be preferred, such as cylindrical filters, that
would allow us to take the `Fingers of God' 
effect \citep{jackson72} into account in the
highest densities. Moreover, the use of an adaptive scale (such as the
distance from the $n^{th}$ nearest neighbour) could help 
reconstruct the highest densities. Thanks to mock galaxy 
catalogues embedded in DM simulations, it is also possible to 
compare the reconstructed density field to the underlying DM 
distribution. Such a comparison can give clues to  
how and why the local environment affects galaxy evolution 
(see e.g. \citealp{haas12,muldrew12,cucciati12_MILL,hirschmann13}).
Building on the results we obtained in the present work, we will 
describe the parameterisation of local density in the VIPERS survey for
environmental studies in a future work (Cucciati et al., in prep.)

We would like to point out that VIPERS has been designed to be a 
cosmological survey,
but thanks to its observational strategy, the density of tracers is
higher than other cosmological surveys at the same redshift (e.g. the WiggleZ survey, 
see \citealp{drinkwater10_wiggleZ}). We have shown that this allows, for the first time, the
reconstruction of environment on scales of $5\mpcoh$ on a very large
field at $0.5\lesssim z \lesssim 1.1$. This will have a strong 
impact in the characterisation of cosmic variance and rare populations
(e.g. brightest galaxies) in environmental studies at this redshift.
Moreover, the results obtained in this work with the ZADE method are 
also promising for environmental studies in future surveys such as EUCLID, 
which is based on a spectroscopic data set complemented by a 
photometric galaxy catalogue, with a photometric redshift error similar 
to the one in VIPERS.


\begin{acknowledgements} We acknowledge the crucial contribution of
the ESO staff in the management of service observations. In
particular, we are deeply grateful to M. Hilker for his constant help
and support of this programme. Italian participation in VIPERS has
been funded by INAF through PRIN 2008 and 2010 programmes. LG and BRG
acknowledge support of the European Research Council through the
Darklight ERC Advanced Research Grant (\# 291521). LAMT
acknowledges support of the European Research Council through the EARLY
ERC Advanced Research Grant (\# 268107). Polish participants have been
supported by the Polish National Science Centre (grants N N203 51 29
38 and 2012/07/B/ST9/04425), the Polish-Swiss Astro Project
(co-financed by a grant from Switzerland, through the Swiss
Contribution to the enlarged European Union), the European Associated
Laboratory Astrophysics Poland-France HECOLS, and a Japan Society for
the Promotion of Science (JSPS) Postdoctoral Fellowship for Foreign
Researchers (P11802). OC, EB, FM, and LM acknowledge the support
from grants ASI-INAF I/023/12/0 and PRIN MIUR 2010-2011. LM also
acknowledges financial support from PRIN INAF 2012. 
\end{acknowledgements}

\bibliographystyle{aa}
\bibliography{biblio}

\appendix


\section{Results of the Tests A, B, C1 and C2}\label{app_testABC}

\subsection{Test A}\label{testA_sec}

\begin{figure*} \centering
\includegraphics[width=4cm]{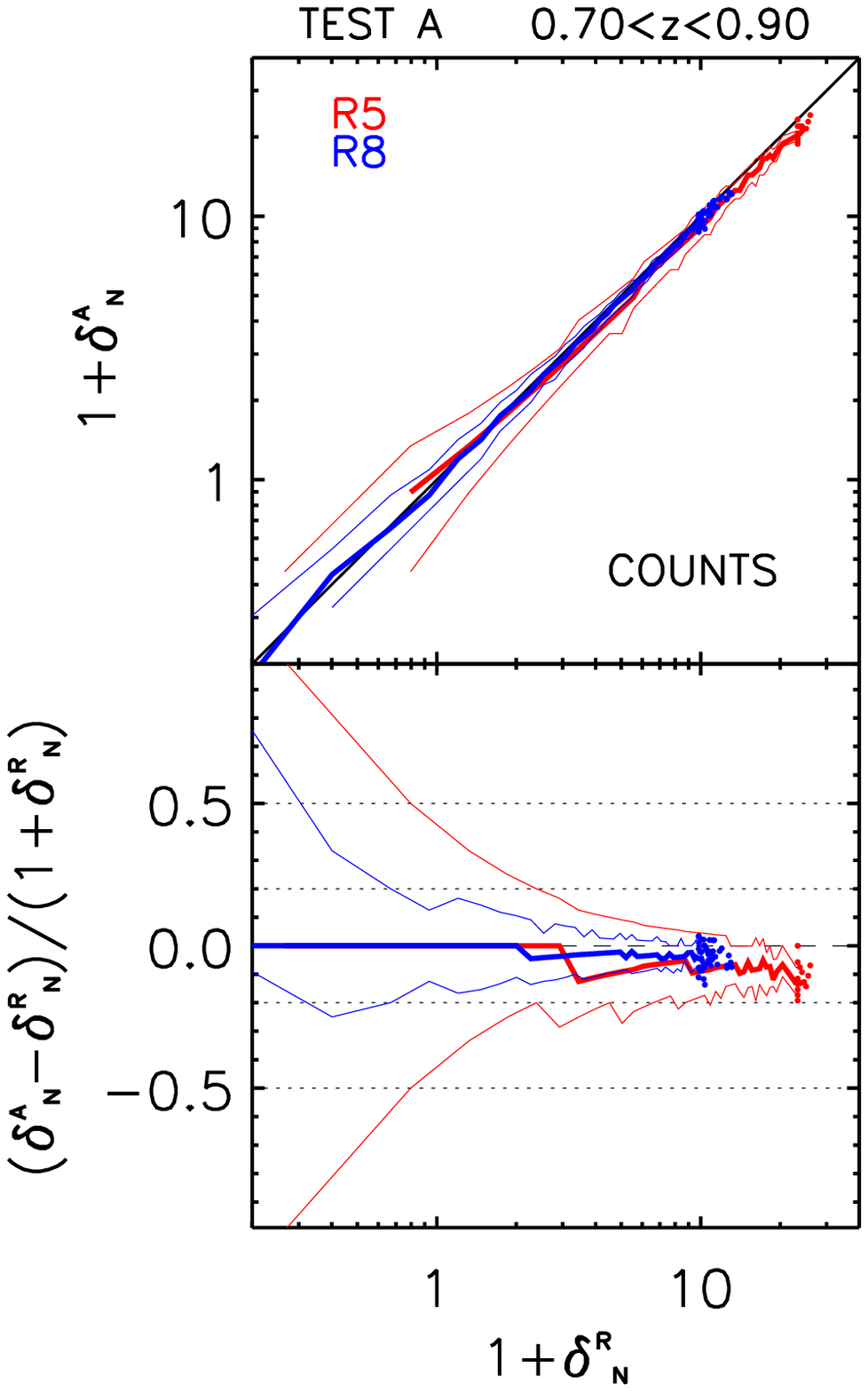}
\includegraphics[width=4cm]{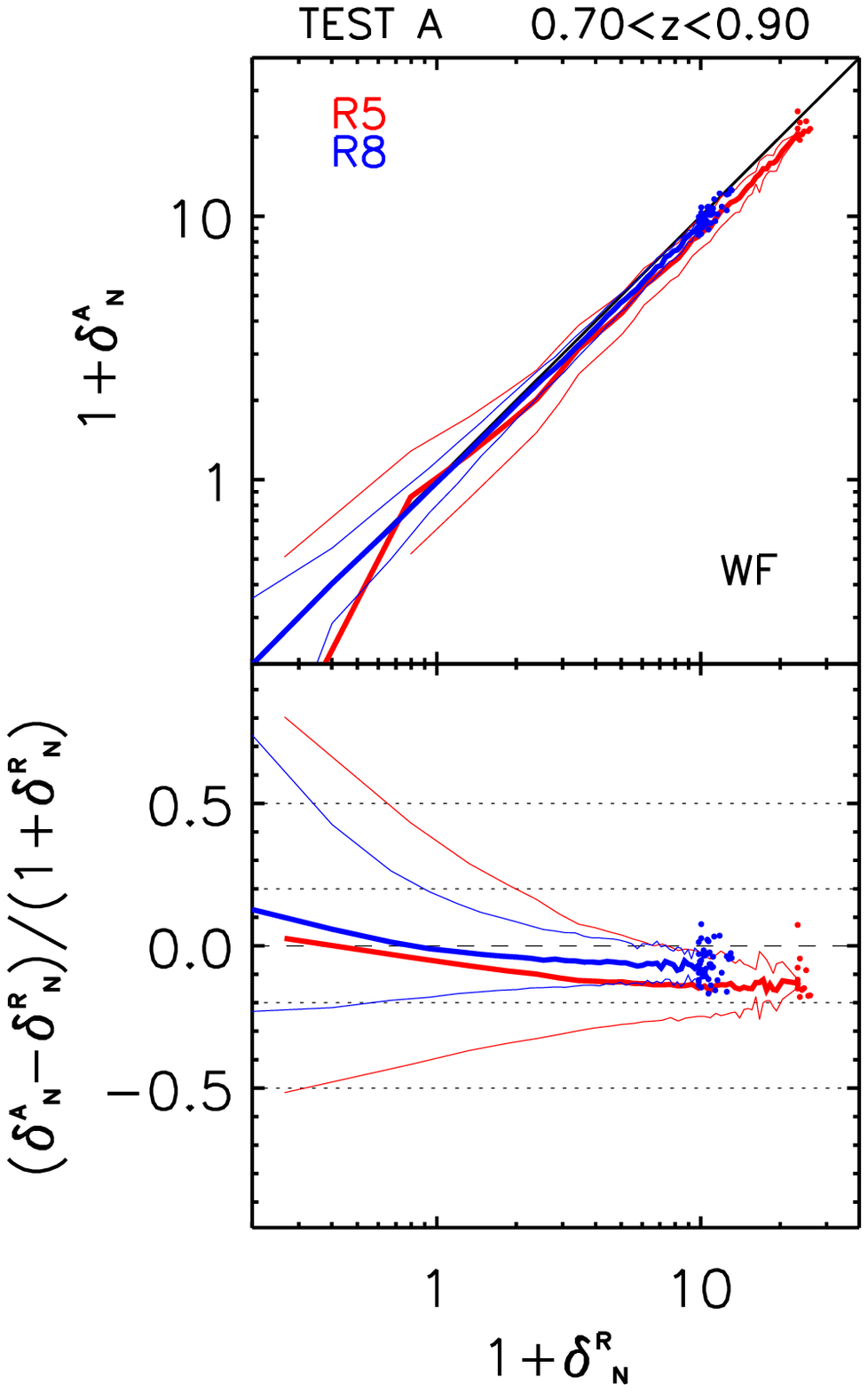}
\includegraphics[width=4cm]{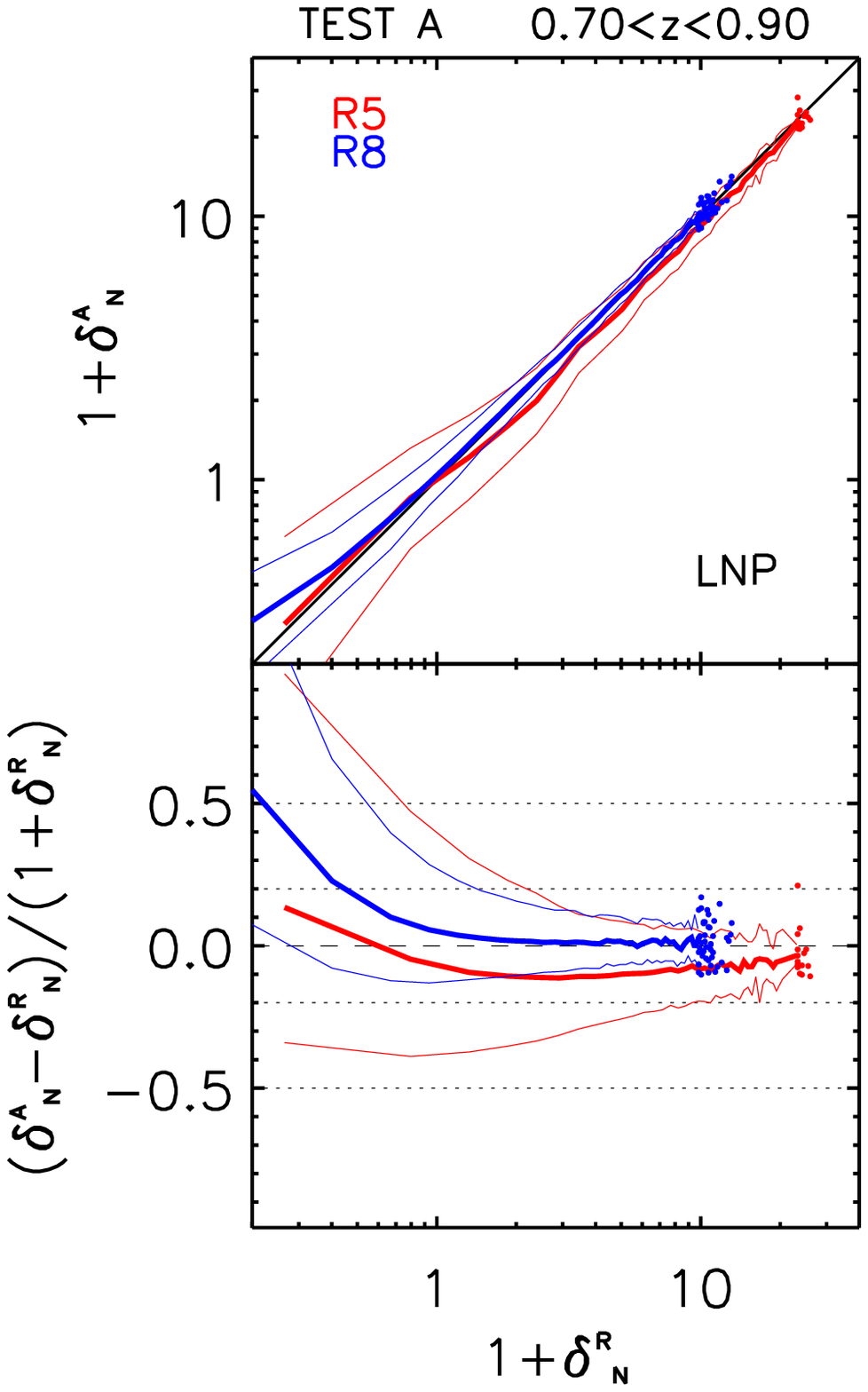}
\caption{As in Fig.~\ref{testD_1940}, but for Test A. ZADE and cloning
are not used. $\delta^A_N$ in the left panel represents counts in cells in Test A without any attempt to correct for the spectroscopic redshift error (see text for details). }
\label{testA_1940} \end{figure*}

The goal of this test is to assess the impact of redshift measurement
errors on the counts in cells. Results are shown in
Fig.~\ref{testA_1940}. The ZADE and cloning methods do not correct for
spectroscopic redshift error, so for this test we only compare the WF
and LNP methods.

\begin{figure*} \centering
\includegraphics[width=4.4cm]{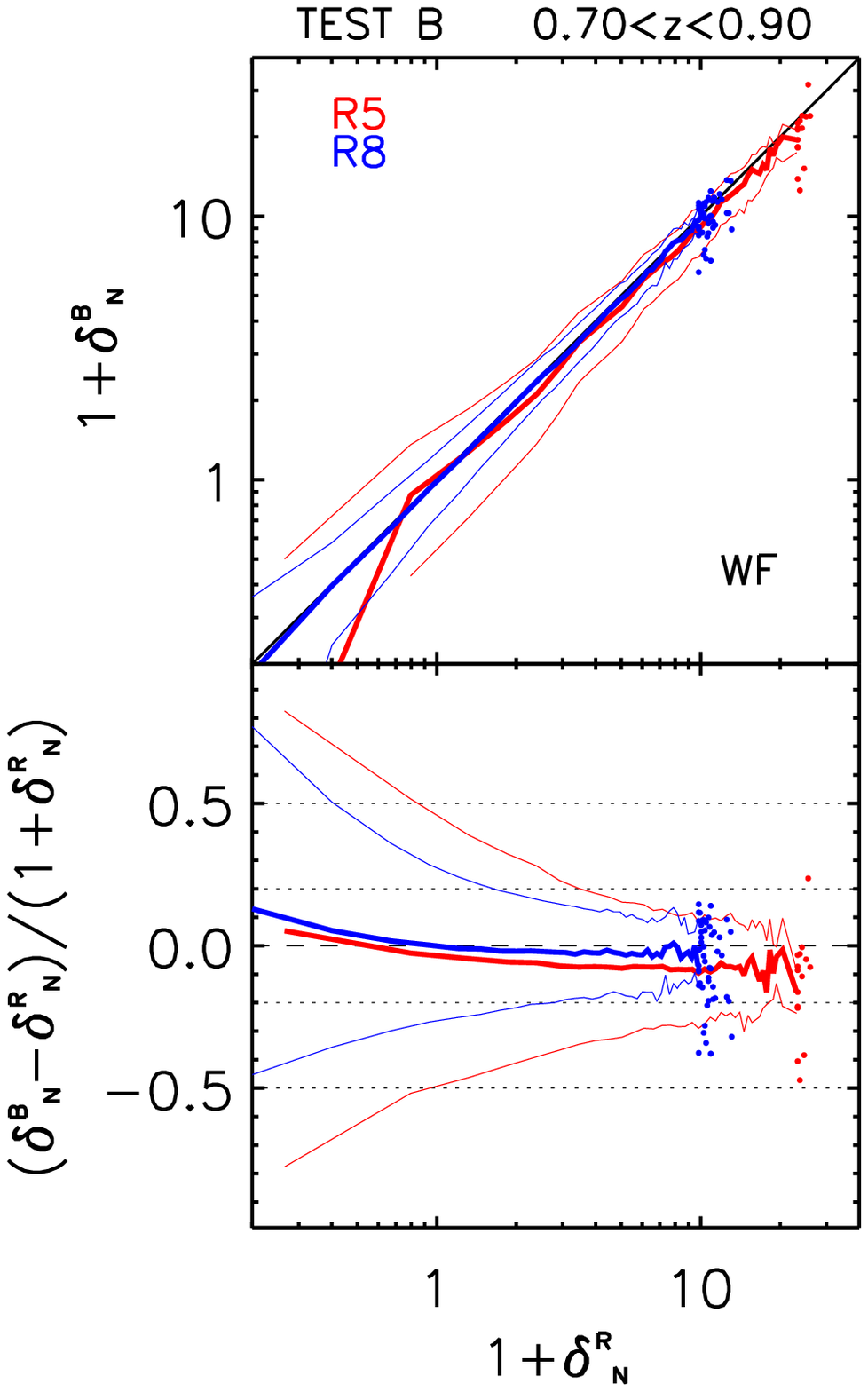}
\includegraphics[width=4.4cm]{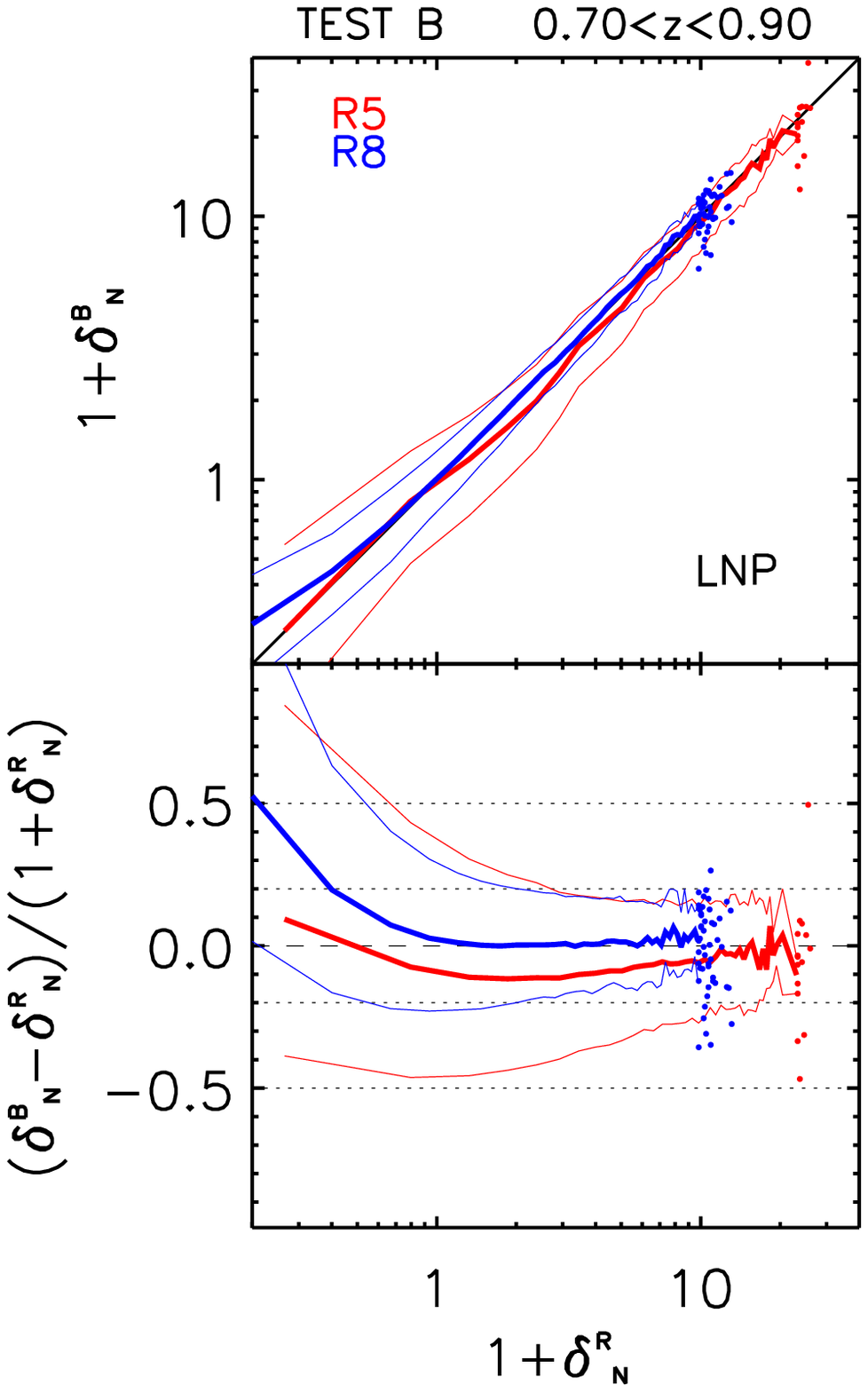}
\includegraphics[width=4.4cm]{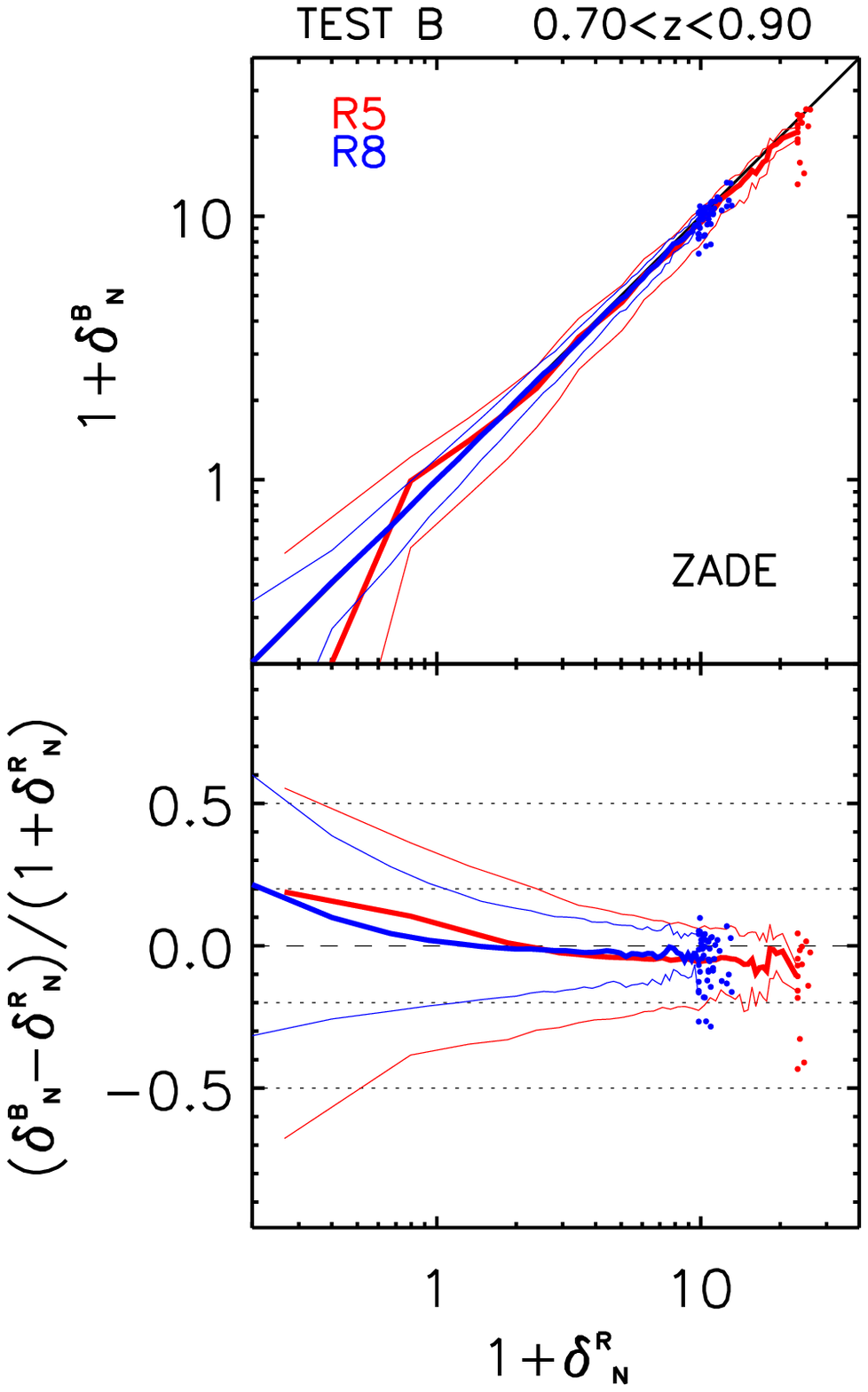}
\includegraphics[width=4.4cm]{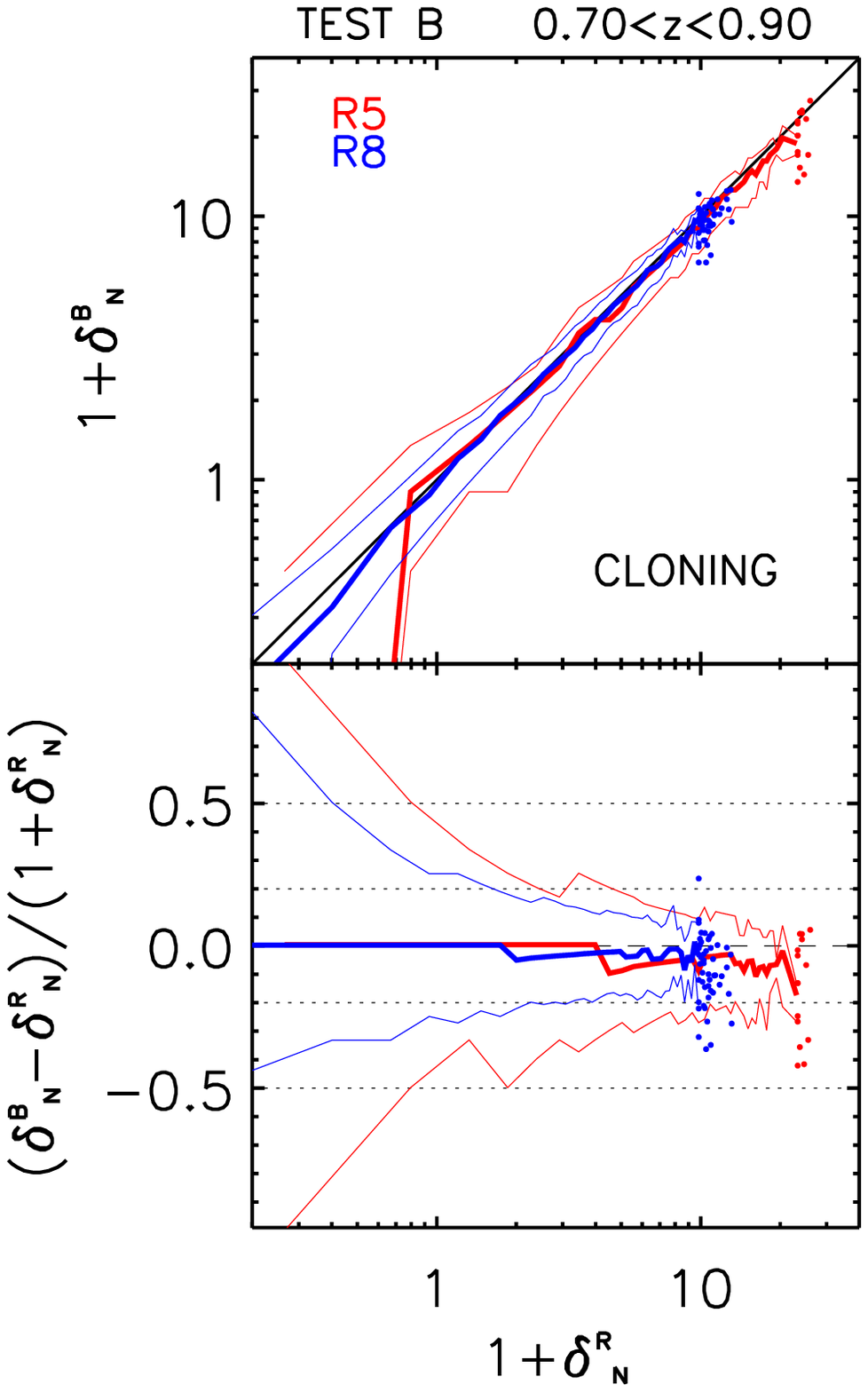}
\caption{As in Fig.~\ref{testD_1940}, but for Test B.}
\label{testB_1940} 
\end{figure*}

\begin{figure*} \centering
\includegraphics[width=4.4cm]{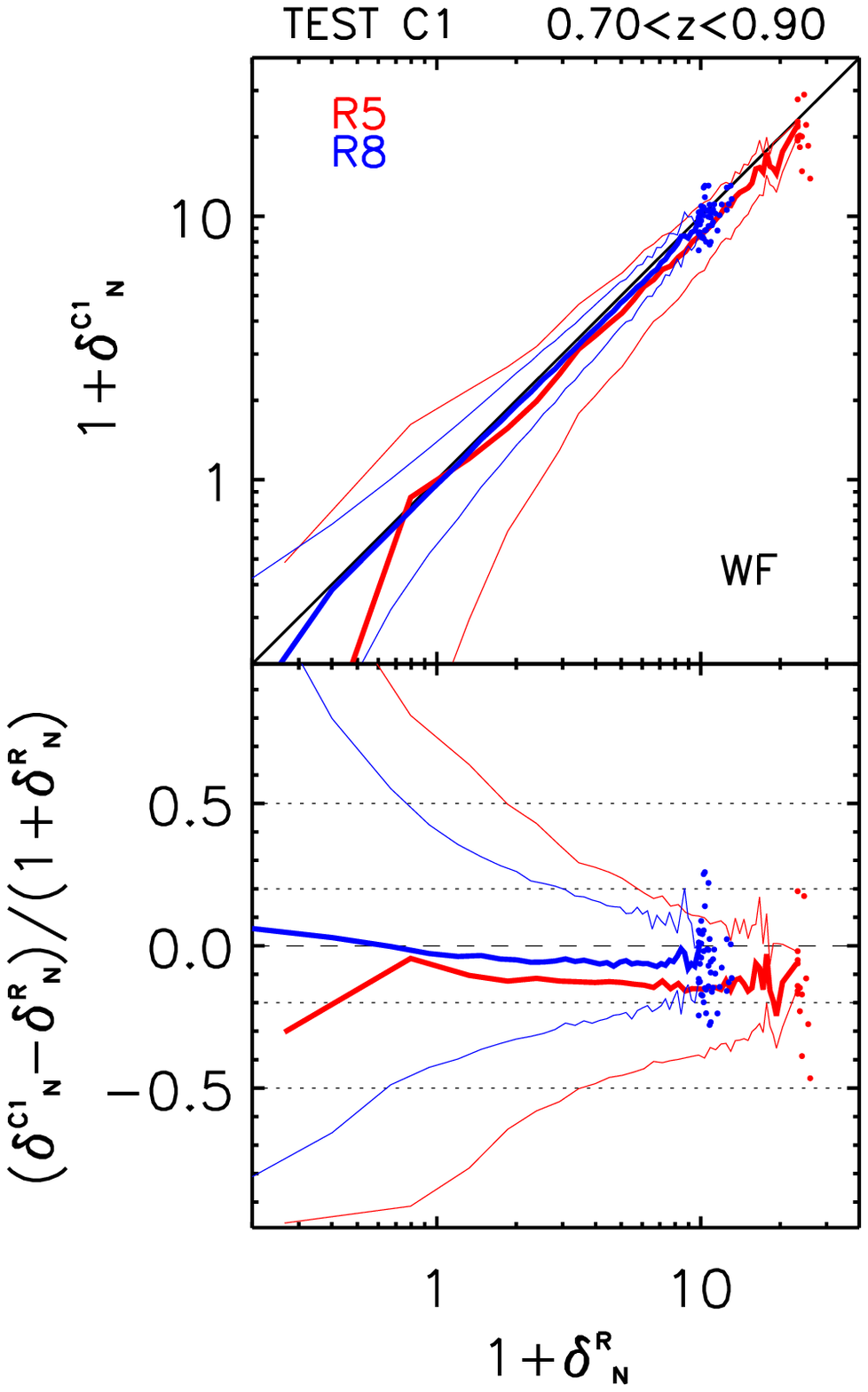}
\includegraphics[width=4.4cm]{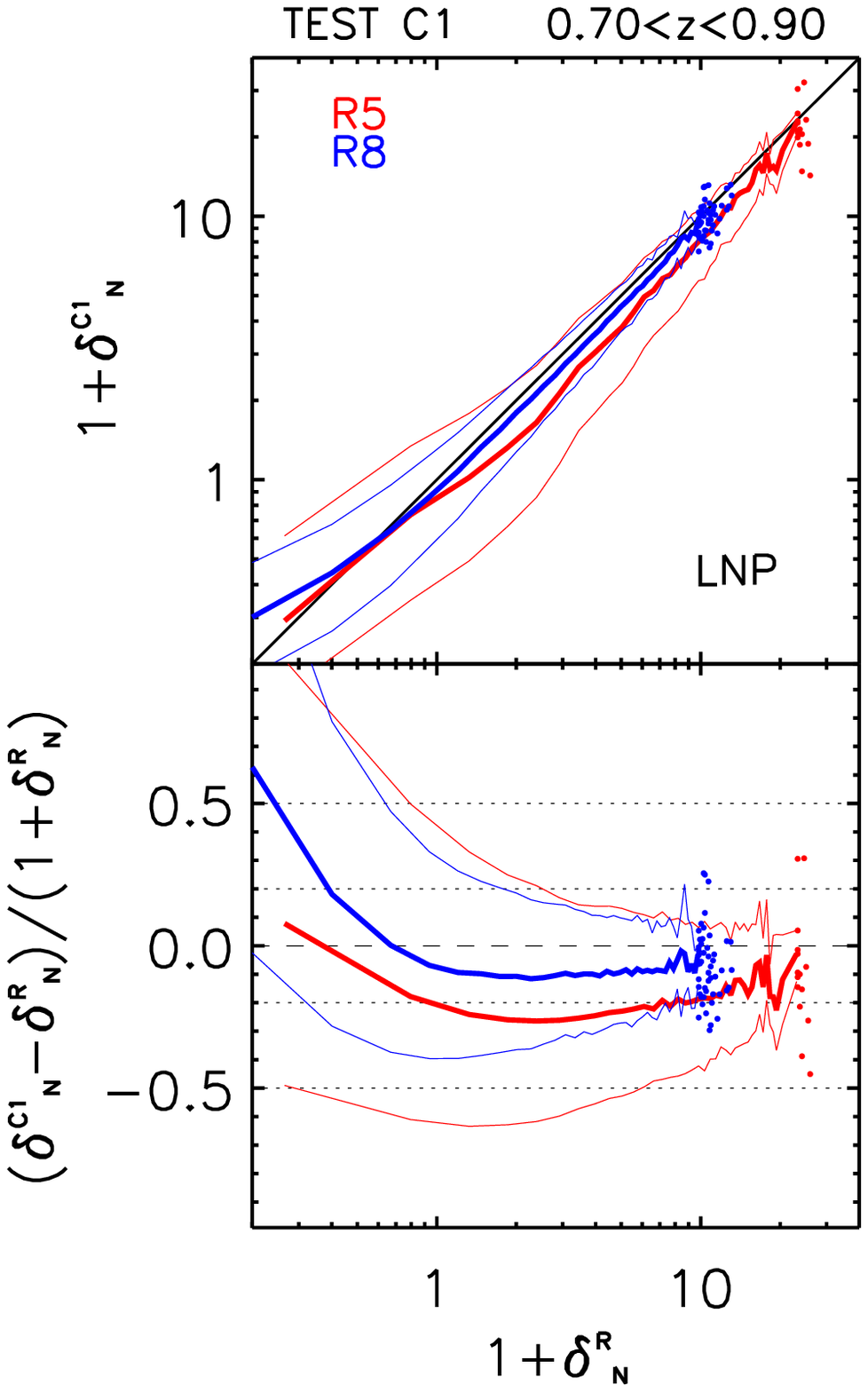}
\includegraphics[width=4.4cm]{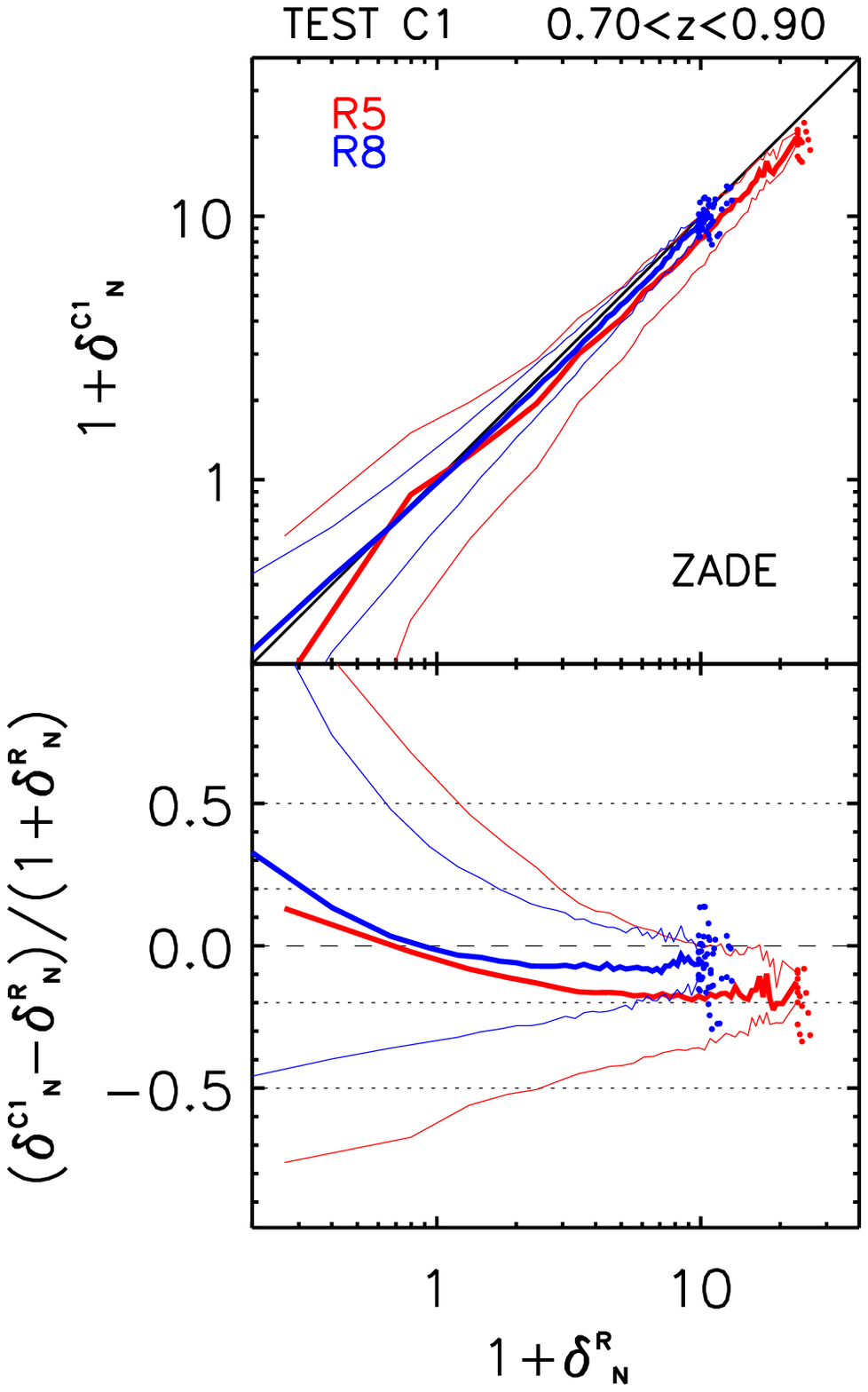}
\caption{As in Fig.~\ref{testD_1940}, but for Test C1. In this test
the cloning method is not used (see text for details).}
\label{testC1_1940} 
\end{figure*}

\begin{figure*} \centering
\includegraphics[width=4.4cm]{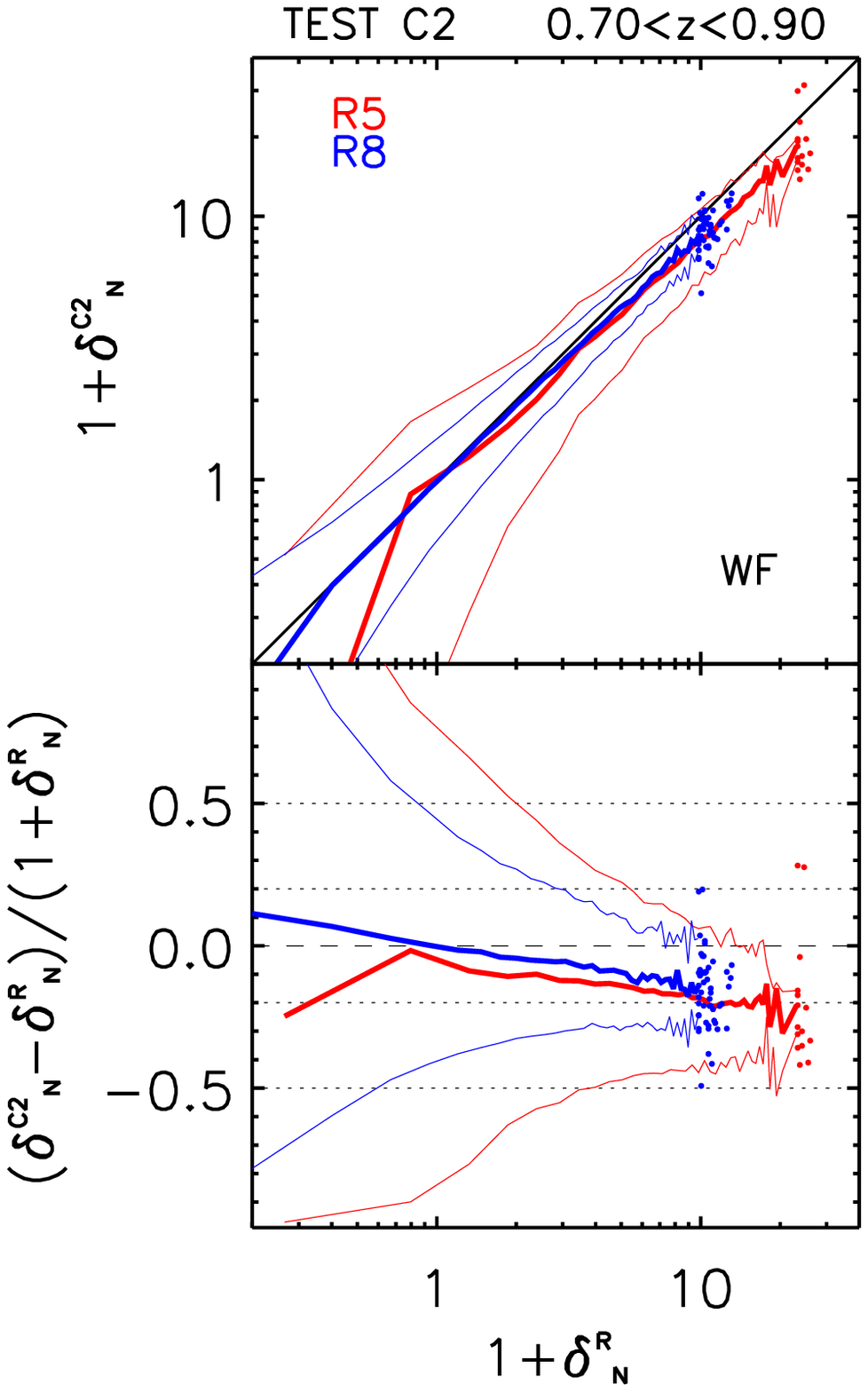}
\includegraphics[width=4.4cm]{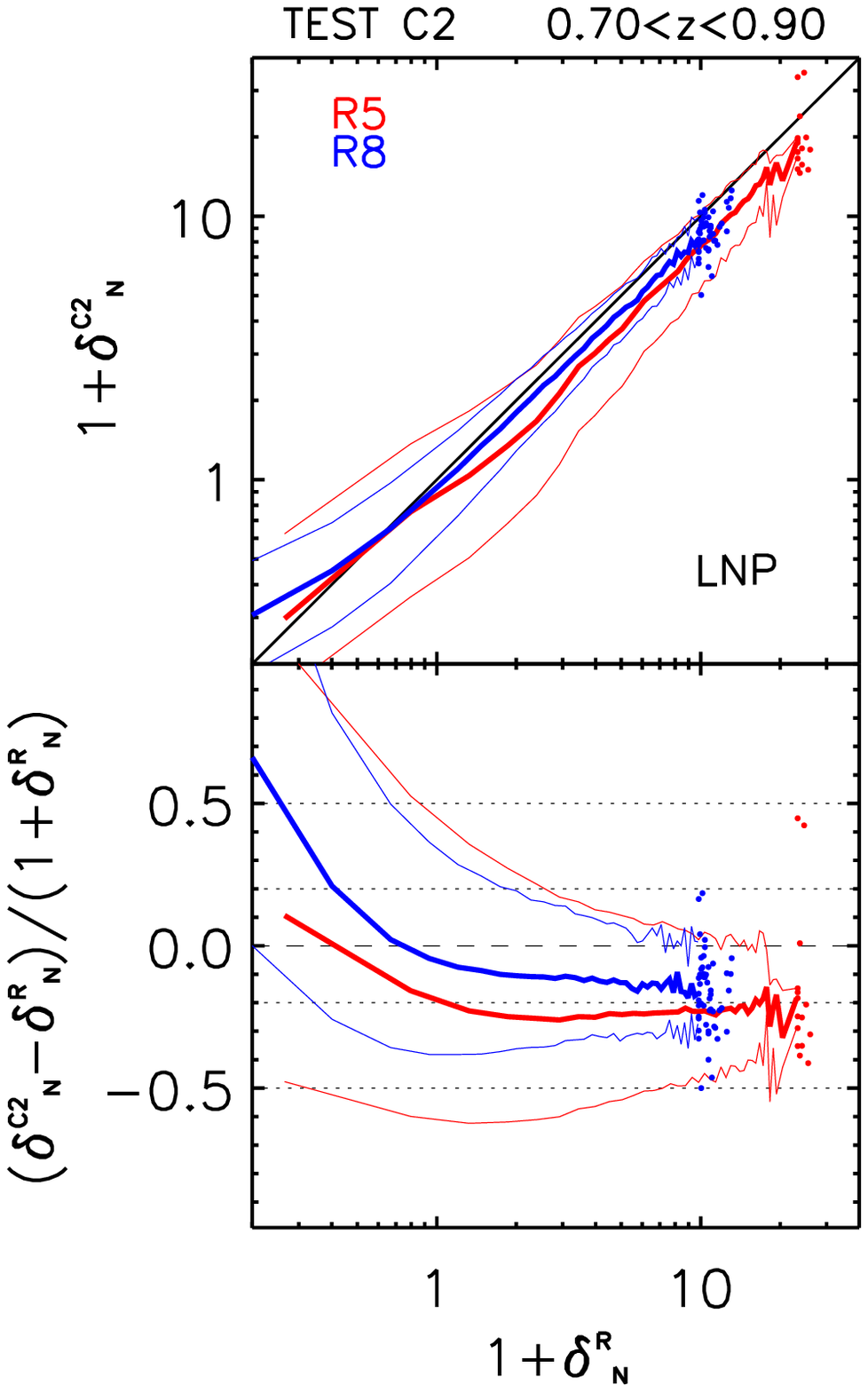}
\includegraphics[width=4.4cm]{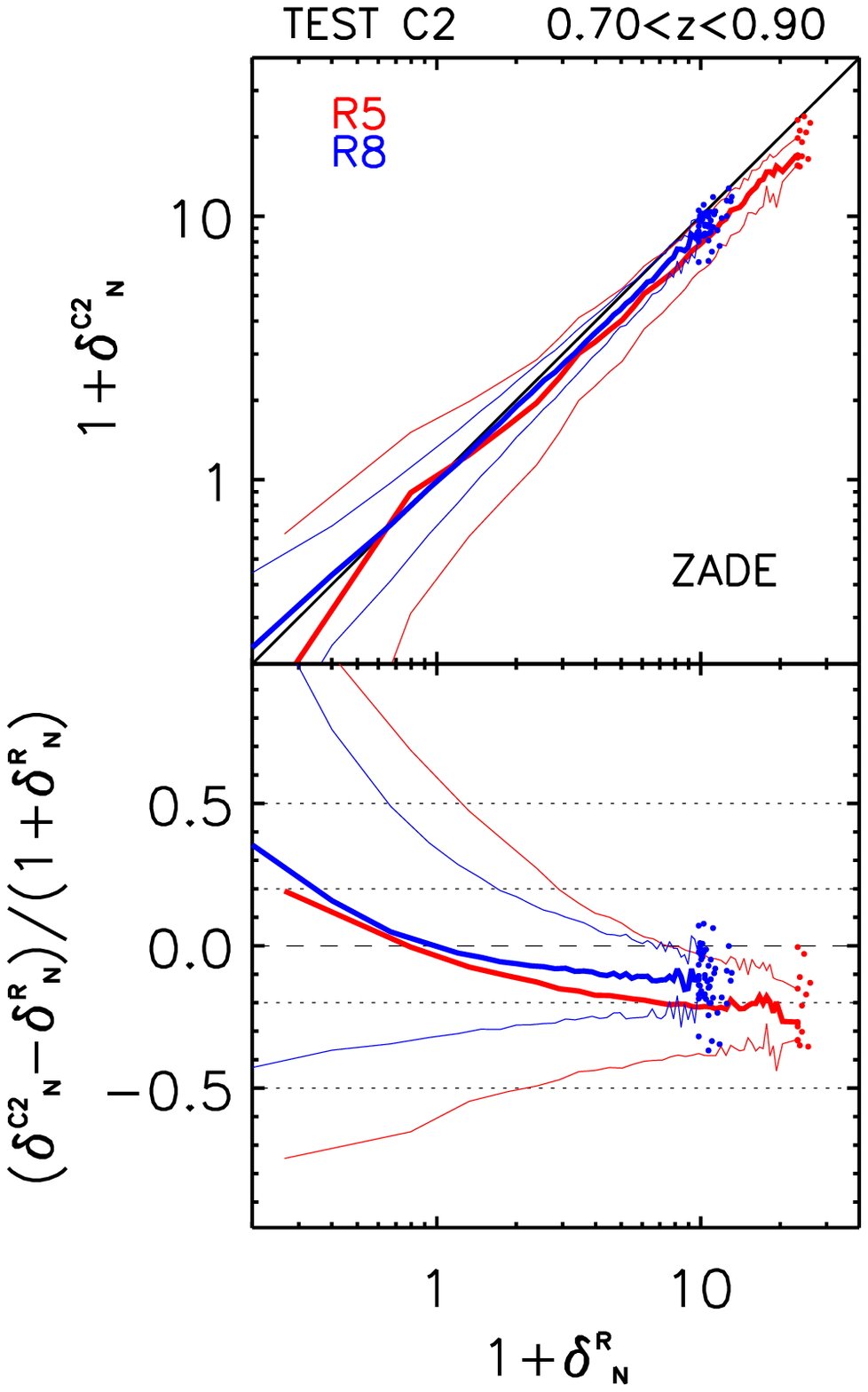}
\caption{As in Fig.~\ref{testD_1940}, but for Test C2. In this test
the cloning method is not used (see text for details).}
\label{testC2_1940} 
\end{figure*}

First, we verified the effects of the spectroscopic redshift error on
the counts in cell when no attempt is made to correct for it (left
panel in Fig.~\ref{testA_1940}): this error does not induce systematic
errors in regions of low/mean density (1+$\delta^R_N \leq 1$), while
it systematically makes us underestimate the counts at high densities
(by $\sim$8\% for $R_5$ and $\sim$3\% for $R_8$, for
$1+\delta^R_N \gtrsim 5$). Random and systematic errors are
significantly smaller for $R_8$ than for $R_5$, i.e. when the size of
the cell is larger than the linear scale associated with the redshift
error. For both radii, the systematic error is comparable to the
scatter.

Applying the WF method to recover the counts in the reference catalogue does
not improve the reconstruction, or even increases (almost double) the
systematic error at high density for $R_5$. Applying the LNP method,
the systematic error for $R_5$ slightly increases at high density
(becoming $\sim10$\%), but disappears for $R_8$ (but at the expense
of a larger random error, that approaches $\sim10$\%).  Both
estimates effectively smooth the density field, and thus the extremes
of the density field are systematically underestimated, especially on
smaller scales. For the WF method, the systematic error is comparable to
the scatter, while for the LNP it is $\sim50$\% smaller, for both
$R_5$ and $R_8$.

Even though our aim is to reconstruct
counts in redshift space, we also compared the counts in Test A with a
reference mock catalogue in real space to check the effect of peculiar
velocities. As expected, with respect to the results of Test A in
redshift space, there is a further under-estimation of high densities,
and the scatter at low densities is larger, because the
cell radii that we use are close to the order of magnitude of peculiar
velocities.

\subsection{Test B}

This test is designed to assess the impact of gaps in the galaxy
distribution.  Our
gaps are a combination of the cross-shaped regions that reflect the
footprint of the VIMOS spectrograph and the empty regions
corresponding to missing quadrants.

We applied all four methods described in Sect.~\ref{methods} to
reconstruct the counts in cells, as shown in Fig.~\ref{testB_1940}.
For all methods, the scatter is larger than found in Test A, while the
systematic error is comparable.  The accuracy of the reconstruction
increases when one considers cells with
$R_8$.

The ZADE method shows the smallest scatter with low systematic error
for both cases $R_5$ and $R_8$. In all cases, the scatter around the
systematic error decreases for higher densities.

In this test, the ZADE method performs better than cloning and
outperforms the WF and LNP reconstructions. We attribute this to the
effective smoothing scale adopted in the WF and LNP methods.
Small-scale structures are lost in the filtered fields even within
quadrants that are sampled at 100\%.  The effect of the smoothing is
greater for $R_5$, and the density is systematically underestimated.
The LNP method shows the largest scatter around the mean, but its
systematic error goes to zero for the highest densities, which does
not happen for the other methods.

It is interesting to notice that for high counts, all the methods tend
to underestimate the counts in the reference catalogue, while all (but
cloning) tend to overestimate it for the lowest counts. The cloning
method is the only one that gives unbiased average counts for the
lowest value of $\delta_N^R$.

\subsection{Test C1 and C2}

With Test C1 we want to assess the effects of a low
sampling rate, homogeneous over the entire VIPERS field. With Test C2,
we implement in the mocks the variation in the sampling rate as a
function of quadrant, keeping the average value as in Test C1. We used
the methods WF, LNP, and ZADE, and the results are shown in
Figs.~\ref{testC1_1940} and \ref{testC2_1940}. We did not use cloning to
correct for low sampling rate for the reasons described in
Sect.~\ref{cloning}.

In the case of Test C1, it is evident from Fig.~\ref{testC1_1940} that
the density in the reference catalogue is overestimated for the lowest
counts and underestimated (up to $\sim20$\% for LNP and ZADE for the
case of $R_5$) for large counts. In general, the scatter is larger than or
comparable to the systematic error, possibly with the exception of the
highest densities, as the scatter decreases for higher densities. For
all three methods, and for both $R_5$ and $R_8$, the systematic
error and the scatter due to low sampling rate are larger than those
due to gaps, and much larger than those due to the spec-z error. The
relative importance of these error sources depends of course on the
survey characteristics. In the case of VIPERS, where the gaps cover
$\sim25$\% of the observed areas while the sampling rate is at the
level of 35\%, the second effect is bound to dominate the error
budget. Spectroscopic redshift errors are marginal on the scales of
the cells considered here. We verified that, keeping the dimension of
the gaps fixed ($\sim25$\%) and progressively reducing the sampling
rate from 100 \% in steps of 10 \%, the systematic error due to low
sampling rate becomes comparable to the one due to gaps (Test B) at a
sampling rate of $\sim60\%$.

Figure \ref{testC2_1940} shows that the results for Test C2 are only
slightly worse than those of Test C1, for both the amplitude of the
systematic error and the scatter. This confirms that in VIPERS the
major source of uncertainty in counts in cells is the low ($\sim35$\%)
sampling rate.

\end{document}